\documentclass[12pt]{article}
\newcommand{\blind}{1}
\usepackage{graphicx}
\usepackage{amssymb}

\usepackage{natbib}
\usepackage{url}
\usepackage{multirow}
\usepackage {booktabs}
\usepackage{subcaption}
\usepackage{xr-hyper}
\makeatletter
\newcommand*{\addFileDependency}[1]{
	\typeout{(#1)}
	\@addtofilelist{#1}
	\IfFileExists{#1}{}{\typeout{No file #1.}}
}
\makeatother
\newcommand*{\myexternaldocument}[1]{
	\externaldocument{#1}
	\addFileDependency{#1.tex}
	\addFileDependency{#1.aux}
}

\usepackage[colorlinks,linkcolor=blue,anchorcolor=black,citecolor=blue,urlcolor=black]{hyperref}
\usepackage{amsthm, amsmath, amsfonts, amssymb,mathrsfs}
\theoremstyle{plain}
\newtheorem{prop}{Proposition}
\newtheorem{theorem}{Theorem}
\newtheorem{lemma}{Lemma}

\theoremstyle{definition}
\newtheorem{assumption}{Condition}

\newtheorem{remark}{Remark}
\newtheorem{example}{Example}
\usepackage[plain,ruled,noend]{algorithm2e}
\usepackage{multirow, multicol}
\usepackage{float, subcaption}
\usepackage{bm, enumerate, xcolor}

\graphicspath{{figures/}}

\newcommand{\T}{{ \mathrm{\scriptscriptstyle T} }}

\newcommand{\E}{\mathbb{E}}
\newcommand{\g}{\bm g}
\newcommand{\R}{\mathbb{R}}
\newcommand{\C}{\mathcal{C}}
\newcommand{\pr}{\mathbb{P}}
\newcommand{\I}{\mathbb{I}}
\newcommand{\X}{\mathbf{X}}
\newcommand{\XY}{\mathbf{X},Y}
\newcommand{\coef}{\mathcal{B}}
\newcommand{\G}{\mathcal{G}}
\newcommand{\tX}{{\mathbf{X}}}
\newcommand{\x}{\bm{x}}

\newcommand{\sgn}{{\rm{sgn}}}
\newcommand{\var}{\mathrm{var}}
\newcommand{\tr}{\mathrm{tr}}
\newcommand{\pilot}{\mathrm{pilot}}
\newcommand{\itm}{\mathcal{T}}

\newcommand{\sub}{\mathrm{sub}}
\newcommand{\indexset}{\mathcal{S}}
\newcommand{\fR}{\mathfrak{R}}
\newcommand{\fu}{\mathcal{F}_{\rm union}}
\newcommand{\fp}{\mathcal{F}_{r_0}^1}

\newcommand{\score}{\Psi}
\newcommand{\V}{\var_a(\Tilde{\bm\theta})}

\newcommand{\fn}{\mathcal{F}_n}
\newcommand{\pirb}{\pi}
\newcommand{\res}{\bm{\aleph}}

\myexternaldocument{supp}

\begin{document}
\def\spacingset#1{\renewcommand{\baselinestretch}%
{#1}\small\normalsize} \spacingset{1}

\if1\blind
{
  \title{\bf Multi-resolution subsampling for large-scale linear classification }

  \author{Haolin Chen$^1$, Holger Dette$^2$ and Jun Yu$^1$\thanks{
  		All the authors have equally contributed to this work. Authors are listed alphabetically by their surname. The corresponding author is Jun Yu (yujunbeta@bit.edu.cn).  }\hspace{.2cm}\\
  	1: School of Mathematics and Statistics, Beijing Institute of Technology\\
  	2: Ruhr-Universit\"{a}t Bochum, Fakult\"{a}t f\"{u}r Mathematik\\
  	}
  \maketitle
} \fi

\if0\blind
{
  \bigskip
  \bigskip
  \bigskip
  \begin{center}
  {\LARGE\bf Multi-resolution subsampling for large-scale linear classification}
\end{center}
  \medskip
} \fi

\bigskip

\bigskip
\begin{abstract}

Subsampling is one of the popular methods to balance statistical efficiency and computational efficiency in the big data era. Most approaches aim at selecting informative or representative sample points to achieve good overall information of the full data.  The present work takes the view that sampling techniques are recommended for the region we focus on and summary measures are enough to collect the information for the rest according to a well-designed data partitioning. We propose a multi-resolution subsampling strategy that combines global information described by summary measures and local information obtained from selected subsample points. We show that the proposed method will lead to a more efficient subsample based estimator for general large-scale classification problems. Some asymptotic properties of the proposed method are established and connections to existing subsampling procedures are explored. Finally, we illustrate the proposed subsampling strategy via simulated and real-world examples.

\end{abstract}
\noindent%
\vfill

\newpage
\spacingset{1.45}\par
\section{Introduction}\label{sec:intro}

Classification is one of the main tasks in big data analysis.
Extensive classification algorithms have been developed in statistics and machine learning fields. 
Readers may refer to \cite{bartlett2006convexity}, \cite{Hastie09ESL}, and \cite{fan2020statistical} for a more detailed introduction and applications.
Despite statistical guarantees that the excess risk will converge to zero as the sample size increases,  it is still a challenge to train a classifier on a large-scale dataset due to the huge computational costs. 
This is also true even for the linear classification problem.
For example, the computational cost for large-scale logistic regression \citep{mccullagh1989generalized} and distance weighted discrimination \citep[DWD,][]{Marron2007DWD} is $O(nd^2)$  where $n$ is the sample size and $d$ the dimension of the predictor.
Moreover, the computational cost increases to $O(n^2d)$ when the support vector machine \citep[SVM,][]{Cortes1995SVM} is adopted.

Handling classification problems on large-scale datasets 
results in a huge computational complexity and ever-increasing demand for high computing power and possible carbon emissions.
To tackle the challenges of computing resources together with the environmental impact of statistical learning algorithms, 
data scientists have to balance a trade-off between statistical accuracy and computational costs \citep{10.1145/3381831}.
One of the ubiquitous solutions is subsampling (or subdata selection) and several authors have demonstrated that subsampling can balance computational complexity (or resource) and statistical efficiency in many real-world applications.
For example, \cite{NEURIPS2021_a51c896c} analyze the click-through rate for ByteDance Apps via nonuniform negative subsampling techniques and \cite{wang2024Sequential} use subsampling to predict customer churn for a securities company in China.

Two kinds of subsampling techniques are routinely adopted to handle massive data.
One approach is to find the data points that enable the researchers to better explore the entire dataset. Typical examples include uniform subsampling (also known as simple random sampling), space-filling-design-based
subsampling \citep{zhao2018efficient,Meng2020lowcon,shi2021model,zhang2023model},  
 distribution representative points 
\citep{mak2018support,Zhang2022an}, and  prediction inference based subdata selection \citep{zou2024prediction}.
Other works exploit the model information and focus on finding data points in the sample that yield more precise parameter estimates. 
For instance, \cite{ma2015} and \cite{ma2020} use the high leverage-score points to improve the estimation efficiency for large-scale linear regressions.
Local case-control sampling \citep{fithian2014local} is proposed to achieve a better estimation efficiency for logistic regression. 
Design inspired subsampling include IBOSS \citep{wang2019information} and OSS \citep{wanglin2021} for linear regression, OSMAC \citep{wang2018optimal,ai2021optimal}, IBOSS  for logistic regression \citep{cheng2020information} and ODBSS  for generalized linear models \citep{2023arXiv230616821D}. Similar ideas have been extended to more general or complex models. Important works include but are not limited to \cite{wang2021optimal},   \cite{Han2023Leverage}, \cite{zhang2024independence}.
We also refer to \cite{yu2023review} for a more comprehensive review of the state of the art.

To the best of our knowledge, most existing works focus on drawing inferences from the selected subsample points.
This is quite natural in the classical statistical literature because statisticians only acquire the information from the survey data.
However, subsampling provides a valuable chance for data scientists to combine data collection and data analysis.
More concretely, data scientists can not only query the individual level information, i.e., the sample points itself, but can also query some group-level or finite population (or full data) level information, such as some summary measures that can be easily obtained by scanning the entire data once during the sampling step with negligible additional computational costs compared to most existing optimal subsampling procedures. 
A single data point provides high-resolution local information but fails to reflect the global information of the entire dataset.
On the contrary, summary statistics provide low-resolution local information but they certainly contain some useful global information.   Therefore, it is of interest to investigate the amount of improvement in estimation efficiency by aggregating the selected subdata and summary measures.

\textbf{Our contribution:} In this article, we present a multi-resolution optimal subsampling (MROSS), which combines the summary measures and selected subdata points, for the construction of a statistically and computationally efficient classifier in a general large-scale linear classification problem.
The improvement of the proposed method comes from the following two aspects.
Firstly, it uses the information from the selected subdata and the summary measures to collect the information from the unselected data points (see Section~\ref{sec:RB} for details).
Secondly, we carefully extricate ourselves from the common point of view that subsampling should reflect the information of the entire data.
According to a well-designed data partitioning, we propose to use sampling techniques for the region we focus on and to use summary measures to collect the information for the rest.
This action provides an efficient way to make the selected data concentrated on the informative region (see Section~\ref{sec:tail} for details) and reduce the risk of selecting data points with a small inclusion probability. As a consequence, the resulting estimators become more efficient and stable compared to other approaches.

A theoretical analysis of the asymptotic properties of the new estimation procedures is presented, and it is demonstrated that the proposed method outperforms the existing optimal subsampling strategy which uses only the selected subdata (see Theorems~\ref{thm:RB-1}-\ref{thm:RB} for more details). Moreover, the additional computational costs due to the calculation of summary measures are extremely small compared with the current optimal subsampling methods. Finally, the theoretical advantages of the new method are confirmed in a simulation study and three data examples.


\section{Problem setups}\label{sec:preliminaries}

  \def\theequation{1.\arabic{equation}}	
  \setcounter{equation}{0}

\subsection{Linear classification}
Consider the binary classification problem where $(\X_1,Y_1),\ldots,(\X_n,Y_n)$ is a sample of 
independent, identically distributed (i.i.d.)~random observations with 
$d$-dimensional covariates $\X_i$ taking values in a subset $\mathcal{X}$ of $\mathbb{R}^d$ and   $Y_i\in\{-1,+1\}$. The classification problem aims to define a function
$\C:\mathcal{X}\to\{-1,+1\}$ from the input space to the output space, which yields a classification rule for any point $\X$ in the input space $\mathcal{X}$.  We also assume 
that $\X$ contains a constant component (without loss of generality taken as the first component of $\X$), which corresponds to the intercept term in the linear predictor $\tX^\T \bm \theta $, where $\bm \theta$ is a $d$-dimensional unknown parameter of interest. 

The two main ingredients of a classification algorithm are a loss function, say $\phi$, and a hypothesis space $\mathcal{H}$.
For the linear classification problem,   $\mathcal{H}$ consists of the class of linear functions of the form  $\X^\T\bm\theta$ with the unknown parameter $\bm\theta$, which needs to be estimated from the data.
The resultant classifiers have the form $\C(\X ) =\sgn(\X^\T\bm\theta)$, where $\sgn(f)=\I(f\ge 0)-\I(f< 0)$ denotes the sign of $f$ and  $\I(\cdot)$ is the indicator function.
To ease the presentation, we restrict ourselves to binary linear classification problems in this work.
 Let $\eta(\X)=\pr(Y=+1|\X)$ denote the conditional probability that $Y$ is in class ``+1" given $ {\X}$. The $\phi$-risk of a given classifier is defined by
 \begin{equation}\label{eq:risk}
R_\phi(\bm\theta)=\E \big \{ \phi(Y\tX^\T\bm\theta) \} =\E\{\eta(\X)\phi(\tX^\T\bm\theta)+(1-\eta(\X))\phi(-\tX^\T\bm\theta)\}.
 \end{equation}
 When the full sample $\fn:=\{(\X_i,Y_i)\}_{i=1}^n$ is observed, the  parameter $\bm\theta$ can be estimated by minimizing the empirical loss
 \begin{equation}\label{eq:hat-risk}
     {\hat{R}}_{\phi}(\bm\theta)=\frac{1}{n}\sum_{i=1}^n\phi(Y_i\tX_i^\T\bm\theta).
 \end{equation}
 To end this subsection, it is worth mentioning that $\phi$ is usually selected as a surrogate function of the $0$-$1$ loss function  $\I(Y\neq \sgn(\tX^\T\bm\theta))$.
 Therefore, we assume that $\phi$ is classification-calibrated throughout this paper, i.e., $\phi$ is a convex non-increasing function with $\dot{\phi}(0)<0$ and $\inf_{z\in\R}\phi(z)=0$ \citep{bartlett2006convexity}.
Typical examples of $\phi$ include  
\begin{align}
    \label{det1} \phi(z) &=\log(1+\exp(-z))~~~~~~~~~~~~~~\text{(logistic loss for logistic regression)} \\
      \label{det2} \phi(z) & =(1-z)_+  =\max\{1-z,0\}
       ~~~\text{(hinge loss )}  \\ 
       \phi(z)& =(1-z)_+^2 ~~~~~~~~~~~~~~~~~~~~~~~~~\text{(squared hinge loss for SVM)} 
        \label{det3} \\
         \label{det4} 
    \phi(z) & =z^{-1}\I(z\ge \gamma)+(2/\gamma-z/\gamma^2)\I(z< \gamma) ~~~(\gamma>0) \\
    & \nonumber ~~~~~~~~~~~~~~~~~~~~~~~~~~~~~~~~~~~~~~~\text{(DWD loss for distance weighted discrimination)} 
\end{align} 
Since $\phi$ is convex, the risk minimizer $\bm\theta_t$ will naturally satisfy systems of equations 
\begin{align}
 \label{det8}   
\score_\phi(\bm\theta)=\E
\Big ( {\partial R_\phi(\bm\theta) \over \partial\bm\theta } \Big ) 
=\E \big ( \dot{\phi}(Y\tX^\T{\bm\theta})Y{{\tX}} \big )=\bm 0.
\end{align}
In many cases the  minimizer 
$\bm\theta_t$ is estimated by the minimizer 
$\hat{\bm\theta}$ of the empirical loss  in \eqref{eq:hat-risk},
 which is obtained as the solution of the equation
 \begin{align}
     \label{det9}
\hat{\score}_{\phi} (\bm\theta ) =n^{-1}\sum_{i=1}^n\dot{\phi}(Y_i\tX_i^\T{\bm\theta})Y_i{{\tX_i}}=\bm 0.
 \end{align}
 Therefore, $\hat{\bm\theta}$  is also called  Z-estimator \citep{van1998asymptotic}.

 \subsection{Ordinary subsampling framework}
 When data size $n$ is huge, learning the classifier by optimizing \eqref{eq:hat-risk} is complicated. 
To balance the computational costs and statistical efficiency, subsampling is usually regarded as one of the feasible solutions. In the following, we briefly recap the ordinary subsampling framework.
 We denote  by  $\bm\pi=(\pi_1,\ldots,\pi_n)^\T$ a vector of inclusion probabilities, where $\sum_{i=1}^n\pi_i=r$ and $r$ is the subsample size.
 If Poisson sampling \citep{sarndal2003model} is adopted, researchers play a Bernoulli trial with probability $\pi_i$ to decide whether the $i$th observation $(\X_i,Y_i ) $ is selected or not.
 Let $\delta_i\in\{0,1\}$ be an indicator identifying if the $i$th data point is selected ($\delta_i = 1$).
 The subsample based classifier is given by $\Tilde{\C} (\X) =\sgn(\tX^\T\tilde{\bm\theta})$, where $\tilde{\bm\theta}$ is the minimizer of the following subsample based empirical loss 
 \begin{equation}\label{eq:subloss}
     \Tilde{R}_\phi (\bm\theta)=\frac{1}{n}\sum_{i=1}^n\frac{\delta_i}{\pi_i}\phi(Y_i\tX_i^\T\bm\theta).
 \end{equation}
 To derive the asymptotic properties of the estimator $\tilde{\bm\theta}$ we make the following regularity assumptions.

 \begin{assumption}\label{ass:1}
The $\bm\theta $ varies in an open convex set $ \Theta_1 \subset  \mathbb{R}^d$.  Let  $\Theta\subset \Theta_1$
denote a compact subset of  $\Theta_1$.   
The parameter $\bm\theta_t \in \Theta$  is the unique minimizer of the risk function $R_\phi(\bm\theta)$ defined in \eqref{eq:risk} over the set $\Theta_1$.
Further assume that $\E \big ( \sup_{\bm\theta\in\Theta_1}\phi^4(Y\tX^{\T}\bm\theta) \big )  < \infty$ 
 and assume that there exists  
at least one non-vanishing slope parameter of $\bm\theta_t$. 
 \end{assumption}
 \begin{assumption}\label{ass:2}
     The loss function $\phi (\cdot)$ is twice   differentiable almost everywhere and $\phi(\cdot)$, and $\dot{\phi}(\cdot)$  are Lipschitz continuous on $\R$, i.e., $|\phi(t_1)-\phi(t_2)|\le L_0|t_1-t_2|$, and
     $|\dot\phi(t_1)-\dot\phi(t_2)|\le L_1|t_1-t_2|$, 
     hold for some constant $L_0,L_1$.
     Further assume that the function, $\bm \theta \to \ddot{\phi}(Y\tX^\T\bm\theta)$
     is almost surely Lipschitz continuous in a sufficiently small 
neighborhood
    $\mathcal{N}_{\upsilon}(\bm\theta_t)$  of $\bm\theta_t$ with a  radius $\upsilon >0$, that is:  
     $|\ddot{\phi}(Y\tX^\T\bm\theta_1)-\ddot{\phi}(Y\tX^\T\bm\theta_2)|\le L_2(\XY)\|\bm\theta_1-\bm\theta_2\|$ for any $\bm\theta_1,\bm\theta_2\in \mathcal{N}_{\delta}(\bm\theta_t)$, where  $\E \big( L_2^4(\XY)\big) <\infty$ almost surely.
 \end{assumption}
 \begin{assumption}\label{ass:3} 
The matrices $\E\big(\dot{\phi}(Y\tX^\T\bm\theta)^2{\tX}{\tX}^\T\big)$ and  
 \begin{align}   
-  H(\bm\theta):=  - { \partial^2 R_\phi(\bm\theta) \over \partial\bm\theta^\T\partial\bm\theta} =\E\big(\ddot{\phi}(Y\tX^\T{\bm\theta}){\tX}{\tX}^\T\big).\label{det6}
 \end{align}
   are     uniformly positive definite matrices, that is 
     	\begin{align*}
	&0<\inf_{\bm\theta\in\Theta_1}\lambda_{\textrm{min}}\big(\E\big(\dot{\phi}^2(Y\tX^\T\bm\theta){\tX}{\tX}^\T\big)\big)\le \sup_{\bm\theta\in\Theta}\big\|\E\big(\dot{\phi}^2(Y\tX^\T\bm\theta){\tX}{\tX}^\T\big)\big\|_s
	<\infty,\\
	& 0<\inf_{\bm\theta\in\Theta_1}\lambda_{\textrm{min}}(-H({\bm\theta}))\le \sup_{\bm\theta\in\Theta}\|-H({\bm\theta})\|_s<\infty,
	\end{align*}
 where 
 $\lambda_{\min}(A)$ and $\|A\|_s$ denote the smallest eigenvalue and the spectral norm of the matrix $A$, respectively.
 \end{assumption}

 \begin{assumption}\label{ass:5} There exists a function $\Pi: \mathcal{X}^d \times \{-1,1\} \longrightarrow \mathbb {R}$ such that the 
     inclusion probabilities have the form
     \begin{equation}
       \label{det301}  \pi_i=\frac{r\Pi(\X_i,Y_i)}{\sum_{l=1}^n \Pi(\X_l,Y_l)},
     \end{equation}
     where $\Pi(\X_i,Y_i)\ge 0$ for all $i = 1, \ldots, n$,  $\E\Pi^4(\X,Y)<\infty$, and $(r/n)\E\big ( \max_{i=1}^n \Pi(\X_i,Y_i) \big) =o(1)$ as $r,n\to\infty$.
 \end{assumption}
 \begin{assumption}\label{ass:6}
     The following moment conditions hold: 
     (i)$\E\|\X\|{^4} <\infty$, (ii)$\E {\|\tX\|^4}/{\Pi(\X,Y)}=o(r)$. 
 \end{assumption}

Condition~\ref{ass:1} is required to guarantee consistency of M-estimation \citep[see][for example]{NEWEY19942111}.
Condition~\ref{ass:2} are naturally satisfied for the logistic regression, DWD, and some smoothed hinge loss given in \cite{Luo2021learning}. 
Condition~\ref{ass:3} guarantees convexity of the population risk \eqref{eq:risk}. 
Condition~\ref{ass:5} is naturally satisfied by many subsampling strategies, such as uniform subsampling, local case control sampling \citep{fithian2014local}, and OSMAC \citep{wang2018optimal}. The moment assumptions on $\Pi(\XY)$ can be achieved by proper scaling of $\Pi(\XY)$. The assumption $(r/n)\E \big ( \max_{i=1}^n \Pi(\X_i,Y_i) \big ) =o(1)$ as $r,n\to\infty$ guarantees that with high probability all  probabilities 
$\pi_i$  in  \eqref{det301}  are contained in the interval $[0,1]$.
Moment assumptions as in 
Condition~\ref{ass:6} are quite common in the literature; see, for example, \cite{wang2018optimal,yu2020quasi} among many others. 
They are used to establish the consistency and asymptotic normality of the subsample based estimator.

It is shown in \cite{LIN2004Anote} that for the classification-calibrated loss $\phi(\cdot)$ the minimizer $\bm\theta_t$ of the
loss function \eqref{eq:risk}
leads to the Bayes optimal classification rule, i.e., $\sgn(\bm x^\T\bm\theta_t)=\sgn(\eta(\bm x)-1/2)$. Therefore,  a desired property of an  estimator $\tilde{\bm\theta}$ 
from the subsample is that it 
should be close to the risk minimizer $\bm\theta_t$. 
Consequently, we need to study the distance between  $\tilde{\bm\theta}$  and  $\bm\theta_t$.
The following lemma presents a Bahadur-type representation of $\tilde{\bm\theta}$ and is proved in the online supplement.

 \begin{lemma}[Bahadur-type representation]\label{lem:bahadur}
 Under Conditions ~\ref{ass:1}--\ref{ass:6}, as $n,r\to\infty$, it holds
     \begin{equation}\label{eq:bahadur}
         \tilde{\bm\theta}-\bm\theta_t= -H^{-1}(\bm\theta_t)\sum_{i=1}^n \delta_i \frac{Y_i\dot{\phi}(Y_i\tX_i^\T\bm\theta_t)}{n\pi_i}\tX_i +o_P(\|\tilde{\bm\theta}-\bm\theta_t\|). 
     \end{equation}
 \end{lemma}

Therefore, the leading term of the asymptotic variance of $\Tilde{\bm\theta}$ can be obtained from the variance of the first term and the right hand side of \eqref{eq:bahadur}, that is
\begin{align}
\label{det500}
\text{var} \Big ( H^{-1}(\bm\theta_t)\sum_{i=1}^n \delta_i \frac{Y_i\dot{\phi}(Y_i\tX_i^\T\bm\theta_t)}{n\pi_i}\tX_i \Big ) 
\end{align}

To end this section, Proposition~\ref{pro:1} gives the explicit form of the limit of $\var(\Tilde{\bm\theta})$, which will guide us in finding the optimal subsampling probabilities. 

\begin{prop}[Asymptotic variance]\label{pro:1}
    Under Conditions~\ref{ass:1}--\ref{ass:6}, as $n,r\to\infty$ with $r/n\to 0$, it holds that 
    \begin{equation} \label{det5}
r \cdot \var (\Tilde{\bm\theta}) \to \var_a(\Tilde{\bm\theta})
:=\E \big( \Pi(\X,Y) \big) H^{-1}(\bm\theta_t)\E\Big ( \frac{\dot{\phi}^2(Y\tX^\T\bm\theta_t)}{\Pi(\X,Y)}\tX\tX^\T\Big )  H^{-1}(\bm\theta_t).
    \end{equation}
\end{prop}

\section{Multi-resolution subsampling}\label{sec:method}
  \def\theequation{2.\arabic{equation}}	
  \setcounter{equation}{0}

In this section, we first present the optimal subsample rule based on the $A$- and $L$-optimality criterion for the ordinary subsampling framework.
Later, we will show how the corresponding estimators can be improved by summary statistics.
Therefore, the method proposed in this section is called \textit{multi-resolution optimal subsampling} since it combines the selected subdata points and summary statistics for the unsampled data.

\subsection{Optimal subsampling probability}
The key to an efficient subsampling strategy is to assign a higher probability to the more informative data points.
Note that $\Tilde{\bm\theta}$ is asymptotically unbiased by Lemma~\ref{lem:bahadur}.
 Therefore, natural to design the sampling function $\Pi(\XY)$ such that the (asymptotic) variance of the resulting estimator $\Tilde{\bm\theta}$ or the variance of some linear combination of the parameters, say $L^\T\Tilde{\bm\theta}$, is minimized, where $L$ is a user-specified matrix with $d$ rows. This is also known as $A$- and $L$-optimal subsampling. In practice, the matrix $L$ is usually selected as $H(\bm\theta_t)$ to simplify the calculation of the resulting optimal inclusion probabilities.
To be precise, $A$- and $L$-optimal subsampling aim for minimizing $\tr(\V)$ and $\tr(H(\bm\theta_t)\V H(\bm\theta_t))$ with respect to the choice of the function $\Pi$, respectively, where the matrices $\V$ and $H(\bm\theta_t)$ are defined in \eqref{det5} and \eqref{det6}, respectively.

\begin{prop}[Optimal sampling probability]\label{pro:2}  Let the 
 matrices  $H(\bm\theta_t)$ and $\V$ are defined in  \eqref{det6} and \eqref{det5}, respectively.  Under Condition~\ref{ass:5}, the following statements hold. \\
(a) $\tr(\V)$ is minimized by   $\Pi(\X,Y)\propto |\dot{\phi}(Y\tX^\T\bm\theta_t)|\|H^{-1}(\bm\theta_t)\tX\|$. \\ 
(b) 
The $\tr(H(\bm\theta)\V H(\bm\theta))$ is minimized by  $\Pi(\X,Y)\propto |\dot{\phi}(Y\tX^\T\bm\theta_t)|\|\tX\|$.    
\end{prop}
Note that under Condition~\ref{ass:5}  the inclusion probabilities satisfy 
\[0\le \pi_i:=\frac{r\Pi(\X_i,Y_i)}{\sum_{l=1}^n \Pi(\X_l,Y_l)}<1,\]
 with high probability if $n$ is sufficiently large. 
 Based on Proposition~\ref{pro:2}, the $A$- and $L$-optimal subsampling probabilities are obtained as
\begin{align}\label{eq:Aopt}
    \pi_i^{A}&=\frac{r|\dot{\phi}(Y_i\tX_i^\T\bm\theta_t)|\|H^{-1}(\bm\theta_t)\tX_i\|}{\sum_{l=1}^n|\dot{\phi}(Y_l\tX_l^\T\bm\theta_t)|\|H^{-1}(\bm\theta_t)\tX_l\|},\quad
    \pi_i^{L}=\frac{r|\dot{\phi}(Y_i\tX_i^\T\bm\theta_t)|\|\tX_i\|}{\sum_{l=1}^n|\dot{\phi}(Y_l\tX_l^\T\bm\theta_t)|\|\tX_l\|}.
\end{align}
respectively. Note that these results cover the optimal subsampling methods in logistic regression \citep{wang2018optimal} and SVM \citep{Han2023Leverage} as special cases.
For example,   the $L$-optimal weights 
 for logistic regression with $\phi$ being the logistic loss in \eqref{det1}
are  obtained  by a direct calculation  as  $\pi_i^L \propto |1-\exp(-\phi(Y_i\tX^\T\bm\theta_t))|\|\X_i\| = | 1 + \exp (Y_i\tX^\T\bm\theta_t)|^{-1} \|\X_i\|$.  Similarly, it follows for SVM that $\pi_i ^L \propto \mathbb{I}(Y_i\tX^\T\bm\theta_t<1)\|\tX\|$. 
One can easily see that the resulting sampling probabilities coincide with 
those in the current literature on specified models. Note that we use a $+1,-1$ labeling, and as a consequence, some expressions may slightly differ if the authors use a $0,1$ labeling system.

Since $\bm\theta_t$ is unknown, we propose to use an initial sample to obtain a pilot estimator, say $\hat{\bm\theta}_{\pilot}$.
In practice, $\pi_i^{A}$s and $\pi_i^{L}$s can thus be implemented as
\begin{align*}
\tilde{\pi}_i^A&= \frac{r|\dot{\phi}(Y_i\tX_i^\T\hat{\bm\theta}_{\pilot})|\|\hat{H}^{-1}(\hat{\bm\theta}_{\pilot})\tX_i\|}{\sum_{l=1}^n|\dot{\phi}(Y_l\tX_l^\T\hat{\bm\theta}_{\pilot})|\|\hat{H}^{-1}(\hat{\bm\theta}_{\pilot})\tX_l\|},\quad
    \tilde{\pi}_i^L= \frac{r|\dot{\phi}(Y_i\tX_i^\T\hat{\bm\theta}_{\pilot})|\|\tX_i\|}{\sum_{l=1}^n|\dot{\phi}(Y_l\tX_l^\T\hat{\bm\theta}_{\pilot})|\|\tX_l\|},
\end{align*}
where 
$$
\hat{H}(\hat{\bm\theta}_{\pilot}) =
{1 \over n} \sum_{i=1}^n\ddot{\phi}(Y_i\tX_i^\T\hat{\bm\theta}_{\pilot})\tX_i\tX_i^\T
$$
denotes an estimate of the Hessian matrix \eqref{det6} at the point $\hat{\bm\theta}_{\pilot}$.
To keep the discussion simple, we assume at this point that the pilot estimate $\hat{\bm\theta}_{\pilot}$ is derived from the initial sample which is independent of the sample $\fn$ from which we construct the subsample. This assumption can be easily satisfied via data splitting.

\subsection{Improved estimator based on summary statistics}
\label{sec:RB}

The $L$-optimal subsampling with probabilities $\bm\pi^L = (\pi ^L_1, \ldots, \pi ^L_n)^\top$ given in \eqref{eq:Aopt}  
plays an important role among various strategies with a focus on minimizing the asymptotic MSE of the estimate $L^\T\hat{\bm\theta}$.
To be precise, define $\tilde{\score}_{\sub}(\bm\theta)=n^{-1}\sum_{l=1}^n {\pi}_i^{-1}\delta_i {Y_i\dot{\phi}(Y_i\tX_i^\T\bm\theta)}\tX_i$
as an estimate of the gradient of the empirical risk \eqref{eq:hat-risk} from the subsample, 
then  for the estimate \eqref{det9}
 from the full sample
 $\fn:=\{(\X_i,Y_i)\}_{i=1}^n$ has the representation $\hat{\score}_{\phi}(\bm\theta) =\E(\tilde{\score}_\sub(\bm\theta)|\fn)$.
It is now easy to see that the $L$-optimal subsampling provides an efficient way to reduce the conditional variance  var$(\tilde{\score}(\bm\theta) | \fn)$ for $\bm\theta = \hat{\bm\theta}_{\pilot}$.
On the one hand, 
adopting $\tilde{\bm\pi}^L$ yields substantial computational advantages. 
On the other hand, the sampling variance of ${\bm\theta}$ depends on $\bm\pi$ only through the conditional variance var($\tilde{\score}(\bm\theta)|\fn$).
Therefore, any inclusion probability $\bm\pi$  with larger conditional variance  $\var(\tilde{\score}(\bm\theta)|\fn)$  with respect to the Loewner-ordering is inadmissible in terms of statistical efficiency. 

{The $L$-optimal subsampling yields informative data points for a good  estimation of  the score function $\score_\phi$ and we will now   construct a more efficient estimator of the gradient of the population risk 
$\score_\phi (\bm\theta) = \E (\tilde{\score}_\sub(\bm\theta)) $  in \eqref{det8}  by  leveraging some  information of  the unsampled data. Our approach is motivated by the fact that in many subsampling problems
 data collection and estimation are intertwined, and  
data scientists can record some summary statistics for the unselected data during the sampling step without increasing too much the computational costs. In such cases, one can expect to improve the estimation by utilizing this information. 
A first approach to implementing this idea would be to condition $\tilde{\score}_\sub(\bm\theta)$ on sufficient statistics. More precisely, if a sufficient statistic, say $T$, would exist, 
 which could be computed with low computational costs, then, by the  Rao–Blackwell theorem, one could certainly improve $\tilde{\score}_\sub(\bm\theta)$ by the statistic $\E ( \tilde{\score}_\sub(\bm\theta)|T ) $ without increasing the computational complexity of the resulting estimate too much.
 However, for logistic regression, the minimal sufficient statistic is equivalent to the data $\mathcal{F}_n$ itself and this is also often true for other regression models  \citep[see, for example,][]{Barber2020TestingGA}. Consequently,  by 
  conditioning on the sufficient statistics $\fn$, we would obtain the gradient of the empirical 
 risk of the full sample, that is  $\hat{\score}_\phi   (\bm\theta)  =  \mathbb{E} (\tilde{\score}_{\rm sub} (\bm\theta)  | \fn)$. Obviously, such an estimator would be more efficient but could not be 
 used due to the huge computational costs.
}

{
 Although  Rao-Blackwellization would yield computationally not feasible statistics, we can utilize its principle idea to obtain more efficient estimates for the gradient of the population risk. To be precise, instead of using sufficient statistics directly, we propose to apply another conditional expectation which is approximated by a linear projection}. 
{To be precise, let $g: \mathcal{X}\times \{-1,1 \} \to \mathbb{R}^{d_g} $ be a deterministic transformation that can be computed with low computational costs.  Define the $d$-dimensional vectors ${\score}_{i}(\bm\theta)=\dot\phi(Y_i\tX_i^{\T}\bm\theta)Y_i\tX_i$ ($i=1, \ldots ,n)$ and the 
$d_g \times n$ 
matrix $\bm G(\mathcal{F}_n) = (g(\X_1,Y_1) , \ldots g(\X_n,Y_n) ) $, then we are  interested in the linear projection of 
the  $n \times d$  matrix   $\bm \score({\fn}) = ({\score}_{1}(\bm\theta) , \ldots , {\score}_{n}(\bm\theta))^\T $ onto  the linear space $\{ \bm G(\mathcal{F}_n)^\T  {\coef} ~ | ~  \coef \in \mathbb{R}^{d_g \times d} \} $, that is in the  least squares problem 
\begin{align*}
\min_{\coef  \in \mathbb{R}^{d_g \times d} } \big \|\bm \score({\fn}) -\bm G(\mathcal{F}_n) ^\T \coef \big \|_F^2
= \min_{\coef  \in \mathbb{R}^{d_g \times d} } \sum_{i=1}^n \|{\score}_{i}(\bm\theta)^\T -\g_i^\T  \coef  \|^2
~, 
\end{align*}
where $\|A \|_F = ({\rm tr} (AA^\T))^{1/2}$ denotes the Frobenius-norm of the matrix $A$ and  we use the notation $\g_i =g(\X_i,Y_i)$ ($i=1, \ldots , n$) for the sake of simplicity.
The solution is given by the $d_g \times d$ matrix 
$$
\hat \coef (\bm \theta ) = \big (  \bm G(\mathcal{F}_n)\bm G(\mathcal{F}_n)^\T 
\big )^{-1}  \bm G(\mathcal{F}_n) \bm \score(\fn)  = 
\Big ( {1\over n} \sum_{i=1}^n  \g_i\g_i^\T 
\Big )^{-1} {1\over n} \sum_{i=1}^n \g_i {\score}_{i}(\bm\theta)^\T ~. 
$$ Since this estimator is computationally not feasible, we replace it with an estimator from the subsample, which is 
\begin{align}
    \label{det12}
\tilde\coef(\bm\theta)=\Big ( {1 \over n} \sum_{i=1}^n{\delta_i \over \pi_i} \g_i\g_i^\T \Big ) ^{-1}\sum_{i=1}^n\g_i\tilde{\score}_{i} (\bm\theta)^\T
\end{align}
with 
\begin{align}
    \tilde{\score}_i(\bm\theta) = \frac{\delta_i}{n{\pi}_i}\dot\phi(Y_i\tX_i^{\T}\bm\theta)Y_i\tX_i, \label{eq:tilde_score}
\end{align} 
 which is the solution of the subsample based least squares problem 
\begin{equation*}
    \sum_{i=1}^n\frac{\delta_i}{n\pi_i}\left\|\score_i(\bm\theta)-\g_i^\T\coef\right\|^2.
\end{equation*}
If the vector  $\g$ contains a constant (corresponding to an intercept term) one obtains the corresponding row in the normal equation 
\begin{align}
    \label{det11}
\sum_{i=1}^n\tilde{\score}_{i}(\bm\theta)^\T-\Big (\sum_{i=1}^{n}\frac{\delta_i}{n{\pi}_i}\g_i\Big )^\T\tilde{\coef}(\bm\theta)=\bm 0^\T~. 
\end{align}
 Now we can use the optimal  predictions $\g_i^\T \tilde{\coef}(\bm\theta)$ of $\score_i(\bm\theta)^\T $ to estimate 
 $  \mathbb{E} (\tilde{\score}_{\rm sub}(\bm\theta)^\T   | \fn) = n^{-1} \sum_{i=1}^n  \score_i (\bm\theta) ^\T$. Observing \eqref{det12} and \eqref{det11}   gives 
 \begin{align} 
   \tilde{\score}_{RB}(\bm\theta)^\T&=\Big (\frac{1}{n}\sum_{i=1}^n \g _i^\T\Big )\tilde{\coef}(\bm\theta)\notag\\
   &=\Big (\frac{1}{n}\sum_{i=1}^n g(\tX_i,Y_i)^\T\Big)\tilde{\coef}(\bm\theta)+ \Big\{\sum_{i=1}^n\tilde{\score}_{i}(\bm\theta)^\T-\Big(\sum_{i=1}^{n}\frac{\delta_i}{n{\pi}_i}\g_i\Big)^\T\tilde{\coef}(\bm\theta)\Big\} 
     \notag \\ 
   &=\sum_{i=1}^{n}\Big \{1-\Big( \sum_{i=1}^{n}\frac{\delta_i}{n{\pi}_i}\g_i - \frac{1}{n}\sum_{i=1}^n \g_i  \Big )^\T \Big (\sum_{i=1}^n\frac{\delta_i}{n{\pi}_i}\g_i\g_i^\T\Big )^{-1}\g_i\Big \}\tilde{\score}_{i}(\bm\theta)^\T.  
    \label{eq:RB1}
 \end{align}
 as an estimate of the gradient of the empirical risk function 
 in \eqref{det9}. 
Note that we use the representation \eqref{eq:RB1} to illustrate how the estimator depends on the individual level information, that is on the non-vanishing  terms $\tilde{\score}_{i}(\bm\theta)$ and $\delta_i \g_i$ (corresponding to the data from the subsample), and 
 on  the mean  $n^{-1}\sum_{i=1}^n \g_i$ of the sample $\g_1, \ldots , \g_n$.  }

We emphasize  that we do not add any assumption on the relationship between $\bm\score(\fn) $ and $\bm G(\fn)$. Especially, we do not require a linear relationship between $\bm\score(\fn)$  and $\bm G(\fn)$. The matrix 
$\bm G(\mathcal{F}_n) ^\T \tilde\coef(\bm\theta)$  is just a linear projection of $\bm\score(\fn)$. 
Importantly, we do not need to calculate and record the values $g(\X _i, Y_i)^\T\tilde\coef(\bm\theta)$ for every data point. Instead, we only need to use the global information $n^{-1}\sum_{i=1}^n g(\X_i,Y_i)$ which can be obtained during scanning the dataset. Therefore, the additional computational cost comes from 
calculating the sample mean  of 
$g(\X_1 ,Y_1) , \ldots , g(\X_n ,Y_n)$.
There is a considerable computational benefit since the corresponding computational cost is of order $O(n)$ and the RAM cost is of order $O(r)$. This implies the total computational cost is the same as for the most existing optimal subsampling methods, such as OSMAC.

Next we analyze the asymptotic properties 
of the the Z-estimator $\tilde{\bm\theta}_{RB}$ solving the equation
$\tilde{\score}_{RB}(\bm\theta)=\bm 0$ by $\tilde{\bm\theta}_{RB}$. For this purpose, we assume  the following regularity condition. 
\begin{assumption}
\label{ass:gx}
     (i) The function  $g: \mathcal{X}\times \{-1,1 \} \to \mathbb{R}^{d_g} $   contains a constant component and  $\E  \big( \sup_{\|\bm u\|=1}|\bm u^\T g(\XY)|^{4} \big) <\infty$; 
    (ii) The matrix $\E \big(  g(\XY)g(\XY)^{\T} \big) $ is positive definite. 
\end{assumption}

\begin{theorem}\label{thm:RB-1}
    Let  Conditions~\ref{ass:1}--\ref{ass:3}, ~\ref{ass:6}(i), and~\ref{ass:gx} be satisfied and   further assume that
    the  inclusion
probabilities satisfy 
    {$\max_{i=1}^n(n\pirb_i)^{-1}\le (\rho r)^{-1}$   for some constant $\rho>0$ almost surely}. As $n,r, r_0  \to\infty$ with $r\log^2(n)/n\to 0$,  it holds that 
    \begin{equation*}
         V_{RB}^{-1/2}(\tilde{\bm\theta}_{RB}-\bm\theta_t)\to N(\bm 0,I_d), 
     \end{equation*}
     in distribution, where $V_{RB}=H^{-1}(\bm\theta_t)V_{RB,c}H^{-1}(\bm\theta_t)$ with 
     $
     V_{RB,c}=\E\{\sum_{i=1}^n{(n^2\pi_i)}^{-1}\res_i\res_i^\T\} 
     $,
 $     \res_i=
    \score_i (\bm\theta_t) -\coef(\bm\theta_t)^\T g(\X_i,Y_i),
      $
      $\score_i(\bm\theta_t)=\dot\phi(Y_i\tX_i^\T{\bm\theta}_{t})Y_i\tX_i$ 
     and 
$\coef(\bm\theta_t) = \G^{-1}\E\big(g(\XY)\score(\bm\theta_t)^{\T}\big)$ with $\G = \E\big(g(\XY)g(\XY)^\T\big)$.
\end{theorem}

The condition $\max_{i=1}^n (n\pirb_i)^{-1}\le (\rho r)^{-1}$ is  required to avoid  that the estimator is sensitive to  extremely small subsampling probabilities.  It is satisfied for the uniform sampling and the  shrinkage-based leverage-score sampling \citep{ma2015}. We only use this assumption for a correct statement of Theorem \ref{thm:RB-1}, which is formulated for general subsampling probabilities. 
The proposed subsampling procedure (which will be presented in Algorithm~\ref{alg:RBS1} below), does not require this assumption as it uses specific optimal weights. 
It is also worth mentioning that  Theorem~\ref{thm:RB-1}  remains correct if   the dimension of the vector $g$ increases with the size of the subsample, such that  $d_g=o({r}^{1/2})$.
If we increase  the dimension of $g$, we  will improve the statistical efficiency, since the variability of $\tilde{\score}(\bm\theta)$ is more likely to be explained by a space consisting of more basis functions. 
 On the other hand, a larger dimension $d_g$ may lead to non-negligible additional computational costs. Thus researchers should find a trade-off between statistical accuracy and computational efficiency in designing the transformation $g(\cdot)$.

 Define  $\var_L(\Tilde{\bm\theta}_{RB})=\tr\left(H(\bm\theta_t) V_{RB} H(\bm\theta_t)\right)$ and $\var_L(\Tilde{\bm\theta})
= \tr(H(\bm\theta_t)\var(\tilde{\bm\theta})H(\bm\theta_t))$.
For uniform subsampling based estimator we obtain from Proposition \ref{pro:1} and Theorem 
\ref{thm:RB-1}, 
$$
    \var_L(\Tilde{\bm\theta}_{RB})
        = \tr(V_{RB,c}) = r^{-1}\E\|\res_i\|^2
           \le   r^{-1} \E\|\score_i(\bm\theta_t)\|^2 =
           \var_L \big ( \tilde{\bm\theta} \big ),
           $$
    which proves  that the $Z$-estimator
$\tilde{\bm\theta}_{RB}$  is an improvement of  $\tilde{\bm\theta}$ with respect to the $L$-optimality criterion. 
Our next result shows that this improvement 
is also obtained (asymptotically) for most of the commonly used subsampling strategies.
 It  will also provide us with a guideline on how to construct the function $g(\cdot)$  defining the projection.

\begin{theorem}\label{thm:2}
    Let the  conditions of Theorem~\ref{thm:RB-1} be satisfied,  further assume that $\bm\pi$ satisfies Condition~\ref{ass:5} 
    and   
    \begin{equation}\label{eq:thm2-1}
       \E \Big ( { \|g(\XY)^\T\coef(\bm\theta_t)\|^2 \over \Pi(\XY)} \Big ) +\E 
    \Big ( {g(\XY)^\T\coef(\bm\theta_t)\mathcal{R}_{\rm RB}(\bm\theta_t) \over \Pi(\XY)} \Big  )
    \ge 0, 
    \end{equation}
  where $\mathcal{R}_{\rm RB}(\bm\theta_t)=\E  \big( \score(\bm\theta_t)|g(\XY) \big) -\coef(\bm\theta_t)^\T g(\XY)$. 
    If
     the $\sigma$-field generated by $\Pi(\XY)$ {is contained} in the $\sigma$-field generated by $g(\XY)$, 
    it holds that 
    \begin{equation*}
      \lim_{r,n\to \infty} r \cdot   \var_L(\tilde{\bm\theta}_{RB})\le \lim_{r,n\to \infty} r \cdot  \var_L(\tilde{\bm\theta}).
    \end{equation*}
\end{theorem}
Interestingly, if  the sampling scheme depends on $\XY$ as in Condition~\ref{ass:5},  then the $Z$-estimator $\tilde{\bm\theta}_{RB}$ improves  (asymptotically)  $\tilde{\bm\theta}$ with respect to the  $L$-optimality criterion if the function  $g(\XY)$ defining the projection contains all the information of $\Pi(\XY)$. 
A straightforward application of the 
Cauchy-Schwarz inequality shows that 
the inequality 
\begin{align} \label{det13}
    \E \left ( {\| \mathcal{R}_{\rm RB}(\bm\theta_t) \|^2 \over \Pi(\XY) }\right  ) \le \E \left  ( {\|g(\XY)^\T\coef(\bm\theta_t)\|^2 \over \Pi(\XY)} \right )
\end{align} 
implies  condition \eqref{eq:thm2-1} in Theorem \ref{thm:2}. Note that this inequality is obviously satisfied 
if  $\E\big(\score(\bm\theta_t)|g(\XY)\big)$ depends linearly on   $g(\XY)$ 
(because  the left  hand side of \eqref{det13} vanishes in this case).   Thus a simple choice for the function $g$ is  
\begin{align}
    \label{det200}
{g(\XY)=(1, Y, \score ^\T (\hat{\bm\theta}_{\pilot}) )^\T = (1,Y,\dot\phi(Y\tX^\T\hat{\bm\theta}_{\pilot})Y\tX^\T)^\T, }
\end{align}
which approximates $(1, Y, \score ^\T ({\bm\theta}_t) )^\T$, and one can expect a variance reduction due to 
an approximate linear relationship. Here we also use the variable  $Y$ to further address the label information.
Note that for  $A$- and  $L$- optimal subsampling  the 
quantities $ \score _i (\hat{\bm\theta}_{\pilot})  = \dot\phi(Y_i\tX_i^\T\hat{\bm\theta}_{\pilot})Y_i\tX_i$ have already been calculated and therefore the extra computational cost is negligible.

\subsection{Dealing with tail inclusion probabilities }\label{sec:tail}

Although the subsample estimator $\tilde{\bm\theta}_{RB}$ obtained by Rao-Blackwell-type conditioning argument is a substantial improvement of the original subsample estimator $\tilde{\bm\theta}$, it might not be stable, if some inclusion probabilities involved in the sampling step are ``close to zero''.  In this case, the corresponding selected subsample points are accompanied by extremely small probabilities in the necessary condition \eqref{eq:RB1} and one can expect variance inflation of the estimators $\Tilde{\bm\theta}$ and $\tilde{\bm\theta}_{\rm RB}$.

 This phenomenon is well known in the literature on optimal subsampling; see  \cite{ma2015}, \cite{ai2021optimal}, \cite{yu2020quasi} among others.   For example, \cite{ma2015} proposes to combine 
leverage-based (non-uniform)   with uniform subsampling. While this approach (and also other methods) yields more stable estimates from the subsample, it comes with a loss in statistical efficiency (according to Proposition~\ref{pro:2}). 

In order to motivate an alternative strategy to deal with 
the stability problem, we note that the $L$-optimal subsampling probabilities are proportional to $|\dot{\phi}(Y_i\tX_i^\T\hat{\bm\theta}_{\pilot})|\|\tX_i\|$.  As the  predictors  $\tX_i$ contain an intercept term,  the norms $\|\tX_i\|$ are bounded away from zero and the inclusion probabilities are ``small'' if $| \dot{\phi}(Y_i\tX_i^\T\hat{\bm\theta}_{\pilot})| $ is ``small''. A similar argument applies for $A$-optimal subsampling because the extra factor  
$\|\hat{H}^{-1}(\hat{\bm\theta}_\pilot)\tX_i\|$  in the optimal weight can be bounded from below by $\lambda_{\min}(\hat{H}^{-1}(\hat{\bm\theta}_\pilot))\|\tX_i\|$.
 Therefore, extremely small $A$- and $L$-optimal  inclusion probabilities 
 arise if $Y\tX^\T\hat{\bm\theta}_{\pilot}$ attains a value in a region where 
 the absolute value of the derivative  of the function $\phi(\cdot)$  is small. 
The following lemma gives some characteristics of such sample points.

\begin{lemma}\label{lem:probregion}
 Let Conditions~\ref{ass:1}--\ref{ass:3} be satisfied.  Suppose that for any $\eta\in[0,1]$ the infimum of the 
function $H: a \to \eta\phi(a)+(1-\eta)\phi(-a)$ is  attained at a unique point  and that the optimal classification function $f^*_\phi=\inf_f\E\phi(Yf(\tX))$ $($with the infimum  taken over all measurable functions of $\X)$ admits a linear form with $f_\phi^*(\x)=\x^\T\bm\theta$ a.e. on $\mathcal{X}$ for some unknown $\bm\theta$. Further assume that for $\varepsilon_0\in(0,1/2)$ there exists   a constant $t>0$ such that $\dot\phi(t)/\dot\phi(-t)= \varepsilon_0$. 
  \begin{itemize}
  \item[(a)]  If $\eta(\x) = \mathbb{P} ( Y=1 | \X = \x ) 
  \le \varepsilon_0/(1+\varepsilon_0)$, then $\x^\T\bm\theta_t\le -t$.
  \item[(b)]  If $\eta(\x) = \mathbb{P} ( Y=1 | \X  = \x )  \ge 1/(1+\varepsilon_0)$,  then $\x^\T\bm\theta_t\ge t$. 
  \end{itemize} 
  Moreover, if $\varepsilon_0/(1+\varepsilon_0)<\eta(\x)<1/(1+\varepsilon_0)$, we have $\dot{\phi}(-t)\le \dot\phi(\x^\T\bm\theta_t)\le \dot\phi(t)$, which implies $\dot\phi(\x^\T\bm\theta_t)/\dot\phi(-\x^\T\bm\theta_t)\ge \varepsilon_0$. 
\end{lemma}

Note that the function $\phi$ is convex which implies   $\dot\phi(t)\le \dot\phi(0)<0$ for $t\le 0$. Therefore, one can conclude that data points with a 
the small gradient $\dot\phi(\tX^\T\bm\theta_t)$  lie in the region where 
 $\dot\phi( \tX^\T \bm \theta_t )/\dot\phi(- \tX^\T \bm \theta_t)$ 
is sufficiently small. 
 The geometric  intuition behind this is that these points are  far away from the separating hyperplane $\x^\T\bm\theta_t=0$ and are likely to be correctly classified. In the following, we take logistic regression and DWD as examples.

\begin{example} 
~~
\begin{itemize}
    \item[(a)]
 For the logistic regression with $\eta(\x)=\exp(\x^\T\bm\theta_t)/(1+\exp(\x^\T\bm\theta_t))$ and $\phi(\cdot)$ being the logistic loss \eqref{det1}, the minimizer of the function 
 $H$ in Lemma \ref{lem:probregion}
 is unique and given by   $a^* (\eta)  = \log \big ( {\eta  \over 1- \eta } \big ) 
 $.  Therefore, the  function $a^*(\eta(\x))$ 
   minimizes 
 $\fR_\phi(a)=\eta(\x)\log(1+\exp(-a))+(1-\eta(\x))\log(1+\exp(a))$
 almost everywhere on $\mathcal{X}$
 and the  function $f^*_\phi(\x)=a^*(\eta(\x))=\x^\T\bm\theta_t$   is the optimal classifier  under the logistic loss.  In this case the assumptions regarding the loss function in Lemma \ref{lem:probregion} are satisfied. 
Note that $\dot{\phi}(t)=-\exp(-t)/(1+\exp(-t))$. With the choice  $t=-\log\varepsilon_0$ one can directly verify the conclusions of Lemma \ref{lem:probregion}.
\item[(b)]
Consider the DWD problem with the odds ratio 
${\eta(\x)\over 1-\eta(\x)}=(2\x^\T\bm\theta_t)^{2}\I({\x^\T\bm\theta_t}>0.5)+(2\x^\T\bm\theta_t)^{-2}\I(\x^\T\bm\theta_t<-0.5)+\I(|\x^\T\bm\theta_t|\le 0.5)$
and let $\phi(\cdot)$ be the DWD loss in  \eqref{det4} with $\gamma = 0.5$. Assume that the low noise condition  $\eta(\x)\neq 0.5$ holds almost everywhere on $\mathcal{X}$. According to Lemma 3 in \cite{wang2017another}, the minimizer of the function 
 $H$ in Lemma \ref{lem:probregion}
 is unique and given by   $a^* (\eta)  = \frac{1}{2}\big ( {\eta  \over 1- \eta } \big )^{1/2}\I(\eta > 0.5) -\frac{1}{2}\big ( {1-\eta  \over \eta } \big )^{1/2}\I(\eta < 0.5)$. Therefore, $f^*_\phi(\x)=|\x^\T\bm\theta_t|(2\I(\eta(\x)>0.5)-1)$ is the optimal classifier under the DWD loss. 
Note that $\sgn(\x^\T\bm\theta_t)=\sgn(\eta(\x)-0.5)$ by the definition of $\eta(\x)$, which means that the Bayesian and linear classifier have the same sign. Therefore, one obtains that $f^*_\phi(\x)=\x^\T\bm\theta_t$  and the assumptions regarding the loss function in Lemma \ref{lem:probregion} are satisfied.  
Note that $\dot\phi(t) = -t^{-2}\I(t\ge 0.5)-4\I(t< 0.5)$ and that $\dot\phi(t)/\dot\phi(-t)= (2t)^2\I(t<-0.5)+\I( |t|\leq 0.5)+(2t)^{-2}\I(t>0.5)$. With the choice  $t={1/(2 \varepsilon_0^{1/2})}$,  one can directly verify that the conclusions of Lemma \ref{lem:probregion}. %

\end{itemize}
\end{example}
The aforementioned discussion motivated us to divide the data into three parts by defining for a threshold $\mathrm{C}_{r_0}>0$, which may increase as $r_0 \to \infty$, the sets 
\begin{align*}
    \chi_{+}(\mathrm{C}_{r_0} ) &
    =\{(\X_i,Y_i) \in \mathcal{F}_n : \tX_i^\T{\bm\theta}_t > \mathrm{C}_{r_0} , ~ Y_i=+1\}, \\
    \chi_{-}(\mathrm{C}_{r_0} )
    & =\{(\X_i,Y_i) \in \mathcal{F}_n : \tX_i^\T{\bm\theta}_t < -\mathrm{C}_{r_0} , ~ Y_i=-1 \},
\end{align*}
and $\chi_s(\mathrm{C}_{r_0} ) = \mathcal{F}_n \setminus (\chi_+(\mathrm{C}_{r_0} )\cup\chi_-(\mathrm{C}_{r_0} ))$
as the rest of the sample points.  Some examples for the choice of $\mathrm{C}_{r_0} $ are given in Lemma~\ref{lem:orderofC} below.
{To visualize the size of the inclusion probabilities in these three regions, we display in Figure \ref{fig.gradient} the $A$-optimal subsampling probabilities for a sample of size $n=10^5$. Here the working model is selected as (a) logistic regression, (b) DWD , and (c) SVM, while the response generated by a standard logistic regression, that is  $\pr(Y = 1|\X) = (1+\exp(-\X^{\T}\bm\theta_t))^{-1}$  with  $\X=(1,X_1,X_2)^\T$ and parameter $\bm\theta_t = (0,0.5,0.5)^{\T}$. We observe that  data points with ``very small'' inclusion probabilities mainly fall in the sets $\chi_{+}(\mathrm{C}_{r_0} )$ and $\chi_{-}(\mathrm{C}_{r_0} )$. The $2$-dimensional predictor $(X_1,X_2)^\T$ is generated from a multivariate normal distribution $\mathcal{N}(\bm 0_2, \sqrt{2}I_2)$.
Consequently, the variance inflation caused by extremely small inclusion probability will be naturally mitigated when we sample on $\chi_s(\mathrm{C}_{r_0} )$ only.
\begin{figure}[htbp]
\centering\spacingset{1}
\begin{subfigure}{.3\textwidth}
  \centering
\includegraphics[width=1\linewidth]{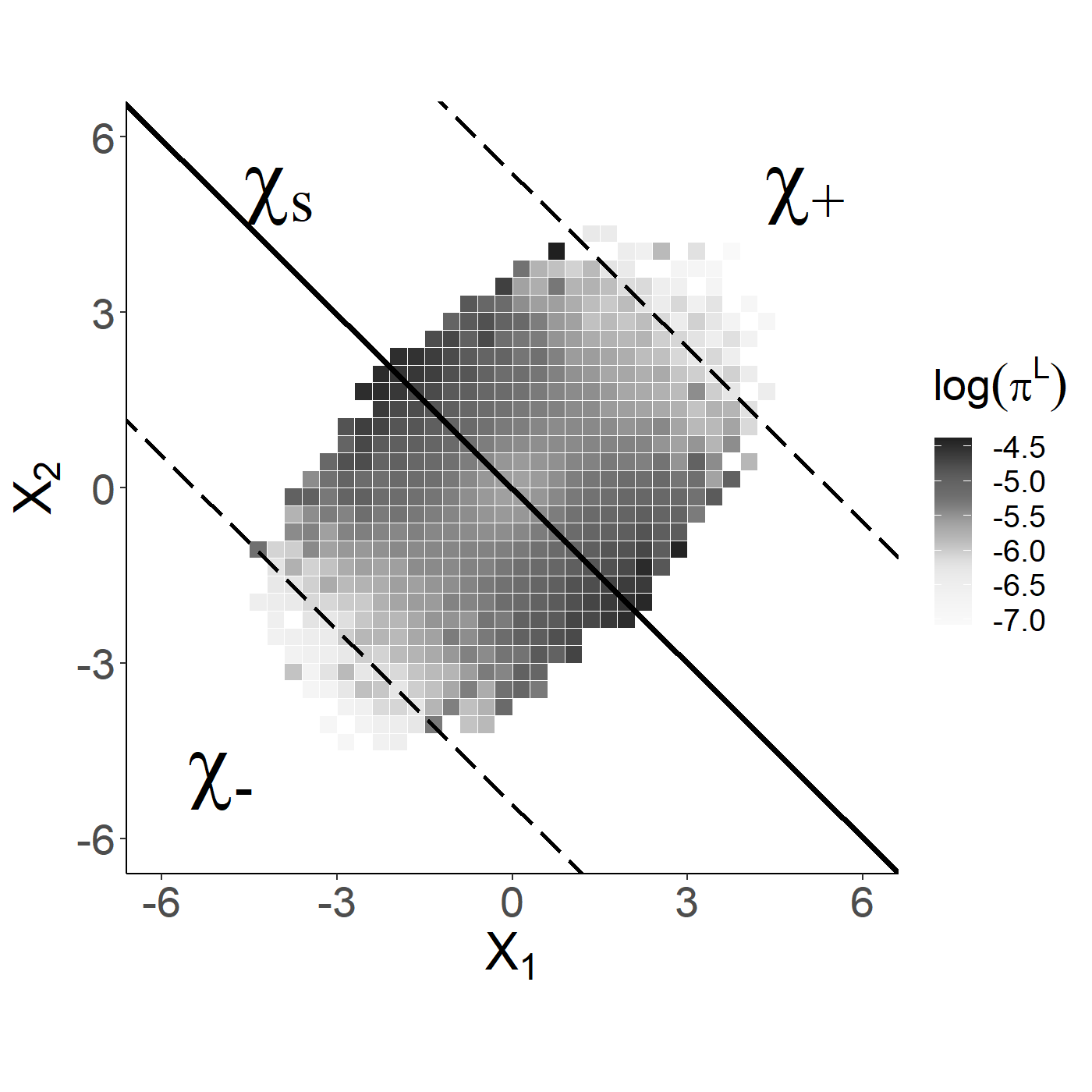}
  \caption{}
\end{subfigure}
\hfil
\begin{subfigure}{.3\textwidth}
  \centering
  \includegraphics[width=1\linewidth]{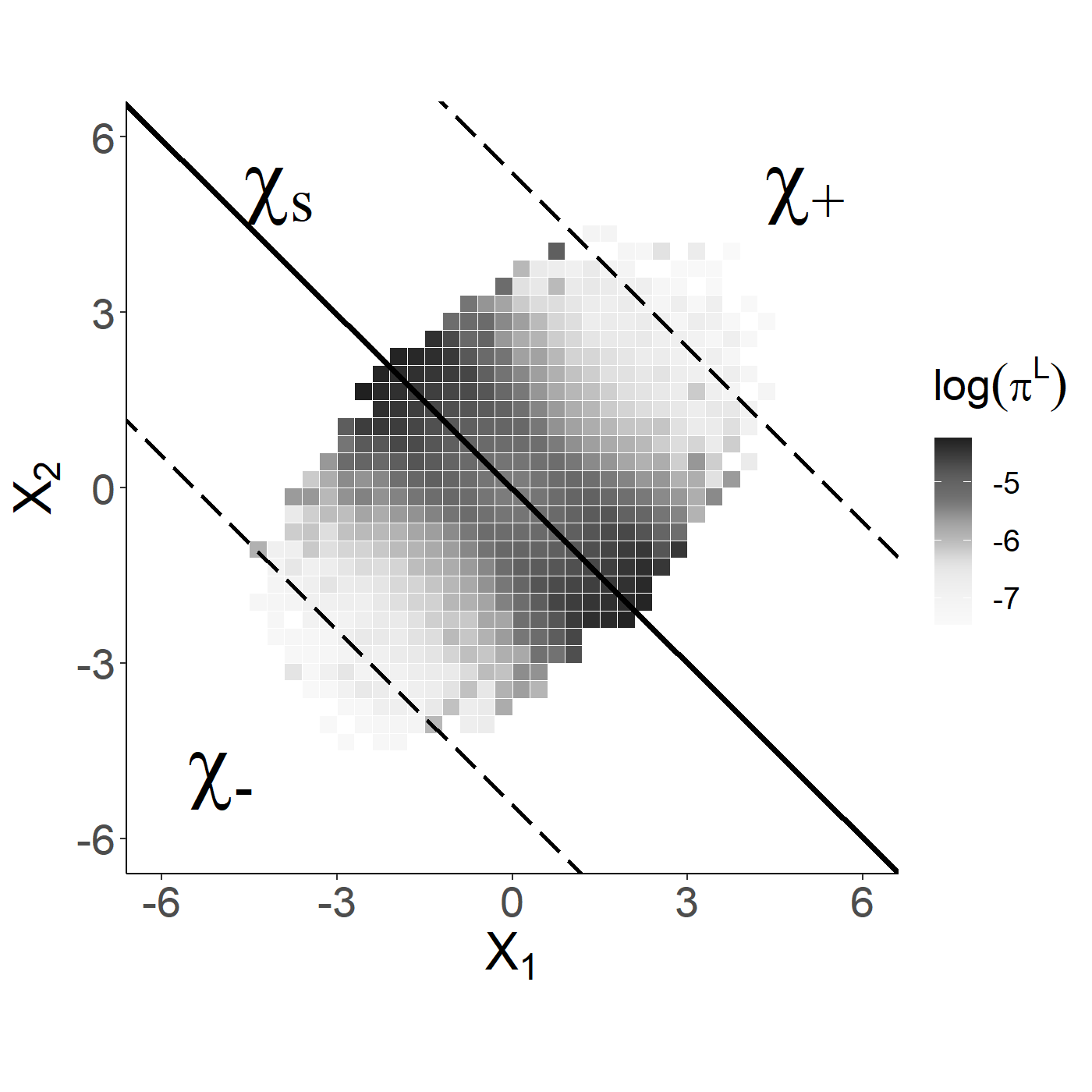}
  \caption{}
\end{subfigure}
\hfil
\begin{subfigure}{.3\textwidth}
  \centering
\includegraphics[width=1\linewidth]{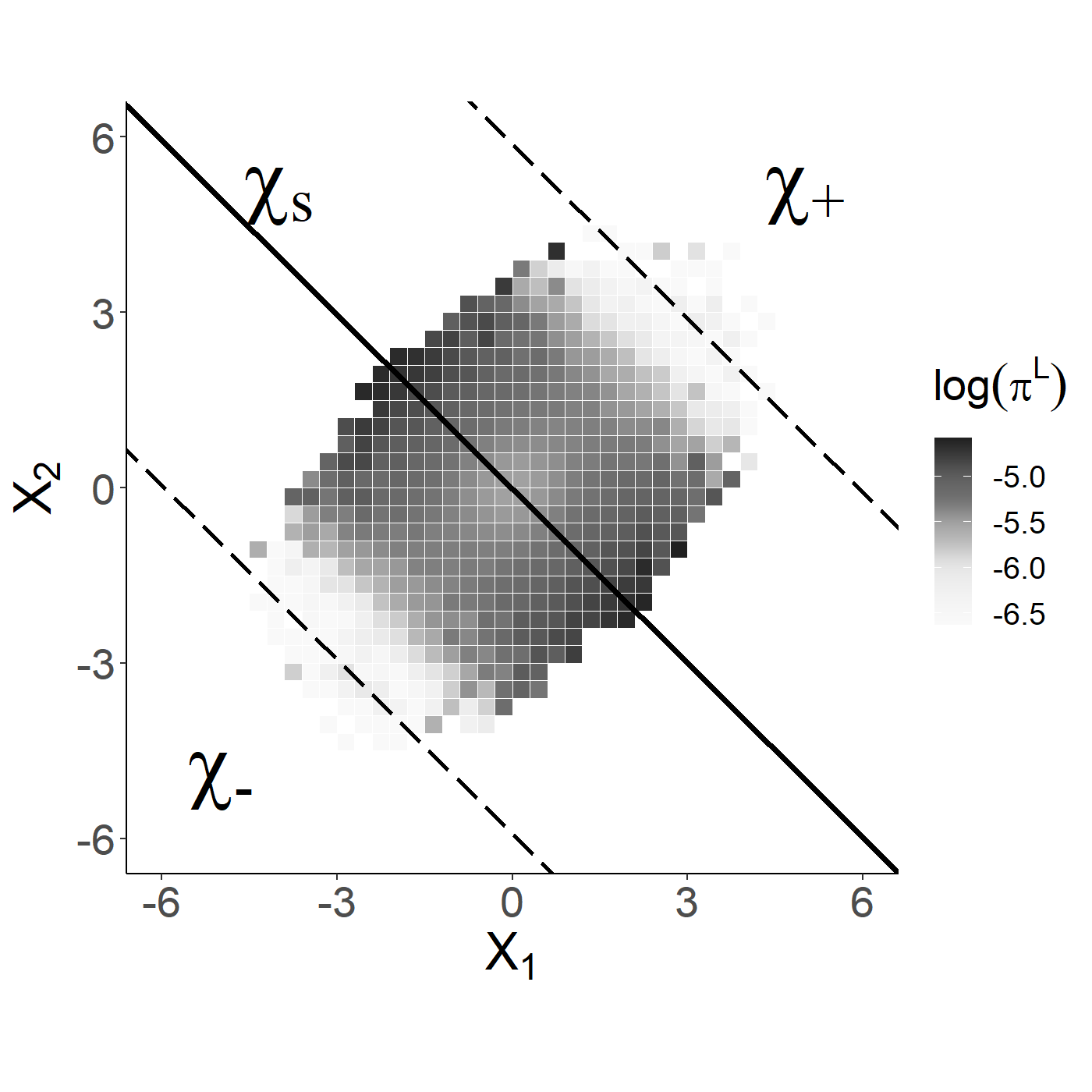}
  \caption{}
\end{subfigure}
    \caption{$A$-optimal subsampling probabilities (after log-transformation) and the sets $\chi_+(\mathrm{C}_{r_0} ),\chi_{-}(\mathrm{C}_{r_0} )$ and $\chi_s(\mathrm{C}_{r_0} )$. The working model is  (a): logistic regression, (b): DWD, and C): SVM (with squared hinge loss). The data is generated by a
standard logistic regression. }
    \label{fig.gradient}
\end{figure}

Recall that 
the minimizer 
$\hat{\bm\theta}$ of the empirical loss  form the full sample $\mathcal{F}_n$ is obtained as the solution of \eqref{det9}. With the new notation, the  equation can be rewritten  as
\begin{equation}\label{eq:d-loss}
   \bm 0=
   \sum_{(\tX_i,Y_i)\in \chi_s(\mathrm{C}_{r_0} )}\frac{{\score}_{i}(\bm\theta)}{n}+\sum_{(\tX_i,Y_i)\in \chi_+(\mathrm{C}_{r_0} )}\frac{{\score}_{i}(\bm\theta)}{n}+\sum_{(\tX_i,Y_i)\in \chi_{-}(\mathrm{C}_{r_0} )}\frac{{\score}_{i}(\bm\theta)}{n},
\end{equation}
where ${\score}_{i}(\bm\theta)=\dot\phi(Y_i\tX_i^{\T}\bm\theta)Y_i\tX_i$ ($i=1, \ldots ,n)$.
We now propose  to use all points in  the  subsample  obtained from  the set $\chi_s(C)$ (with the optimal subsampling probabilities)
to approximate $n^{-1}\sum_{(\tX_i,Y_i)\in \chi_s(\mathrm{C}_{r_0} )}{{\score}_{i}(\bm\theta)}$ via $\tilde{\score}_{RB}(\bm\theta)$ as described 
in 
Section~\ref{sec:RB} and to approximate the two remaining terms in \eqref{eq:d-loss} using summary statistics.

To be precise, note that 
Lemma~\ref{lem:probregion} tells us that  data  points in  the sets $\chi_+(\mathrm{C}_{r_0} )$ or $\chi_{-}(\mathrm{C}_{r_0} )$   have relatively low noise (in the sense that $\eta(\tX)$ deviates substantially from $0.5$). Therefore, these points are easy to classify and a more efficient choice is to use the centroids of the two regions as a summary statistic for this data. More precisely, we define    $\bar{\tX}_+=n_+^{-1}\sum_{\X_i\in\chi_+(\mathrm{C}_{r_0} )}\tX_i$ and $\bar{\tX}_-=n_-^{-1}\sum_{\X_i\in\chi_-(\mathrm{C}_{r_0} )}\tX_i$ with $n_+$ and $n_-$ denoting the sample size of the sets. $\chi_+(\mathrm{C}_{r_0} )$ and $\chi_-(\mathrm{C}_{r_0} )$, respectively. We will show in  
Lemma S.1 of the online supplement that
\begin{equation}
\label{eq:taylor-center}
\sum_{\chi_+(\mathrm{C}_{r_0} )\cup \chi_-(\mathrm{C}_{r_0} )}\frac{{\score}_{i}(\bm\theta)}{n} = 
\sum_{\chi_+(\mathrm{C}_{r_0} )}\frac{\dot\phi(\bar{\tX}_+^\T\bm\theta)}{n}\bar{\tX}_+ + \sum_{\chi_-(\mathrm{C}_{r_0} )}\frac{-\dot\phi(-\bar{\tX}_-^\T\bm\theta)}{n}\bar{\tX}_-
 +  o_P(1) , 
\end{equation}
uniformly with respect to  $\bm\theta $
for a sufficient large  $\mathrm{C}_{r_0}$ under some mild conditions.  
As a result, an improved estimator $\Tilde{\bm\theta}_I$ 
can be obtained by solving the following system of equations: 
\begin{align} \nonumber
\notag  { L ( \bm \theta )} :=& \sum_{
{(\tX_i,Y_i)\in\chi_s(\mathrm{C}_{r_0} )}}\Big \{1-\Big ( \sum_{ (\tX_l,Y_l)\in\chi_s(\mathrm{C}_{r_0} )}\frac{\delta_l}{n{\pi}_l}  \g_l - \frac{1}{n}\sum_{(\tX_l,Y_l)\in{ \chi_s(\mathrm{C}_{r_0} )}}  \g_l \Big )^\T  \\
& ~~~~~~~~~~~~~~~~~~~~~~~~~~~~~~~~~~~~~~~~~
\nonumber 
\times 
\Big (\sum_{(\tX_l,Y_l)\in  \chi_s(\mathrm{C}_{r_0} )}\frac{\delta_l}{n{\pi}_l}\g_l\g_l^\T\Big )^{-1}\g_i\Big \}\tilde{\score}_{i}(\bm\theta)\\
    &+ \frac{n_+\dot\phi(\bar{\tX}_+^\T\bm\theta)}{n}\bar{\tX}_+-\frac{n_-\dot\phi(-\bar{\tX}_-^\T\bm\theta)}{n}\bar{\tX}_-  = \bm 0,
    \label{eq:sub-ee}
\end{align}
where the weights $\pi_l$ satisfy 
$\sum_{(\tX_l,Y_l) \in \chi_s(\mathrm{C}_{r_0} )} \pi_l=r$, 
and the quantities 
$\tilde{\score}_{i}(\bm\theta)$ are defined in \eqref{eq:tilde_score}. 
In particular,  we will further improve this 
estimator using  the optimal  sampling probabilities 
 \begin{equation}\label{eq:chi_s}
  \tilde{\pi}_i^A= \frac{r|\dot{\phi}(Y_i\tX_i^\T\hat{\bm\theta}_{\pilot})|\|\hat{H}^{-1}(\hat{\bm\theta}_\pilot)\tX_i\|
 }{\sum_{\chi_s(\mathrm{C}_{r_0} )}|\dot{\phi}(Y_l\tX_l^\T\hat{\bm\theta}_{\pilot})|\|\tX_l\|},\quad \textrm{or}\quad
   \tilde{\pi}_i^L= \frac{r|\dot{\phi}(Y_i\tX_i^\T\hat{\bm\theta}_{\pilot})|\|\tX_i\|
 }{\sum_{\chi_s(\mathrm{C}_{r_0} )}|\dot{\phi}(Y_l\tX_l^\T\hat{\bm\theta}_{\pilot})|\|\tX_l\|}~,  
 \end{equation}
 for {$(\tX_i,Y_i) \in \chi_s(\mathrm{C}_{r_0} )$}.
Conducting the optimal subsampling procedure only on the $\chi_s(\mathrm{C}_{r_0} )$ certainly mitigates the potential variance inflation caused by the inverse probability weighting scheme.
The additional computations 
for calculating 
the last two terms in \eqref{eq:sub-ee}
can be done during the sampling step.
Thus the proposed procedure is still computationally tractable. 

\begin{remark}    
Considering the logistic regression as a special case,  it is interesting to see some connection with optimal design based subdata selection algorithms such as the  IBOSS \citep{cheng2020information} and ODBSS \citep{2023arXiv230616821D}.
 Note that IBOSS chooses points such that $\tX_i^\T\bm\theta_t$ is close to the maximizers, say $-c^*$ and $c^*$,  of the function $c\to I(c) := c^2\exp(c)^d(1+\exp(c))^{-2d} $, which converges to $0$ at an exponential rate if $|c| \to \infty $. Therefore, for a sufficient large threshold $\mathrm{C}_{r_0} $, one can expect an extremely small value of 
the function $I(c)$ for data points  from  the sets $\chi_+(\mathrm{C}_{r_0} )$ and $\chi_-(\mathrm{C}_{r_0} )$. {Therefore, these points will most likely also not be sampled  by IBOSS.} 
 Moreover, 
such points also have a small directional derivative, and 
 the equivalence theorem \citep{atkinson2015designs} implies that
 these points likely have a large distance from the  support points
 of the optimal design.
Thus, points from the sets $\chi_+(\mathrm{C}_{r_0} )$ and $\chi_-(\mathrm{C}_{r_0} )$ are also likely to be ruled out by  ODBSS.

Note that our approach benefits from the advantages of optimal subsampling and additionally uses information from less informative points in terms of summary statistics. We still conduct subsampling on the most informative part of the sample, but 
in contrast to  IBOSS and ODBSS which always focus on data points with high prediction variance,  the proposed method  also uses  some geometric information of the points that are easy to classify via the centroids of the 
region $\chi_+(\mathrm{C}_{r_0} )$ and $\chi_-(\mathrm{C}_{r_0} )$. 
Therefore, one can expect the new method to keep the sample in a more diverse set of directions, which will certainly make the estimator more stable.

\begin{figure}[htbp]
\centering\spacingset{1}
\begin{subfigure}{.35\textwidth}
  \centering
\includegraphics[width=1\linewidth]{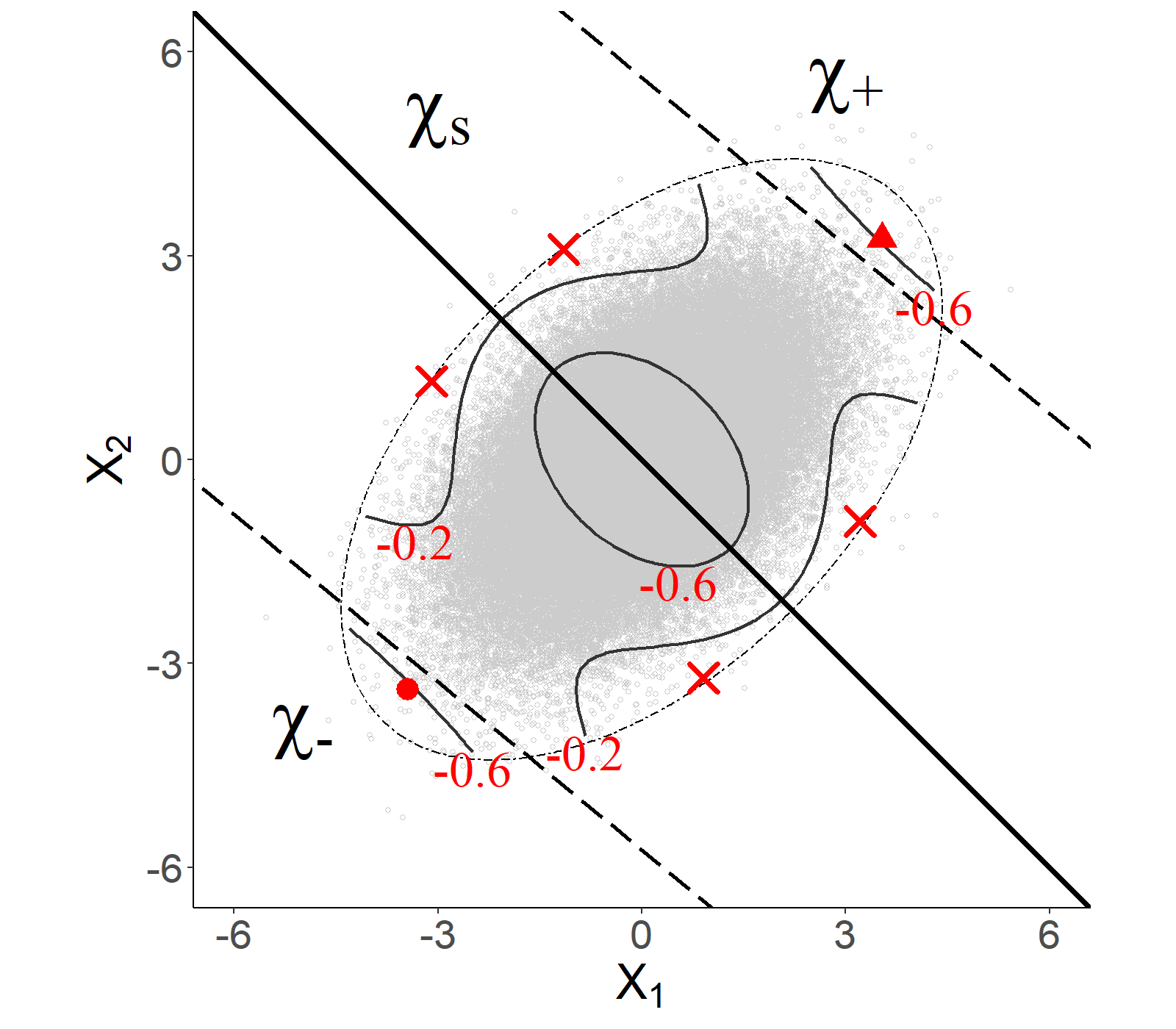}
  \caption{}
\end{subfigure}
\begin{subfigure}{.35\textwidth}
  \centering
\includegraphics[width=1\linewidth]{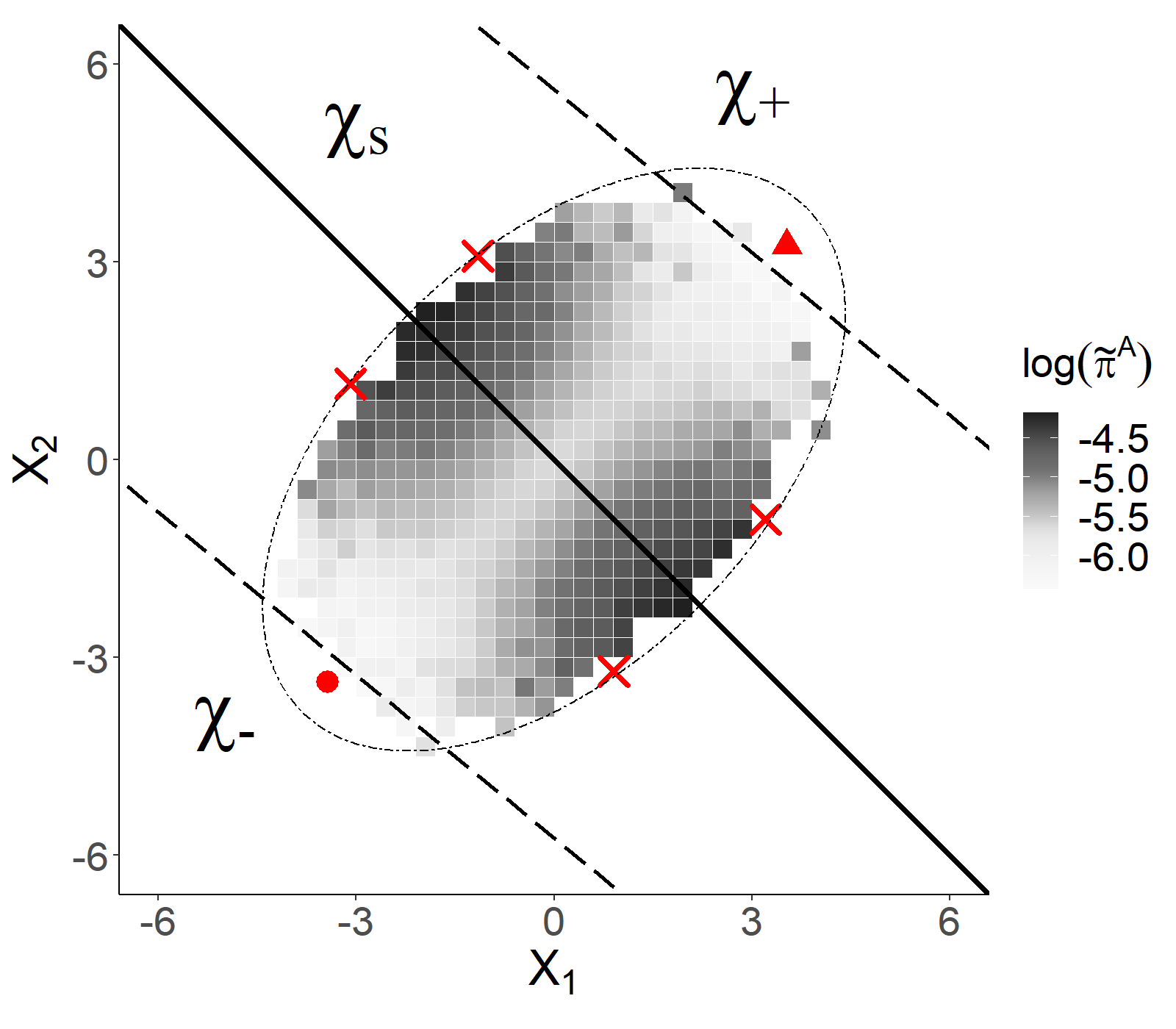}
  \caption{}
\end{subfigure}
    \caption{An illustration of the proposed partition from the perspective of optimal design theory with covariates from a centered Gaussian distribution.
    The red spot and red triangle represent the centroids $\bar{\tX}_-$, and $\bar{\tX}_+$ respectively. In (a), the curves with red numbers are contour lines of the directional derivative of the A-optimality criterion at the optimal design. In (b), we display a heat map  to show the values of the optimal sampling probabilities 
    $\tilde\pi^A_i$ after a log transformation. The crosses mark the support points of the A-optimal design in (a) and (b).}  
    \label{fig.design}
\end{figure}

To further illustrate the advantage of involving the centroids of the $\chi_+(\mathrm{C}_{r_0} )$ and $\chi_-(\mathrm{C}_{r_0} )$, we consider a logistic model with the same data generating process described in Figure~\ref{fig.gradient}.
In  Figure~\ref{fig.design}(a),  we display the data points from the full sample of $n=10^5$ observations  (in grey) and the support points of the optimal design with respect to the A-optimality criterion (marked by crosses),
which was calculated  by the R package ``OptimalDesign'' 
\citep{Optimaldesign2019}.
We also plot contour lines of the directional derivative of the A-optimality criterion at the optimal design. One observes that the data in the sets $\chi_+(\mathrm{C}_{r_0})$ and $\chi_-(\mathrm{C}_{r_0})$ lies in directions roughly parallel to the vector $\bm\theta_t$. Thus such data can be easily predicted by the classifier and is therefore not that informative when estimating the parameter $\bm\theta$.
Thus exploiting only data in $\chi_s(\mathrm{C}_{r_0})$ will naturally mitigate the variance inflation caused by small inclusion probabilities without sacrificing too much information.
In  Figure~\ref{fig.design}(b), we show 
the optimal subsampling  probabilities 
$\tilde{\pi}_i^A$  given in \eqref{eq:chi_s}.
One can observe that the proposed subsampling method assigns larger probabilities to data with a small distance to the support points of the optimal design.
Since the   sampling ratio for the points from $\chi_s(\mathrm{C}_{r_0} )$ is  $r/n_s\ge r/n$, where  $n_s$ in the sample size of $\chi_s(\mathrm{C}_{r_0} )$,
performing sampling only on $\chi_s(\mathrm{C}_{r_0} )$ yields a sample more concentrated around the support of the optimal design.
Furthermore,  using the centroid enables us to explore more directions of a data set and thus will certainly make the estimator more stable.

\end{remark}

\section{MROSS - multi-resolution optimal subsampling}\label{sec:procedure}

  \def\theequation{3.\arabic{equation}}	
  \setcounter{equation}{0}

In this section, we carefully define our proposed subsampling and estimation procedure, which will be called MROSS 
(multi-resolution optimal subsampling) in the following discussion. 
For ease of presentation, we adopt a unified notation $\tilde\pi^{\rm os}_i$ to denote the optimal subsampling probabilities based on $\hat{\bm\theta}_{\pilot}$, 
that is 
\begin{equation}\label{eq:pi-prac}
    \tilde\pi_i^{\rm os}=\frac{r|\dot{\phi}(Y_i\tX_i^\T\hat{\bm\theta}_{\pilot})|h(\tX_i)}{\sum_{l=1}^n|\dot{\phi}(Y_l\tX_l^\T\hat{\bm\theta}_{\pilot})|h(\tX_l)\mathbb{I}(|\tX_l^\T\hat{\bm\theta}_{\pilot}|<\mathrm{C}_{r_0} )}\wedge 1,
\end{equation}
with $h(\tX_i)=\|\tX_i\|$ for $L$-optimal subsampling, and $h(\tX_i)=\|\hat{H}^{-1}(\hat{\bm\theta}_{\pilot})\tX_i\|$ for $A$-optimal subsampling. 
The details of the subsampling and estimation procedure are summarized in Algorithm \ref{alg:RBS1}. Note that we solve the equation \eqref{det201} instead of \eqref{eq:sub-ee} to make full use of the information from the pilot sample.

\begin{algorithm}[H]
{\SetAlgoLined
\spacingset{1}
\small
\LinesNumbered
\medskip 

{\bf Input.} {The entired dataset $\fu$ of size $r_0+n$, threshold $\mathrm{C}_{r_0}  >0 $, and $g(\XY)$ (contains a constant component)}.  

 {\bf Pilot information.} Randomly spilt data $\fu$ into two parts of  size  to $r_0$ and $n$. Use the first part $\mathcal{F}_{r_0}^1$ to obtain the pilot estimator $\hat{\bm\theta}_{\pilot}$, 
and the Hessian matrix $\hat{H}(\hat{\bm\theta}_{\pilot})$, if $A$-optimal subsampling is adopted. Then conduct the subsampling on $\fn:=\fu\setminus\fp$.

{\bf Initialize.} {$\bar{\tX}_+=\bar{\tX}_-=\bar{\g}=n_+=n_-=0$ and $\indexset=\emptyset$.}

\For{$i=1,\ldots,n$}{
Calculate $\itm=\tX_i^\T \hat{\bm\theta}_\pilot$.

\If{$\itm>\mathrm{C}_{r_0} $, $Y_i = 1$}{Update $\bar{\tX}_+=\frac{n_+}{n_++1}\bar{\tX}_++\frac{1}{n_++1}\tX_i$;

Update $n_+=n_++1$}
\ElseIf{$\itm<-\mathrm{C}_{r_0} $, $Y_i = -1$}{Update $\bar{\tX}_-=\frac{n_-}{n_-+1}\bar{\tX}_-+\frac{1}{n_-+1}\tX_i$; 

Update $n_-=n_-+1$}

\Else{
Calculate the inclusion probability $\tilde\pi_i^{\rm os}$ according to \eqref{eq:pi-prac}.

Generate $\delta_i\sim\textrm{Bernoulli}(1,\tilde\pi_i^{\rm os})$

\If{$\delta_i=1$}{
Update $\indexset=\indexset\cup\{(Y_i,\tX_i,\tilde\pi_i^{\rm os})\}$;

Update $\bar{\g}=\frac{i-1}{i}\bar{\g}+\frac{1}{i}g(\X_i,Y_i)$;
}
\Else{
Update $\bar{\g}=\frac{i-1}{i}\bar{\g}+\frac{1}{i}g(\X_i,Y_i)$;
}}}
\medskip

{Determine  the estimator $\tilde{\bm\theta}_I$ based on $ \mathcal{F}_{r_0}^1 \cup  \indexset $ and the summary statistics $\bar{\tX}_+,\bar{\tX}_-,\bar{g},n_+$ and $n_-$
as the solution of  the equation 
\begin{equation}
\label{det201}
\bm 0 = \frac{1}{n+r_0} \sum_{{
(\X_i , Y_i) \in \mathcal{F}_{r_0}^1} }
{\score}_{i}(\bm\theta) +\frac{n}{n+r_0} {L ( \bm \theta )},
\end{equation} 
 where ${L ( \bm \theta )}$ is defined in  \eqref{eq:sub-ee} { with  $\pi_i=\tilde\pi_i^{\rm os}$ and the parameter 
 $\bm \theta_t$  in the definition of the sets
 $ \chi_{+}(\mathrm{C}_{r_0} )$, $ \chi_{-}(\mathrm{C}_{r_0} )$ and $ \chi_s(\mathrm{C}_{r_0} )$ is replaced by the pilot estimate $\hat{\bm \theta}_{\pilot} $.
}
}
}
\caption{\small Multi-resolution optimal subsampling (MROSS)} 
\label{alg:RBS1}
\end{algorithm}

\medskip

\noindent 
In the following discussion, we determine 
the asymptotic properties of the estimator
$\tilde{\bm\theta}_I$ calculated by Algorithm \ref{alg:RBS1}. For this purpose, we need a further assumption.

\begin{assumption}\label{ass:convex}
(i) $\E\|\hat{\bm\theta}_{\pilot}-\bm\theta_t\|^2=O(r_0^{-1})$;
(ii) $\E\|\tX\|^4<\infty$, $(r/n)\E \big( \max_i^n\|\tX_i\|
\big) =o(1)$ as $r,n\to\infty$;
(iii) The threshold $\mathrm{C}_{r_0}  > 0 $  (depending on $r_0$) satisfy that $\dot\phi(\mathrm{C}_{r_0} )<0$ and $|\sqrt{r_0}\dot\phi(\mathrm{C}_{r_0} )|^{-1}=o(1)$  as $r_0 \to \infty$ almost surely; 
 (iv) $\sqrt{r}\E\big(\|\bm{X}_1 \|^{2}\mathbb{I}(|\hat{\bm \theta}_{\pilot}^\T\bm{X}_1 |\ge \mathrm{C}_{r_0} )|\hat{\bm \theta}_{\pilot}\big)=o_P(1)$ as $r,r_0 \to \infty$; (v)
$\E \big(\|g(\mathbf{X}_1,Y_1)\|^4|\hat{\bm\theta}_\pilot\big)<\infty$ in probability;  the  $\sqrt{r}\E\big(\|g(\mathbf{X}_1,Y_1) \|^{2}\mathbb{I}(|\hat{\bm \theta}_{\pilot}^\T\bm{X}_1 |\ge \mathrm{C}_{r_0} )|\hat{\bm \theta}_{\pilot}\big)=o_P(1)$;  
$
\E\big(g(\mathbf{X}_1,Y_1 )g(\mathbf{X}_1,Y_1 )^\T\mathbb{I}((\mathbf{X}_1,Y_1 )\in\chi_s(\mathrm{C}_{r_0} )) |\hat{\bm\theta}_\pilot,\mathrm{C}_{r_0} \big) $
 is a positive definite (almost surely) as $r,r_0 \to \infty$.
\end{assumption}

Here we use $\mathbf{X}_1,Y_1 $ to emphasize the randomness that comes from the data generating process of $\fn$ in the second stage of subsampling. 
Due to data splitting, $\mathbf{X}_1,Y_1 $ is independent of $\fp$, $\mathrm{C}_{r_0} $, and $\hat{\bm\theta}_\pilot$. However 
the function $g(\XY)$ usually depends on $\hat{\bm\theta}_{\pilot}$ even though this not indicated by out notation; see for example the choice suggested in \eqref{det200}.
Condition~\ref{ass:convex} essentially adds some constraints on the subsampling probability distribution $\Tilde{\bm\pi}$. 
For most of the commonly used loss functions, 
{part (i) 
is satisfied}
\citep[see, for example][]{Yoichi2010moment}. 
The first part in Condition~\ref{ass:convex}(ii) is the same moment condition as in Condition~\ref{ass:6}(i).  The second part, $(r/n)\E \big( \max_i^n\|\tX_i\|
\big) =o(1)$,  guarantees $\tilde\pi_i^{\rm os} \in [0,1]$  with high probability for  the sampling 
probabilities   in \eqref{eq:pi-prac}. 
As    a consequence  $\E \big(\sum_{l=1}^n\tilde{\pi}_l^{\rm os} \big) $ 
concentrates around subsampling budget $r$.
If  $\tX_i$'s are sub-Guassian random variables, they can be replaced by $r\log^{1/2}(n)/n\to 0$. 
This assumption can be avoided by replacing the probabilities in \eqref{eq:pi-prac} by 
\begin{equation}\label{eq:pi-prac-M}
    \tilde\pi_i^{\rm os}=r\frac{|\dot{\phi}(Y_i\tX_i^\T\hat{\bm\theta}_{\pilot})|h(\tX_i)\wedge M}{\sum_{l=1}^n|\dot{\phi}(Y_l\tX_l^\T\hat{\bm\theta}_{\pilot})|h(\tX_l)\mathbb{I}(|\tX_l^\T\hat{\bm\theta}_{\pilot}|<\mathrm{C}_{r_0} )\wedge M},
\end{equation}
where  $a\wedge b=\min(a,b)$ and 
$M$ is the maximum number of satisfying  
$$r(|\dot{\phi}(Y_i\tX_i^\T\hat{\bm\theta}_{\pilot})|h(\tX_i)\wedge M)\le \sum_{\chi_s(\mathrm{C}_{r_0} )}(|\dot{\phi}(Y_l\tX_l^\T\hat{\bm\theta}_{\pilot})|h(\tX_l)\wedge M),$$ for all $(\tX_i,Y_i)\in\chi_s(\mathrm{C}_{r_0} )$. As shown in \cite{yu2020quasi}, the corresponding probabilities still minimize the asymptotic variance of  the estimators $\tilde{\bm\theta}_I$ and  $H(\bm\theta_t)\tilde{\bm\theta}_I$
under the constraints $\pi_i\in[0,1]$ for all $i$, $\sum_{l=1}^n\pi_i=r$.  When  $r/n \to c > 0$,  $M$ can be estimated as the empirical  quantile of  the sample  $\{|\dot{\phi}(Y_i\tX_i^\T\hat{\bm\theta}_{\pilot})|h(\tX_i)~\text{for}~ (\tX_i,Y_i)\in\chi_s(\mathrm{C}_{r_0} )\}$ to further reduce the computational costs.
Condition~\ref{ass:convex}(v) 
guarantees that the matrix  $\sum_{  \chi_s(\mathrm{C}_{r_0} )}{\delta_l}/({n{\pi}_l})\g_l\g_l^\T$ 
is positive definite  with high probability 
and its inversion is numerically stable. 
A similar assumption is also made in Condition~\ref{ass:gx}.

A clear trade-off between the pilot sample size and subsample size can be found in Conditions~\ref{ass:convex}(iii) and (iv).
Namely, a more accurate pilot estimator can tolerate a smaller inclusion probability since the numerator and
denominator in $(\tilde\pi_i^{{\rm os}})^{-1}\delta_i {\dot{\phi}(Y_i\tX_i^\T\bm\theta_t)}$ has the same convergence rate.
On the other hand, a larger subsample size requires a small ``truncation'' otherwise the bias caused by the approximation in \eqref{eq:sub-ee} can not be ignored. 

The following lemma provides some examples where  Condition~\ref{ass:convex} holds. 
Throughout this paper, the notation $A\asymp B$ means  that $A$ and $B$  have the same order.

\begin{lemma}
\label{lem:orderofC}
Let $\tX_{1,-1}$ be a sub-Gaussian random vector with mean-zero and finite sub-Gaussian norm, where $\tX_{1,-1}$ is obtained from of $\tX_{1}$ removing the intercept term.  
Assume that  one of the following setups holds.
    \begin{enumerate}
         \item[(i)]  $\phi(\cdot)$ is  the logistic loss, $\mathrm{C}_{r_0}  = \kappa( \log r_0- \log\log r_0)$  for some   $\kappa\in(0,1/2]$ and $r \asymp r_0^{\alpha }/\log^{\alpha+5}(r_0)$ for some   $\alpha  > 1$.  
        \item[(ii)]  $\phi(\cdot)$ is the DWD loss $\mathrm{C}_{r_0}  =\kappa r_0^{1/4-\gamma}$ 
        for some  $\gamma\in(0,1/4), \kappa\in(0,+\infty)$ 
        and $r \asymp r_0^{\alpha }$ for some 
        $\alpha >1$.
    \end{enumerate}
    Then part (iii) and (iv) of  Condition~\ref{ass:convex} are satisfied.
\end{lemma}
 It is worth mentioning that the  (i) and (ii)  in Lemma~\ref{lem:orderofC} are of asymptotic nature  and usually conservative. In practice, a more intuitive way  to choose $\mathrm{C}_{r_0} $ is to estimate $\eta(\x)$ using  the working model and the pilot sample. Then one can choose $\mathrm{C}_{r_0} $ according to Lemma~\ref{lem:probregion}. For example, in the logistic regression one can estimate $\hat\eta(\x)=1/(1+\exp(-\x^\T\hat{\bm\theta}_\pilot)$ from the data $\mathcal{F}_{r_0}^1$. Then one  can choose $\mathrm{C}_{r_0} =\x_0^\T\hat{\bm\theta}_\pilot$ with some $\x_0$ satisfying $\hat\eta(\x_0)=0.99$. Here $0.99$ can be replaced by any other user-specified level to ensure that the gradient norm lies in a reasonable region implied by  Lemma~\ref{lem:probregion}.
This rule was implemented in the numerical study in the following Section \ref{sec:sim}.

Our final result provides the asymptotic distribution of the estimator $\tilde{\bm \theta}_I$ calculated by Algorithm \ref{alg:RBS1}.
For the sake of a simple notation, we denote the conditional expectation  with respect to ${\cal F}_{r_0}^1$ by $\E_{{\cal F}_{r_0}^1}$

\begin{theorem}\label{thm:RB}
  Let  Conditions~\ref{ass:1}--\ref{ass:3},  and~\ref{ass:convex} be satisfied 
  and assume that  $r_0/r\to 0$, $r\log^2(n)/n\to 0$ as $n,r,r_0\to\infty$. Then, 
    conditional on $\fp$,  the estimator $\tilde{\bm\theta}_{I}$ defined by Algorithm~\ref{alg:RBS1} 
   satisfies 
    \begin{equation*}
         V_I^{-1/2}(\tilde{\bm\theta}_{I}-\bm\theta_t)\to N(\bm 0,I_d), 
     \end{equation*} 
      in probability,
     where $V_I=H^{-1}(\bm\theta_t)V_{I,C}H^{-1}(\bm\theta_t)$,
     \[
     V_{I,C}=\E_{\fp}\big(|\dot{\phi}(Y_1 \bm{X}_1 ^\T\hat{\bm\theta}_{\pilot})|\|\bm{X}_1 \|\mathbb{I}(|\bm{X}_1 ^\T\hat{\bm\theta}_{\pilot}|<\mathrm{C}_{r_0} )\big)\E_{\fp}\Big(\frac{\bm{\aleph}_1\bm{\aleph}_1^\T\mathbb{I}(|\bm{X}_1 ^\T\hat{\bm\theta}_{\pilot}|<\mathrm{C}_{r_0} )}{r|\dot\phi(Y_1 \bm{X}_1 ^\T\hat{\bm\theta}_{\pilot})|\|\bm{X}_1 \|}\Big),
     \]
       $\bm{\aleph}_1={\score_1(\bm\theta_t)} -\coef_{\fp}(\bm\theta_t)^\T g(\mathbf{X}_1,Y_1 )$, {$\score_1(\bm\theta_t) = \dot\phi(Y_1\tX_1^\T\bm\theta_t)Y_1\tX_1$},  and    
       \begin{align}
           \coef_{\fp}(\bm\theta_t) =& \G_{\fp}^{-1}\E \big(g(\mathbf{X}_1,Y_1 )\score_1(\bm\theta_t)^\T\I((\mathbf{X}_1,Y_1  ) \in\chi_s(\mathrm{C}_{r_0} )) | \fp \big),
           \label{hol400}
       \end{align}
       with $\G_{\fp}=\E_{\fp} \big(g(\mathbf{X}_1,Y_1 )g(\mathbf{X}_1,Y_1 )^\T\I((\mathbf{X}_1,Y_1  ) \in\chi_s(\mathrm{C}_{r_0} ))\big)$.   
\end{theorem}
If the function $g(\cdot)$ is selected as suggested in Theorem~\ref{thm:2} and subsampling by  $\tilde{\bm\pi}^L$ is adopted, we will show 
in  {Section~S.10} of  the supplementary material that 
\begin{equation}
\label{det202}
\tr(V_{I,C})
    \le \E_{\fp}\big(|\dot{\phi}(Y_1 \bm{X}_1 ^\T\hat{\bm\theta}_{\pilot})|\|\bm{X}_1 \|\big)
\E_{\fp}\big({|\dot\phi(Y_1 \bm{X}_1 ^\T\hat{\bm\theta}_{\pilot})|\|\bm{X}_1 \|}\big)+o_P(1)<\infty
\end{equation}
in probability. Consequently, with high probability, the estimator $\tilde{\bm\theta}_{I}$ will not suffer from the variance inflation caused by extremely small inclusion probabilities.
Furthermore, one can expect a variance reduction by $\tilde{\bm\theta}_{I}$ compared with the ordinal subsample based estimator $\Tilde{\bm\theta}$ when  $A$- or $L$-optimality subsampling strategies are  applied.
{We will demonstrate these advantages by means of  simulation studies in the following section.}

\section{Numerical studies}\label{sec:sim}
  \def\theequation{4.\arabic{equation}}	
  \setcounter{equation}{0}

In this section, we evaluate the performance of  MROSS (Algorithm~\ref{alg:RBS1}) using both synthetic and real datasets. We take logistic regression and DWD as examples to illustrate the performance of the proposed method and also compare it with other methods for linear classification problems. 
All methods are implemented with the R programming language. 

\subsection{Simulation studies}\label{subsec:sim}
In this section, we generate full data of size $n = 5\times10^5$ with $d=21$ dimensional features (including an intercept term) under the following six scenarios.

\begin{description}
\spacingset{1.5}
    \item[Case 1. (Logistic regression)] The response $Y$ is generated by a standard logistic regression model: $\pr(Y=1|\X) = (1+\exp(-\tX^{\T}\bm\theta_t))^{-1}$ with
     $\bm\theta_t = (0,0.5, \ldots , 0.5)^\T \in \mathbb{R}^{21}$.  The covariate (except the intercept term) follows a $20$-dimensional multivariate normal distribution, $\mathcal{N}(\bm 0_{20}, \Sigma_1)$ with $\Sigma_1 = 
    \big ( 0.5^{|i-j|}
     \big )_{i,j=1, \ldots , 20} $. 
 
    \item[Case 2. (Logistic regression)] This scenario is the same as Case 1 except  that the covariate follows a mixture of two  normal distributions, that is $0.5\mathcal{N}(\bm 0_{20}, \Sigma_1)+0.5\mathcal{N}(\bm 0_{20}, \Sigma_2)$, where $\Sigma_1$ is defined in Case 1 and $(\Sigma_2) = \big ( 0.5^{\mathbb{I}(i\neq j)}\big )_{i,j=1, \ldots , 20} $.
    \item[Case 3. (Logistic regression)] This scenario is the same as Case 1 except that the covariate  follows a multivariate $t$ distribution with $3$ degrees of freedom, that is $\X_i \sim t_3(\bm 0_{20}, \Sigma_1)$, where $\Sigma_1$ is defined in Case 1.
    \item[Case 4. (Discriminative model)] The data generated from a discriminative model that labels ($Y$) cause covariates ($\tX$). To be precise,  $\X|Y=1\sim\mathcal{N}(\bm\mu_1, \Sigma_1)$  and $\X|Y=-1\sim\mathcal{N}(\bm\mu_2, \Sigma_2)$ with $\bm\mu_1 = {0.5}\cdot\bm 1_{20}$, $\bm\mu_2 = -{0.5}\cdot{\bm 1}_{20}$. Here $\bm 1_d$ is a $d$-dimensional vector with all entries being 1 and $\Sigma_1$ and $\Sigma_2$ are defined as in Cases 1 and 2, respectively. The number of instances that $Y=1$ is the same as $Y=-1$.
    \item[Case 5. (Discriminative model)] This scenario is the same as Case 4 expect $\X|Y=1\sim 0.5\mathcal{N}(\bm\mu_{11}, \Sigma_1) + 0.25\mathcal{N}(\bm\mu_{12}, \Sigma_1) + 0.25\mathcal{N}(\bm\mu_{13}, \Sigma_1)$ and  $\X|Y=-1\sim 0.5\mathcal{N}(\bm\mu_{21}, \Sigma_1) + 0.25\mathcal{N}(\bm\mu_{22}, \Sigma_1) + 0.25\mathcal{N}(\bm\mu_{23}, \Sigma_1)$. Here $\bm\mu_{11} = (\bm 0_{d/2}^{\T},\bm 1_{d/2}^{\T})^{\T}$, $\bm\mu_{12} = (-\bm 1_{10}^{\T},2\cdot\bm 1_{10}^{\T})^{\T}$, $\bm\mu_{13} = -\bm 1_{20}^{\T}$, $\bm\mu_{21} = (\bm 0_{10}^{\T},-\bm 1_{10}^{\T})^{\T}$, $\bm\mu_{22} = (\bm 1_{10}^{\T},-2\cdot\bm 1_{10}^{\T})^{\T}$, $\bm\mu_{23} = (\bm 1_{10}^{\T},2\cdot\bm 1_{10}^{\T})^{\T}$ and $\Sigma_1$ is defined in Case 1.
    \item[Case 6.(Discriminative model)] In this scenario we consider $\X|Y=1\sim t_3(\bm\mu_{1}, \Sigma_1)$ and  $\X|Y=-1\sim t_3(\bm\mu_{2}, \Sigma_1)$. Here $\bm\mu_{1} = (\bm 0_{10}^{\T},\bm 1_{10}^{\T})^{\T}$,  $\bm\mu_{2} = (\bm 0_{10}^{\T},-\bm 1_{10}^{\T})^{\T}$, and $\Sigma_1$ is defined in Case 1. In addition, this is a moderately imbalanced case in which 80\% instances belong to the class with $Y=1$.
\end{description}

When the logistic regression is taken into account, Cases~1-3 stand for a  scenario where the model is correctly specified and Cases 4-6 stand for model misspecification. 
The performance of a sampling strategy is evaluated by the empirical MSE from $S = 500$ replications. More precisely,
$
    \mathrm{MSE}(\tilde{\bm\theta}) = {S}^{-1}\sum_{s = 1}^S\|\tilde{\bm\theta}^{(s)} - \bm\theta_t\|^2,
$
where $\tilde{\bm\theta}^{(s)}$ is the subsample based estimate from the  $s$th simulation run.  
In  cases where  the risk minimizer
$\bm\theta_t$ defined in \eqref{eq:risk} cannot be calculated explicitly we determine it as the  empirical $M$ estimator on a larger data set with size $10n = 5\times 10^6$.

\textbf{Logistic regression.} In the following, we train the linear classifier based on logistic loss.
We implemented MROSS with $r_0=1000$ and $\mathrm{C}_{r_0} =6.9$  using the  $L$-optimal subsampling with $h(\tX_i)=\|\tX_i\|$ in \eqref{eq:pi-prac} for computational efficiency.
For the sake of comparison, the following subsampling methods are also considered. 

\noindent 
1) Uniform subsampling (UNIF) with sampling probability $\pi_i=(r_0+r)/n$. 

\noindent 
2) OSMAC: optimal subsampling with  $L$-optimal 
sampling probabilities, that is  $\pi_i\propto r|\dot{\phi}(Y_i\tX_i^\T\hat{\bm\theta}_{\pilot})|\|\tX_i\|$. 

\noindent 
3) MSCLE: the maximum sampled conditional likelihood estimator proposed in \cite{wang2022maximum}  with the sampling probabilities  as in  OMASC. 

\noindent 
4) IBOSS: information based subdata selection for logistic regression proposed in \cite{cheng2020information}.

To avoid scanning the data twice, we use the first $r_0$ data points to obtain the  pilot estimate $\hat{\bm\theta}_\pilot$ and conduct the subsampling on the rest $n-r_0$ data points. In 
Figure~\ref{fig:simu.logistic.20} we present (the logarithm of) the empirical MSE where the size of the subsample varies from $2000$ to $5000$. 
 Note that the sampling rate   $ r/n$ is at most  $0.01$ for all cases under consideration.

\begin{figure}[htbp]
\centering\spacingset{1}
\begin{subfigure}{.32\textwidth}
  \centering
\includegraphics[width=1\linewidth]{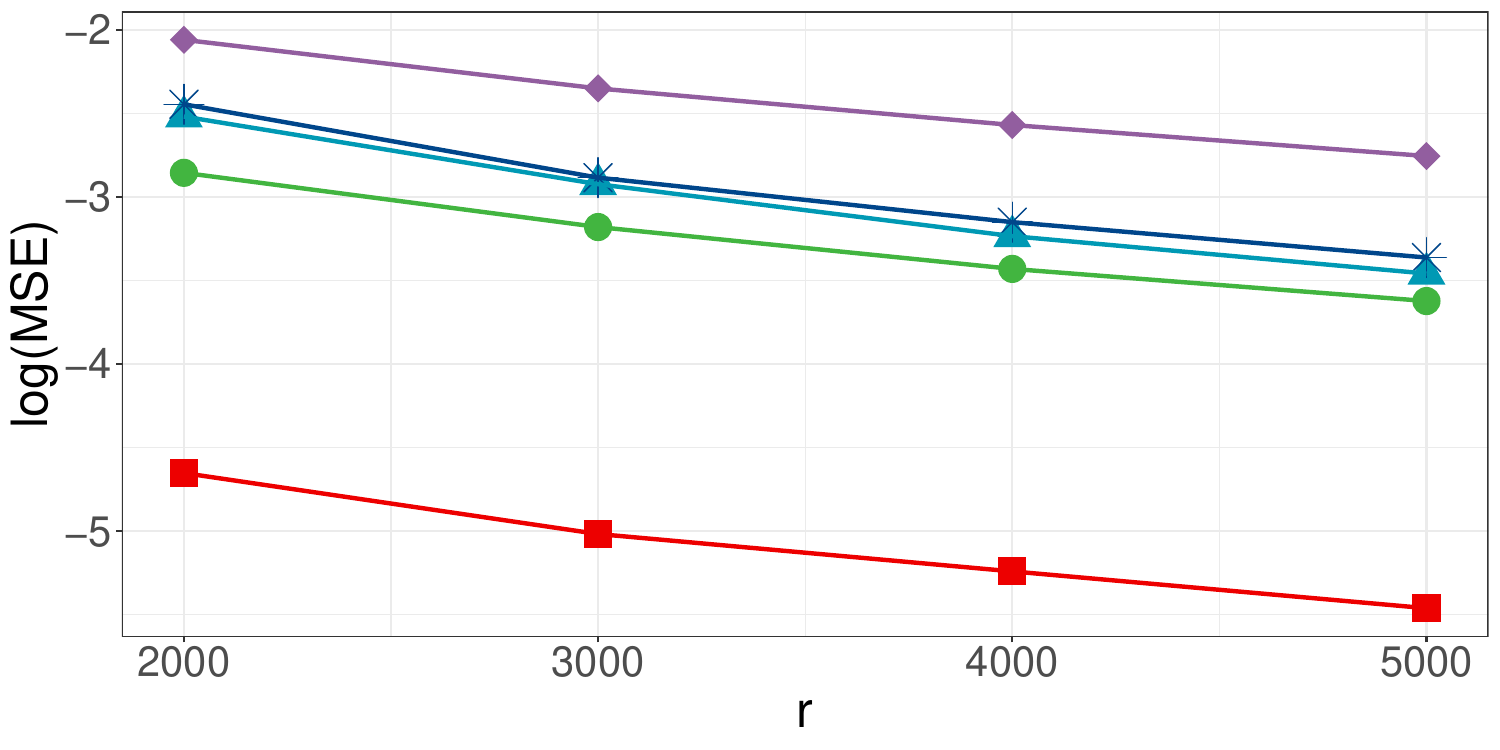}
  \caption{Case 1}
\end{subfigure}
\begin{subfigure}{.32\textwidth}
  \centering
\includegraphics[width=1\linewidth]{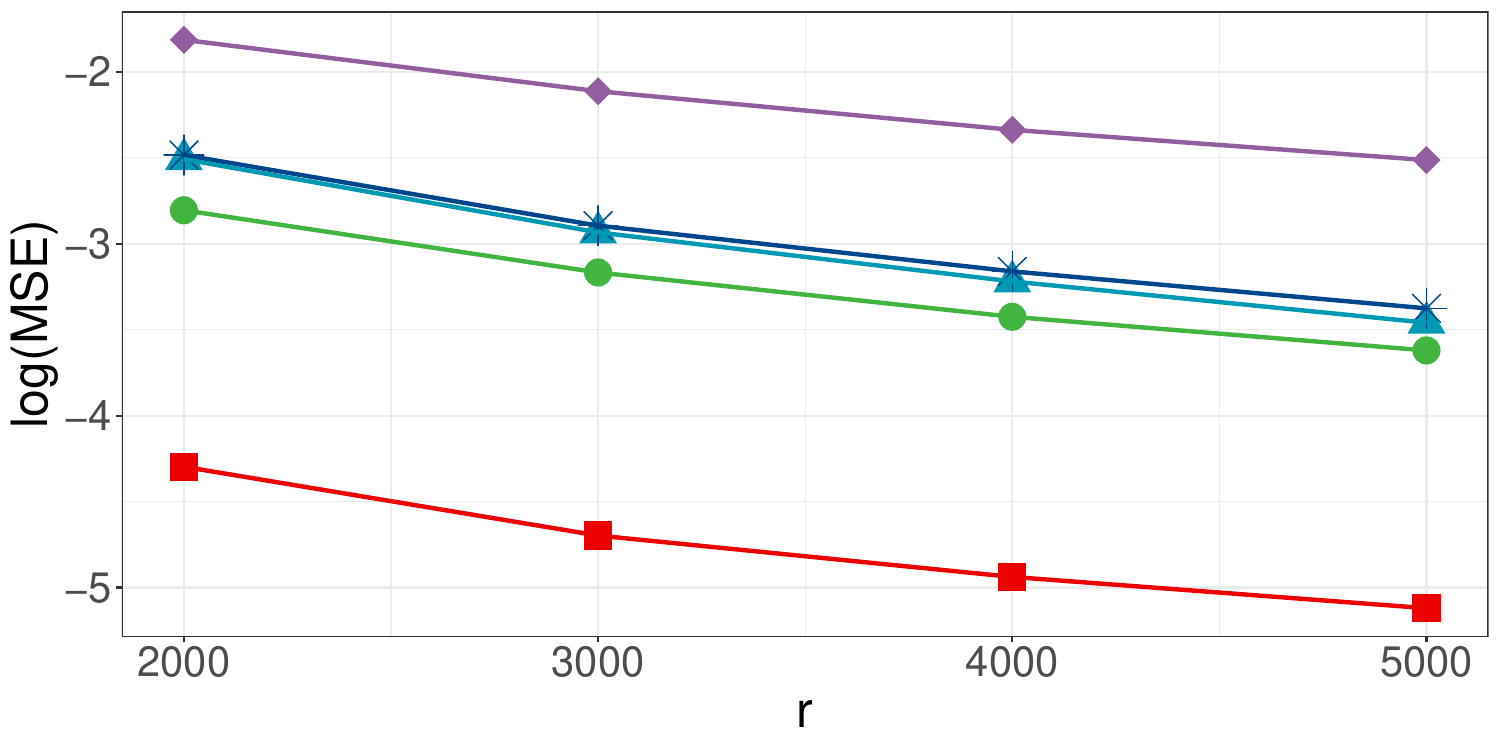}
  \caption{Case 2}
\end{subfigure}
\begin{subfigure}{.32\textwidth}
  \centering
\includegraphics[width=1\linewidth]{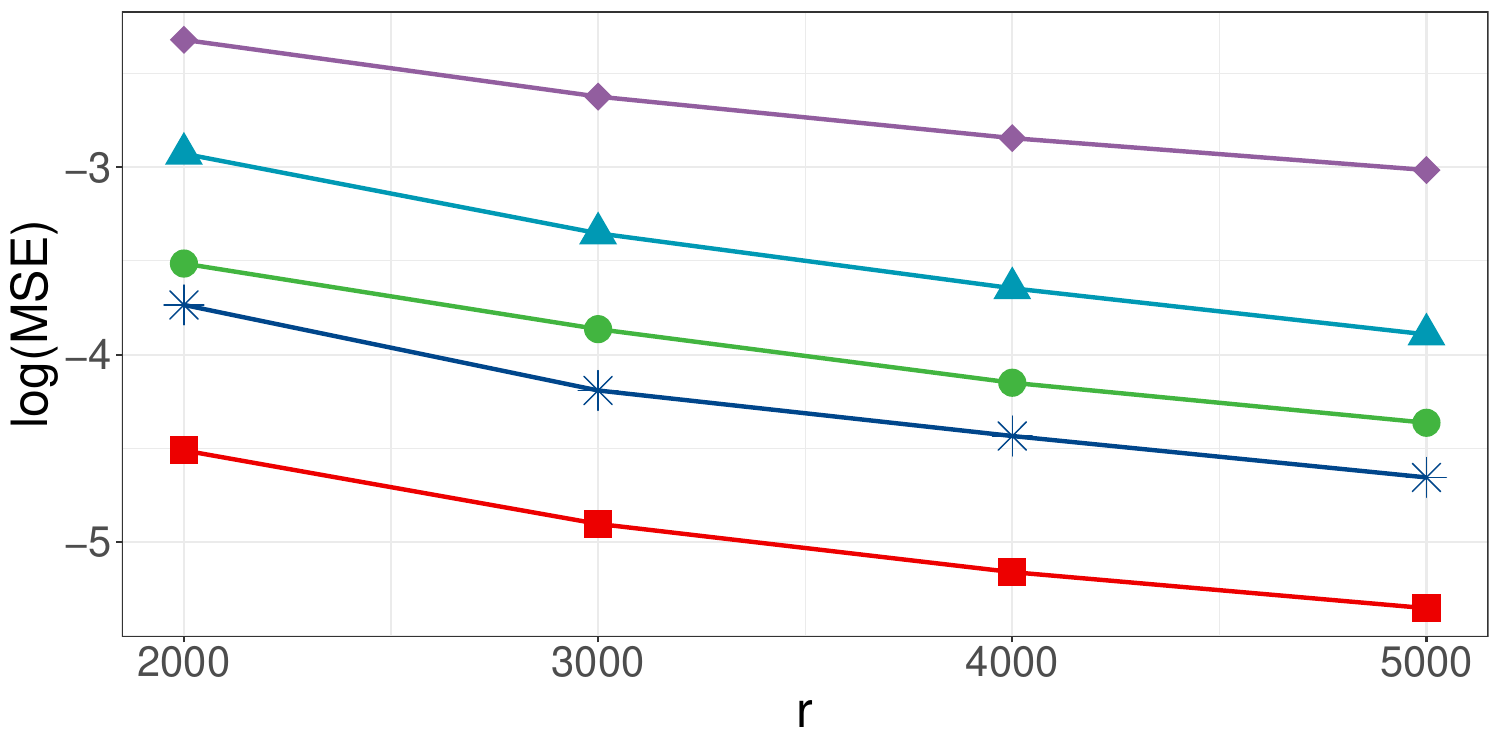}
  \caption{Case 3}
\end{subfigure}\\
\begin{subfigure}{.32\textwidth}
  \centering
\includegraphics[width=1\linewidth]{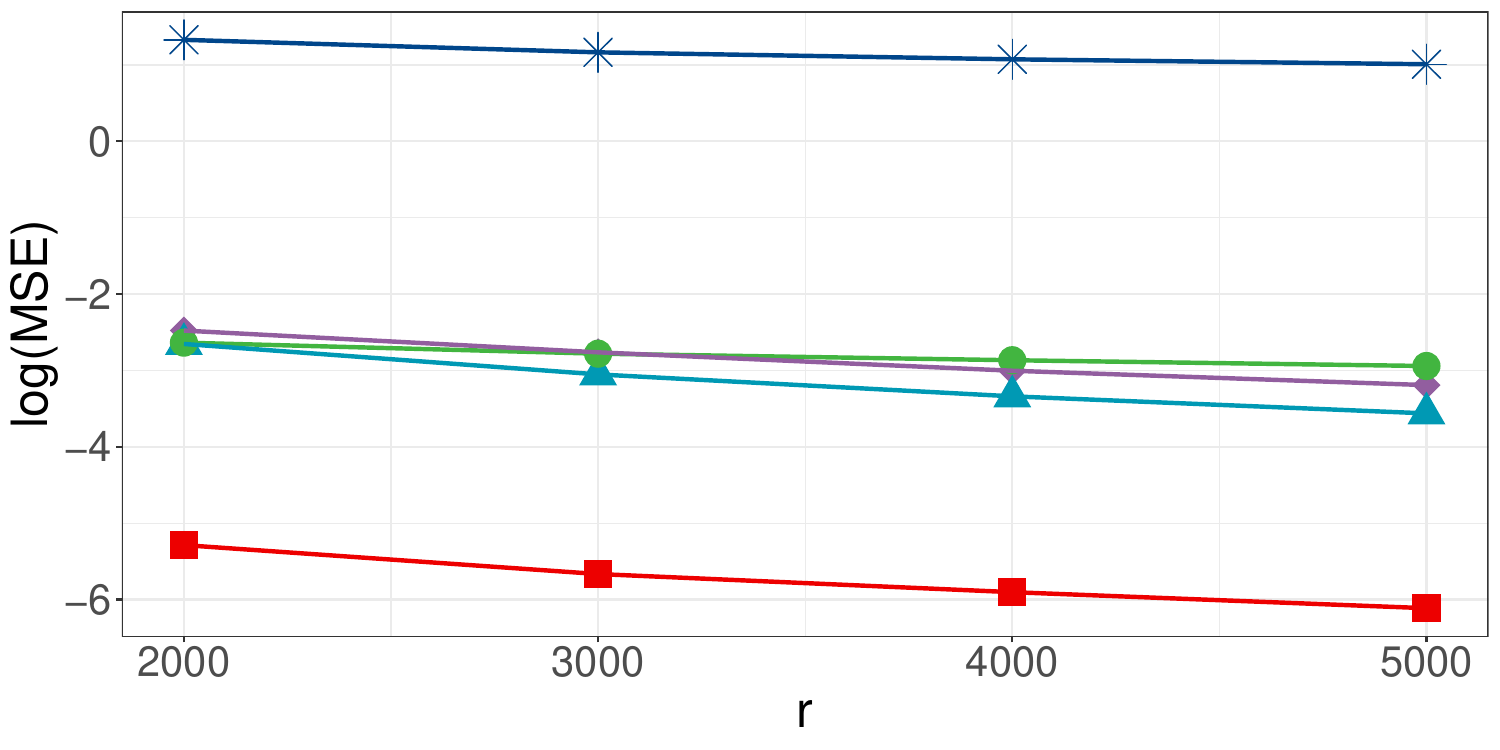}
  \caption{Case 4}
\end{subfigure}
  \begin{subfigure}{.32\textwidth}
  \centering
\includegraphics[width=1\linewidth]{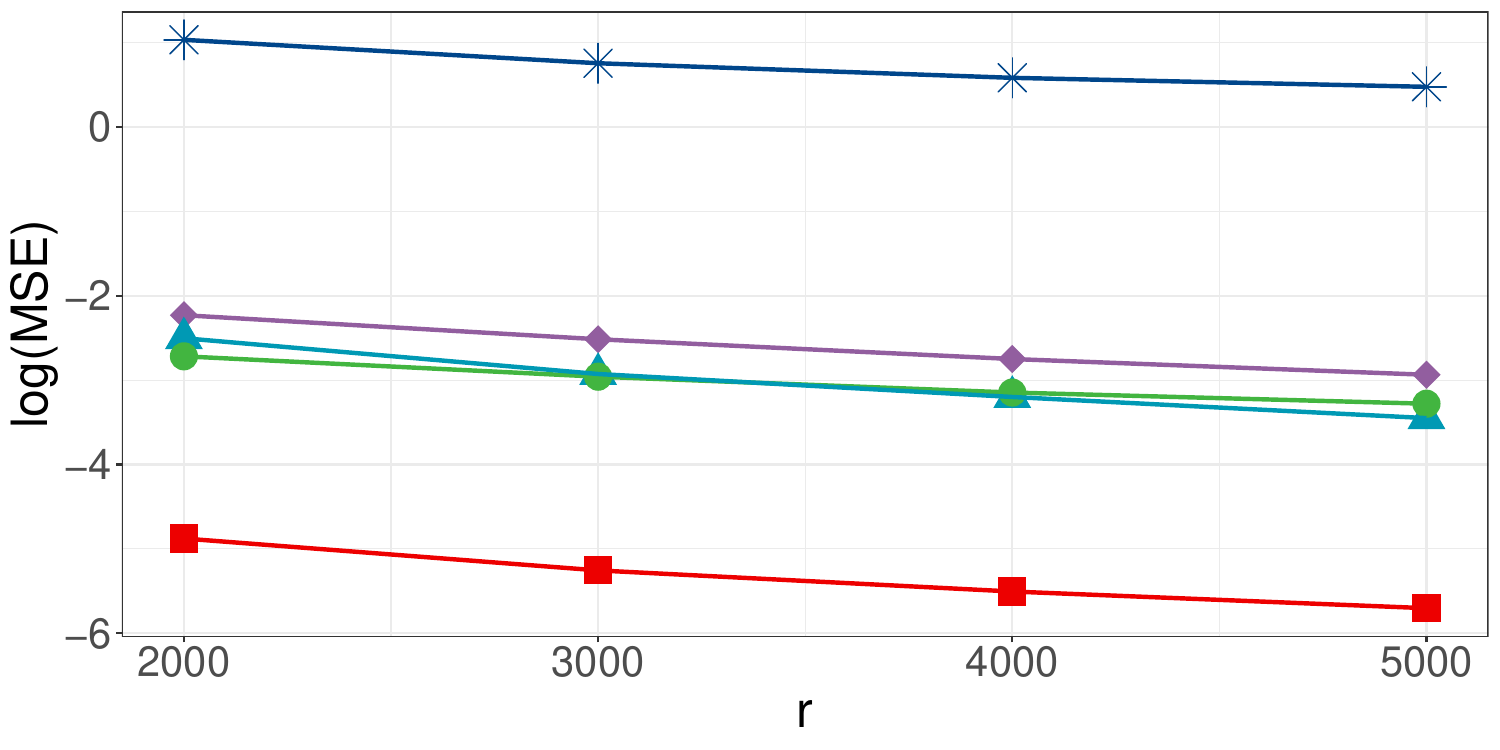}
  \caption{Case 5}
\end{subfigure}
\begin{subfigure}{.32\textwidth}
  \centering
\includegraphics[width=1\linewidth]{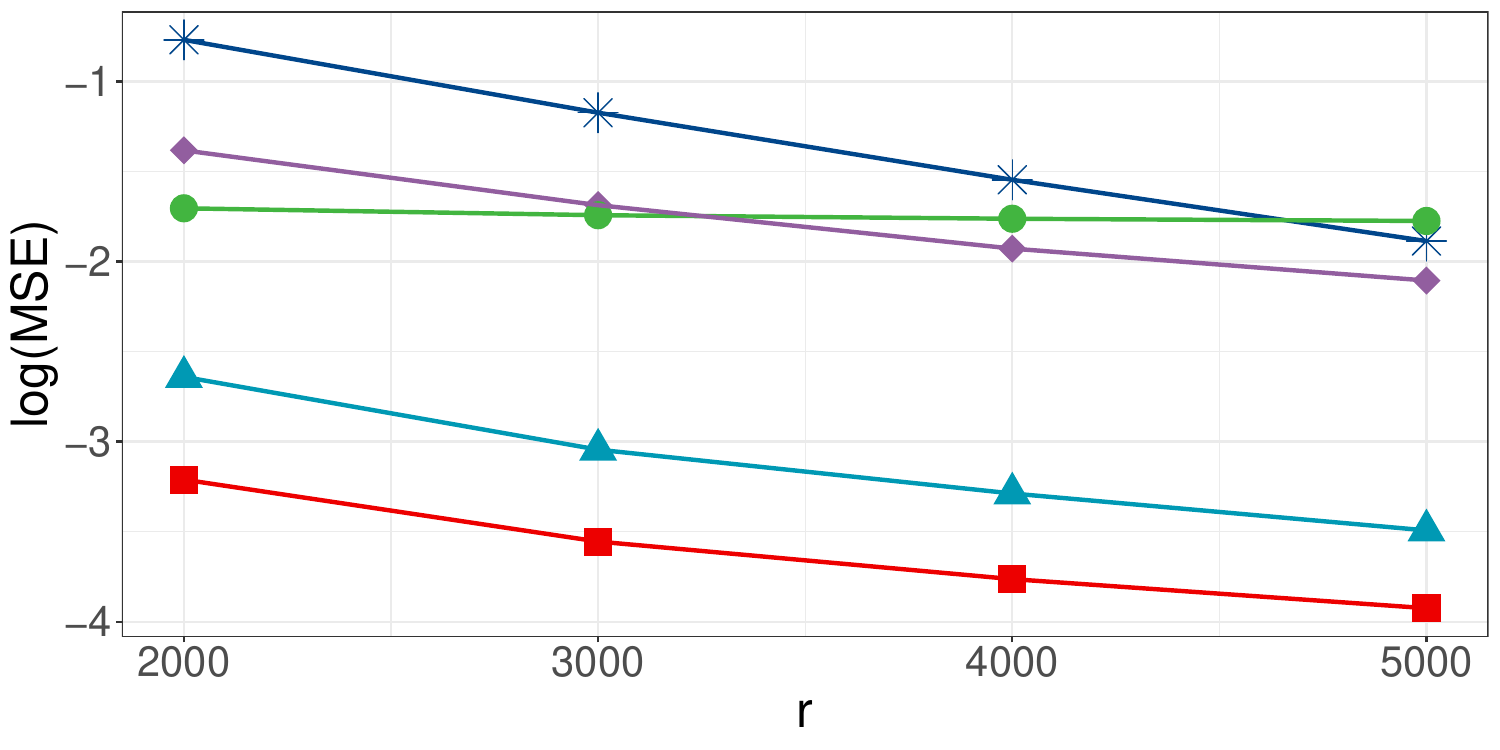}
  \caption{Case 6}
\end{subfigure}
    \caption{{The log MSE with different subsample size ${r}$
    for logistic regression based on UNIF ($\blacklozenge$), OSMAC ($\blacktriangle$), MSCLE ($\bullet$), IBOSS ($\ast$) and MROSS (Algorithm~\ref{alg:RBS1}) ($\blacksquare$).}}
    \label{fig:simu.logistic.20}
\end{figure}

We observe that the proposed method outperforms the four competitors in all six cases.
The MSE decreases as $r$ increases which echoes the results in Theorem~\ref{thm:RB}.
Compared with OSMAC a significant variance reduction as implied in Theorem~\ref{thm:2} can be observed in all six cases.
Note that in Cases 4-6 the logistic model is no longer  the true data-generating process. Therefore, it  is not  surprising   that  IBOSS and MSCLE do not perform well in these scenarios since the optimal design and conditional likelihood inference heavily rely on a correct model specification. In contrast,  UNIF, OSMAC, and MROSS  proposed in  this paper work well no matter whether the model is correctly specified or not.
The reason for this observation is that these methods use the  Thomason-Hurkovitz weighting scheme to provide some robustness against model misspecification \citep{Hansen1983Evaluation}.

In practice,  subsampling is usually applied once for big data analysis.
Therefore, it is important to investigate  the stability of  the  different
subsampling methods. If the methods are not stable the results can be worse or 
better just  by chance.  To address this issue 
we display in Figure~\ref{fig:simu.logistic.boxplot} boxplots of the empirical MSE  for the five subsampling strategies under consideration, where the size of the subsample is 
$r=2,000$ and $5,000$. We observe that MROSS  uniformly outperforms the four competitors in Cases $1$, $2$, $4$, and $5$.
For the more heavy-tailed cases $3$ and $6$, the improvement by MROSS is  smaller but still  visible. Note that in these  cases only  a  part of the  conditions 
for our theoretical considerations are satisfied (for example, $\E\|\tX\|^4=\infty$).
Nevertheless,  the  median of the MSE  of the estimate obtained by MROSS  is still  smaller than $0.25$-quantile of the four other  methods.

\begin{figure}[H]
\centering\spacingset{1}
\begin{subfigure}{.25\textwidth}
  \centering
\includegraphics[width=1\linewidth]{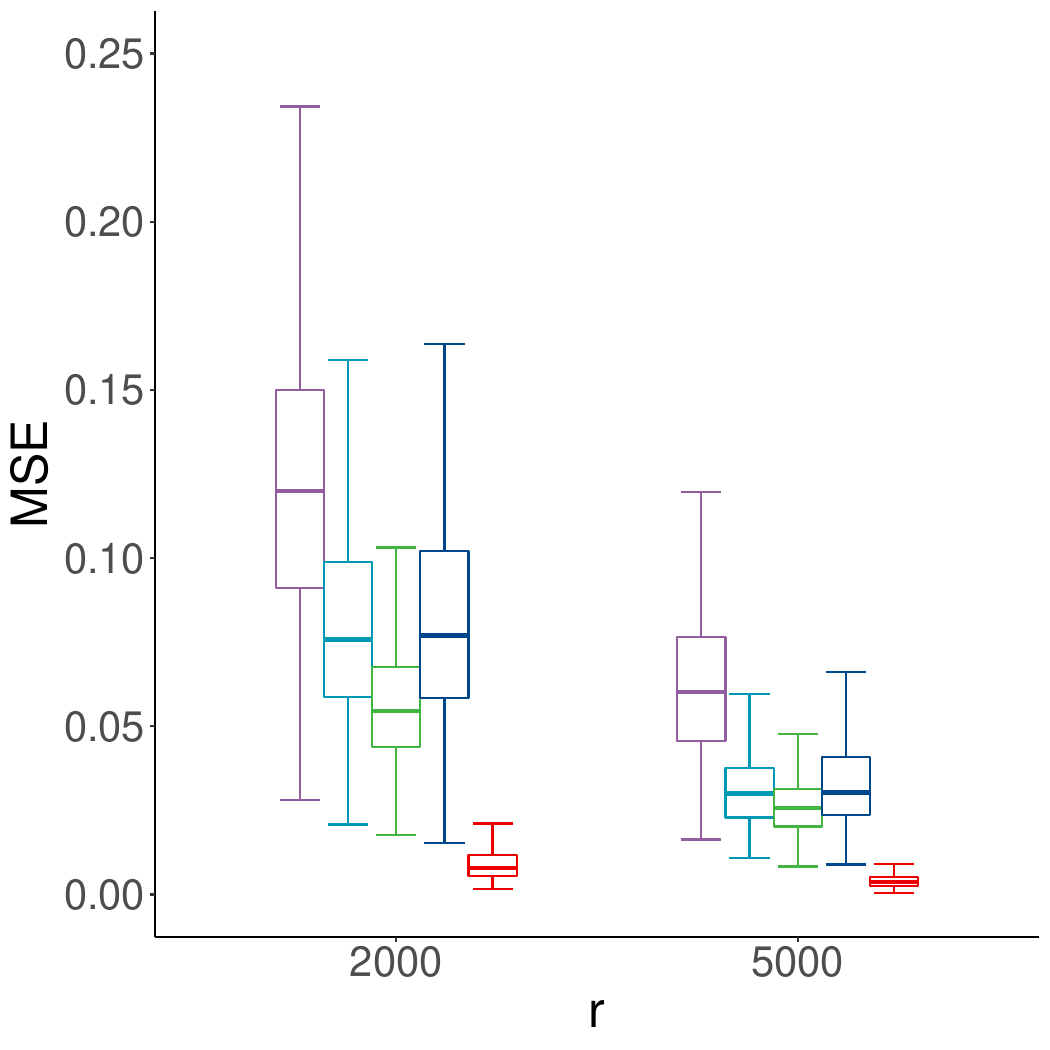}
  \caption{Case 1}
\end{subfigure}
\begin{subfigure}{.25\textwidth}
  \centering
\includegraphics[width=1\linewidth]{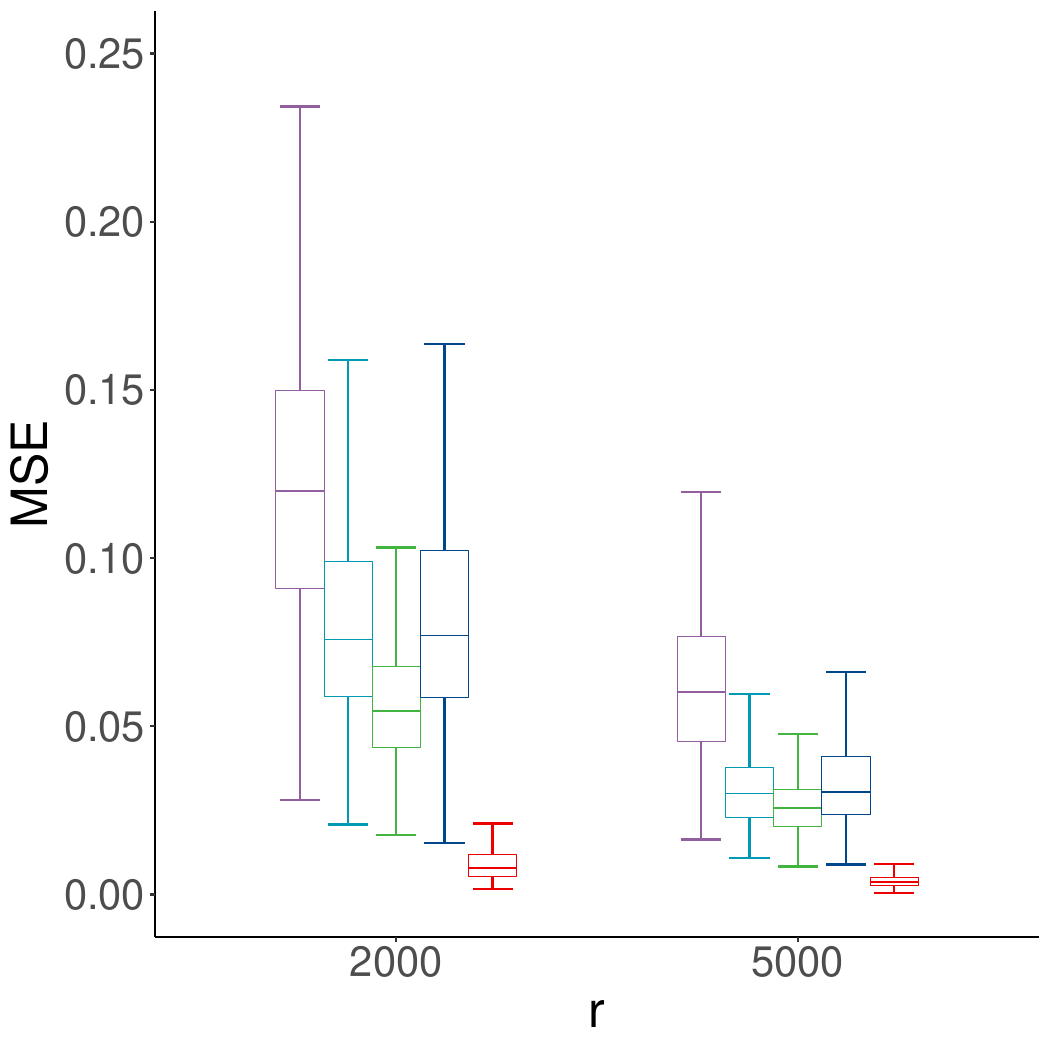}
  \caption{Case 2}
\end{subfigure}
\begin{subfigure}{.25\textwidth}
  \centering
\includegraphics[width=1\linewidth]{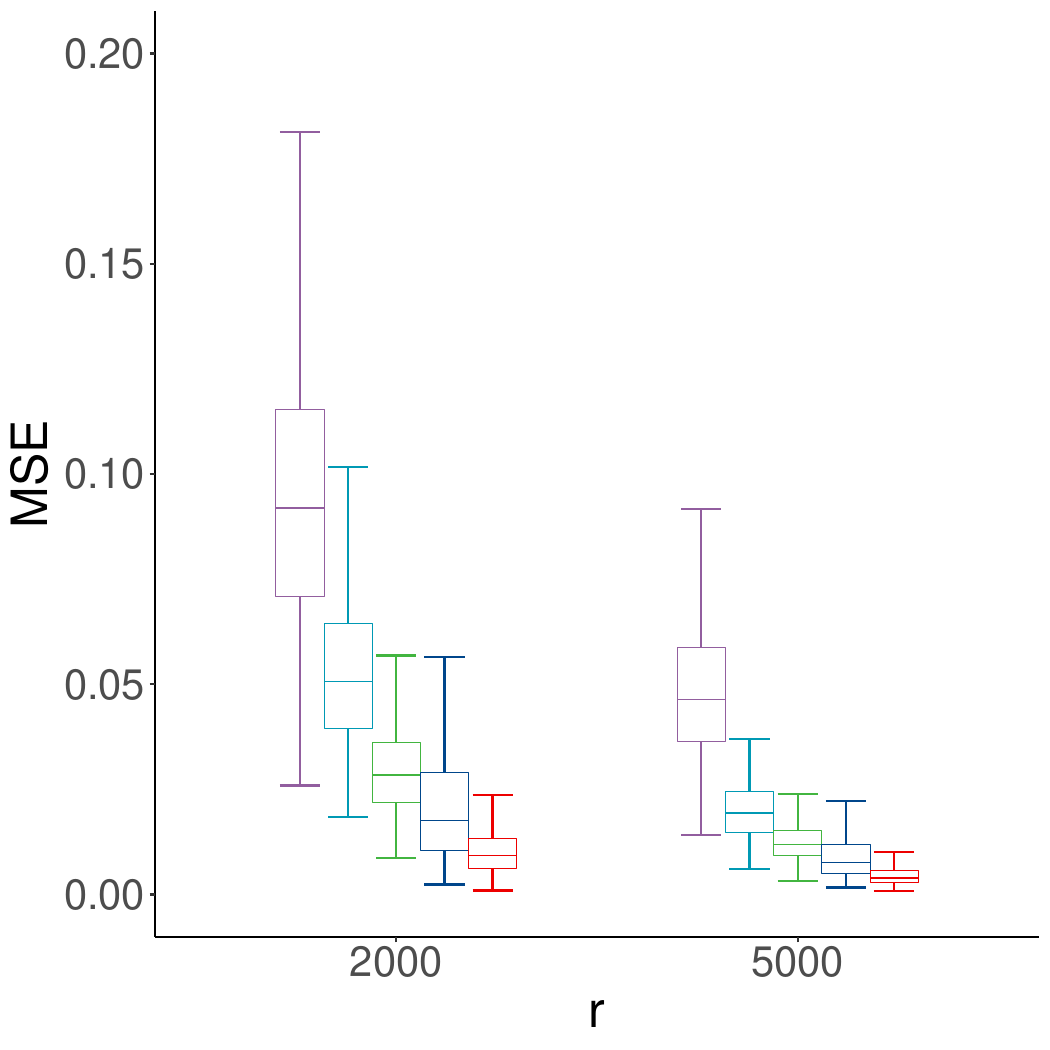}
  \caption{Case 3}
\end{subfigure}

\begin{subfigure}{.25\textwidth}
  \centering
\includegraphics[width=1\linewidth]{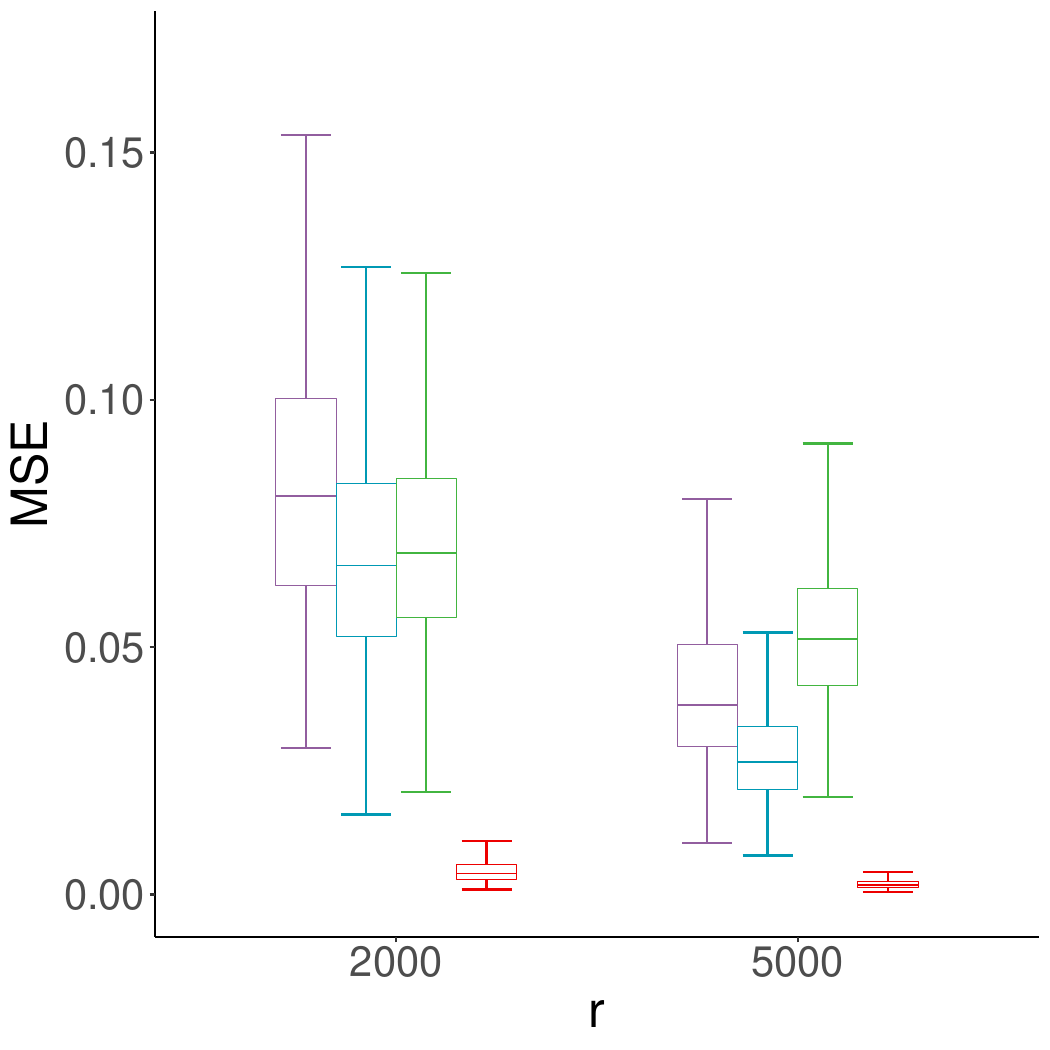}
  \caption{Case 4}
\end{subfigure}
  \begin{subfigure}{.25\textwidth}
  \centering
\includegraphics[width=1\linewidth]{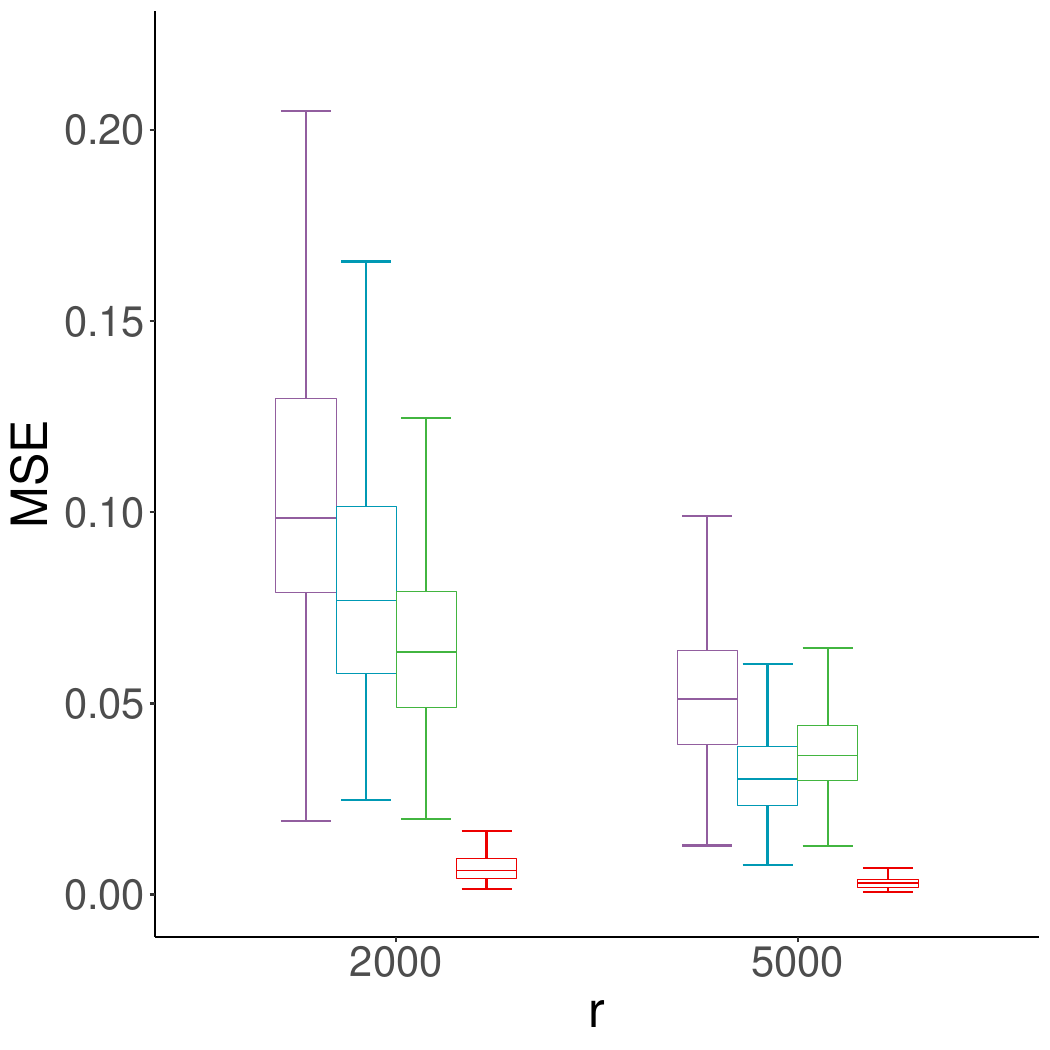}
  \caption{Case 5}
\end{subfigure}
\begin{subfigure}{.25\textwidth}
  \centering
\includegraphics[width=1\linewidth]{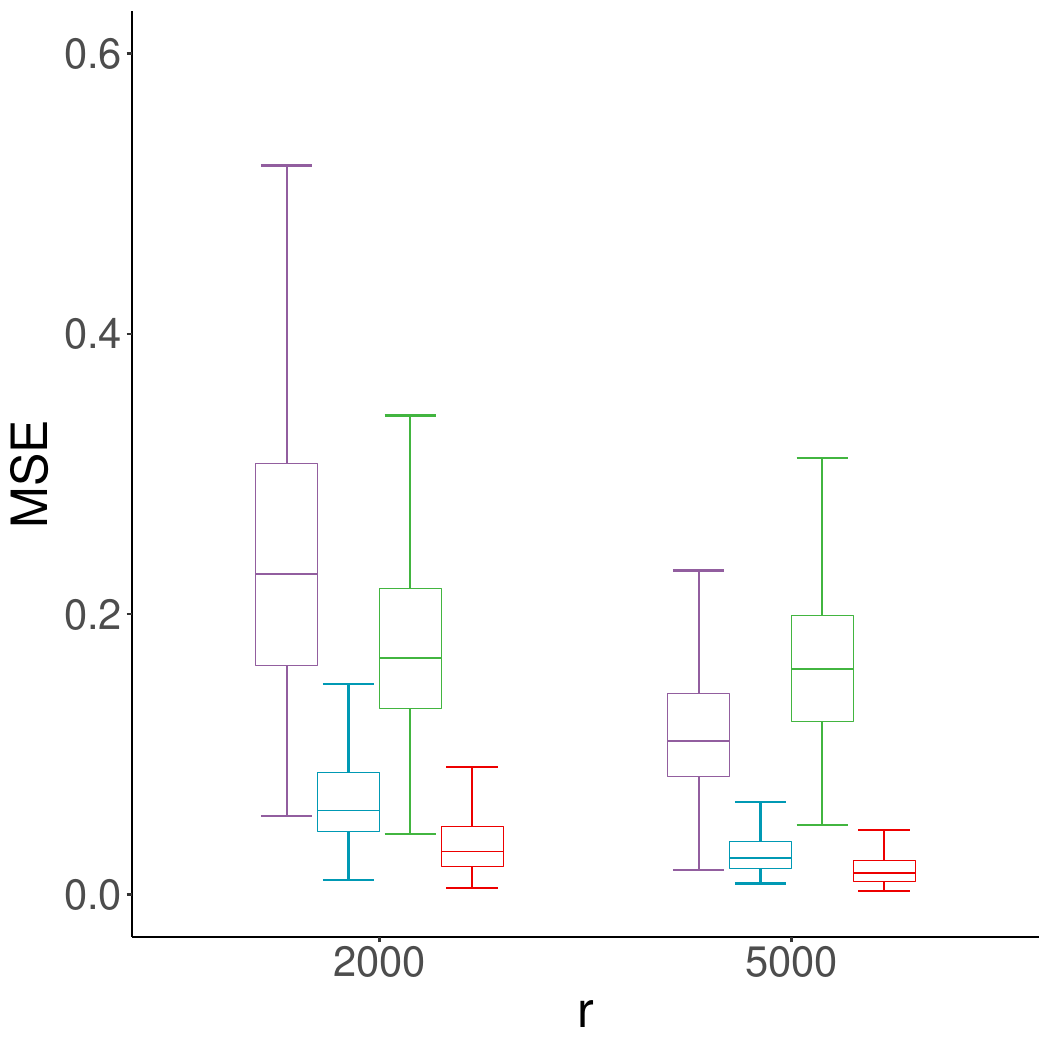}
  \caption{Case 6}
\end{subfigure}
    \caption{{Boxplot  of the MSE of different subsampling methods 
    for logistic regression ($r=2,000$ and $r=5,000$).
    From left to right in each group:   UNIF, OSMAC, MSCLE, IBOSS, and MROSS  (Algorithm~\ref{alg:RBS1}).} The results of IBOSS in Cases 4-6 are omitted since the corresponding MSE is too large.}
    \label{fig:simu.logistic.boxplot}
\end{figure}

When the model is correctly specified, it is interesting to evaluate the performance of MROSS  with respect to statistical inference.
In the following, we construct a $95\%$ confidence interval for the intercept and 
a  slope parameter using the estimators obtained by the different methods and the corresponding
asymptotic theory. Note that the asymptotic distribution for the IBOSS estimator is not available.
 Therefore, we cannot construct a  confidence interval for this method.
For the sake of brevity, we only take Case 1 as an example and  report  the 
empirical coverage probability  and the   average length 
of the confidence interval for the intercept and for 
the parameter corresponding to the first component of the predictor in Table \ref{tab:95CI}. Results for Cases $2$ and $3$  and other slope parameters are similar.
From Table~\ref{tab:95CI}, one can see that all the coverage probabilities are concentrated around 95\% which echoes the theoretical results on asymptotic normality (Theorem~\ref{thm:RB}). We observe that the proposed method produces the shortest intervals. For example, the average length of the confidence interval constructed by MROSS 
is about $4$ times smaller than the length of the confidence interval constructed by uniform subsampling. This means that uniform subsampling would require a $16$ times larger  subsample size
to achieve the same accuracy as MROSS  (note that the convergence rate of the four subsample based estimators is  of order $\sqrt{r}$). Moreover, the new method also yields a substantial improvement compared to OSMAC and MSCLE (here the length of the confidence interval is about $2$ to $3$ times smaller).

\begin{table}[htbp]
\spacingset{1}
  \centering
  \caption{Empirical coverage probabilities (CP) and the average length  of a $95\%$ confidence interval for the intercept ($\theta_0$) and  the 
  parameter $\theta_1$  corresponding to the first component of the predictor
  (Case 1).}
    \begin{tabular}{clcccccccc}
    \hline
          & \multicolumn{1}{c}{\multirow{2}[2]{*}{Method}} & \multicolumn{2}{c}{$r=2000$} & \multicolumn{2}{c}{$r=3000$} & \multicolumn{2}{c}{$r=4000$} & \multicolumn{2}{c}{r=5000} \\
          &       & CP & Length & CP & Length & CP & Length & CP & Length \\
\cmidrule{1-10}    \multirow{4}[2]{*}{$\theta_0$} & UNIF  & 0.944  & 0.299  & 0.942  & 0.242  & 0.946  & 0.208  & 0.950  & 0.186  \\
          & OSMAC & 0.942  & 0.195  & 0.938  & 0.158  & 0.960  & 0.136  & 0.954  & 0.121  \\
          & MSCLE & 0.960  & 0.169  & 0.952  & 0.142  & 0.938  & 0.124  & 0.960  & 0.112  \\
          & MROSS  & 0.952  & \textbf{0.075}  & 0.932  & \textbf{0.061}  & 0.936  & \textbf{0.053}  & 0.934  & \textbf{0.047}  \\
\cmidrule{1-10}    \multirow{4}[2]{*}{$\theta_1$} & UNIF  & 0.944  & 0.343  & 0.960  & 0.278  & 0.956  & 0.239  & 0.946  & 0.213  \\
          & OSMAC & 0.952  & 0.220  & 0.968  & 0.178  & 0.956  & 0.154  & 0.960  & 0.137  \\
          & MSCLE & 0.940  & 0.193  & 0.952  & 0.161  & 0.950  & 0.141  & 0.940  & 0.127  \\
          & MROSS  & 0.954  & \textbf{0.077}  & 0.938  & \textbf{0.062}  & 0.950  & \textbf{0.054}  & 0.940  & \textbf{0.048}  \\
\cmidrule{1-10}    \end{tabular}%
  \label{tab:95CI}%
\end{table}%

\textbf{Distance Weighted Discrimination.} DWD is another popular classifier and is regarded as a blend of the SVM and logistic regression. In the following, we train the linear classifier based on DWD loss with $\gamma=1/2$ as adopted in \cite{wang2017another}.
Here we implemented  MROSS  with $r_0=1,000$ and {$\mathrm{C}_{r_0} =5.9$} and also adopt $L$-optimal subsampling.
The UNIF and OSMAC subsampling methods are also implemented for comparison.
As in the previous example, we use the first $r_0$ data points to obtain the initial estimate $\hat{\bm\theta}_\pilot$ and conduct  subsampling on the remaining  $n-r_0$ data points.
Figure~\ref{fig:simu.dwd.20} presents the empirical MSE for different  subsample size ranging  from 2,000 to 5,000 under the six cases listed at the beginning of this section. 
We observe again that MROSS  outperforms the competing  other subsampling methods. Interestingly the differences between uniform subsampling and OSMAC are rather small in Cases $2$, $4$, and $5$.

To show the stability of the proposed method, we present 
box plots of $\|\tilde{\bm\theta}^{(s)} - \bm\theta_t\|^2$ for the three subsampling strategies under consoderation in Figure~\ref{fig:simu.dwd.boxplot}.
We observe a  significant variance reduction  by MROSS   in all six cases compared with OSMAC and UNIF. Note  that the difference between OSMAC and UNIF is small in the case of light tail covariates. A potential explanation of this observation is that the gradient norm of DWD is relatively small compared with logistic regression. 
{Therefore, small inclusion probabilities have been assigned to  more data points in the subsample, which causes variance inflation in the corresponding estimates (see 
the  discussion  in Section~\ref{sec:tail}).}

\begin{figure}[htbp]
\centering\spacingset{1}
\begin{subfigure}{.32\textwidth}
  \centering
\includegraphics[width=1\linewidth]{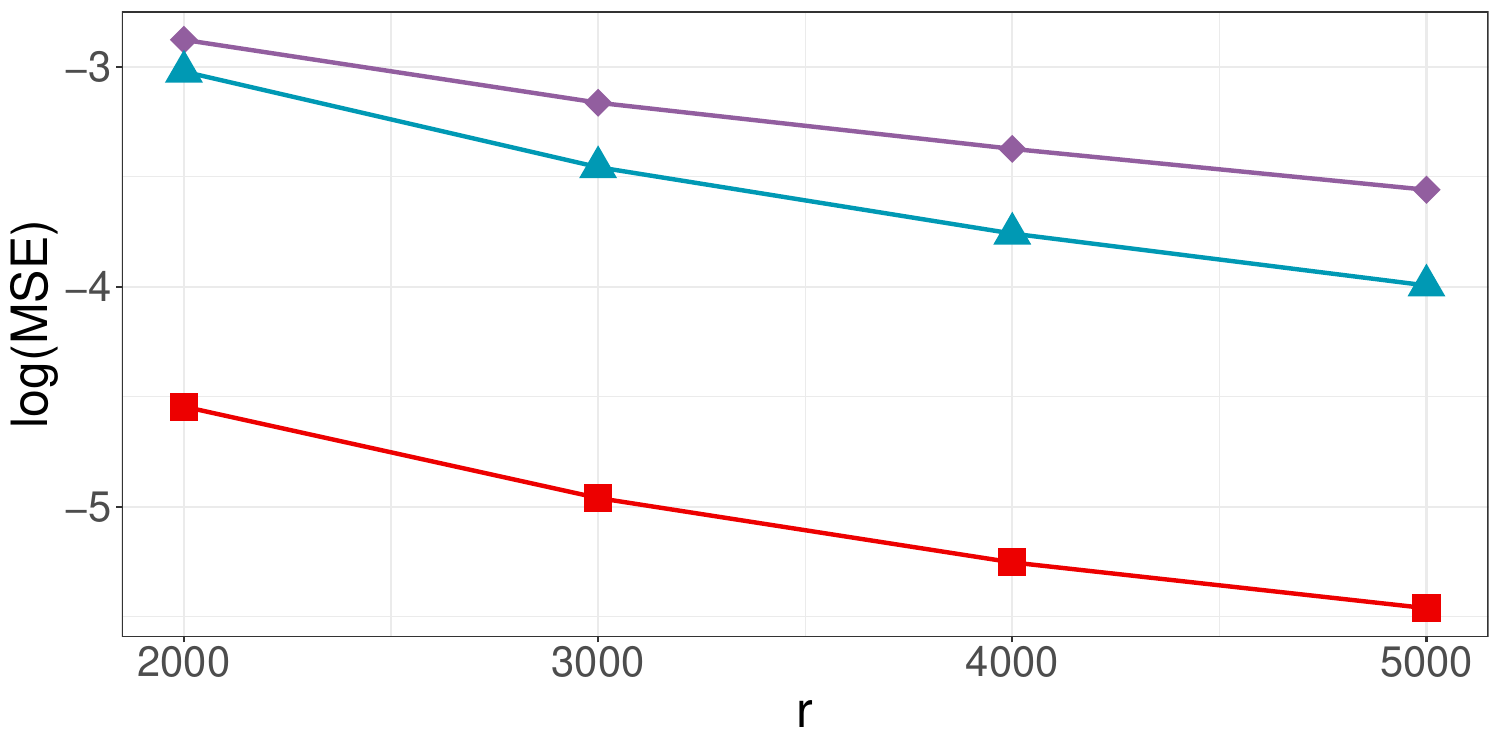}
  \caption{Case 1}
\end{subfigure}
\begin{subfigure}{.32\textwidth}
  \centering
\includegraphics[width=1\linewidth]{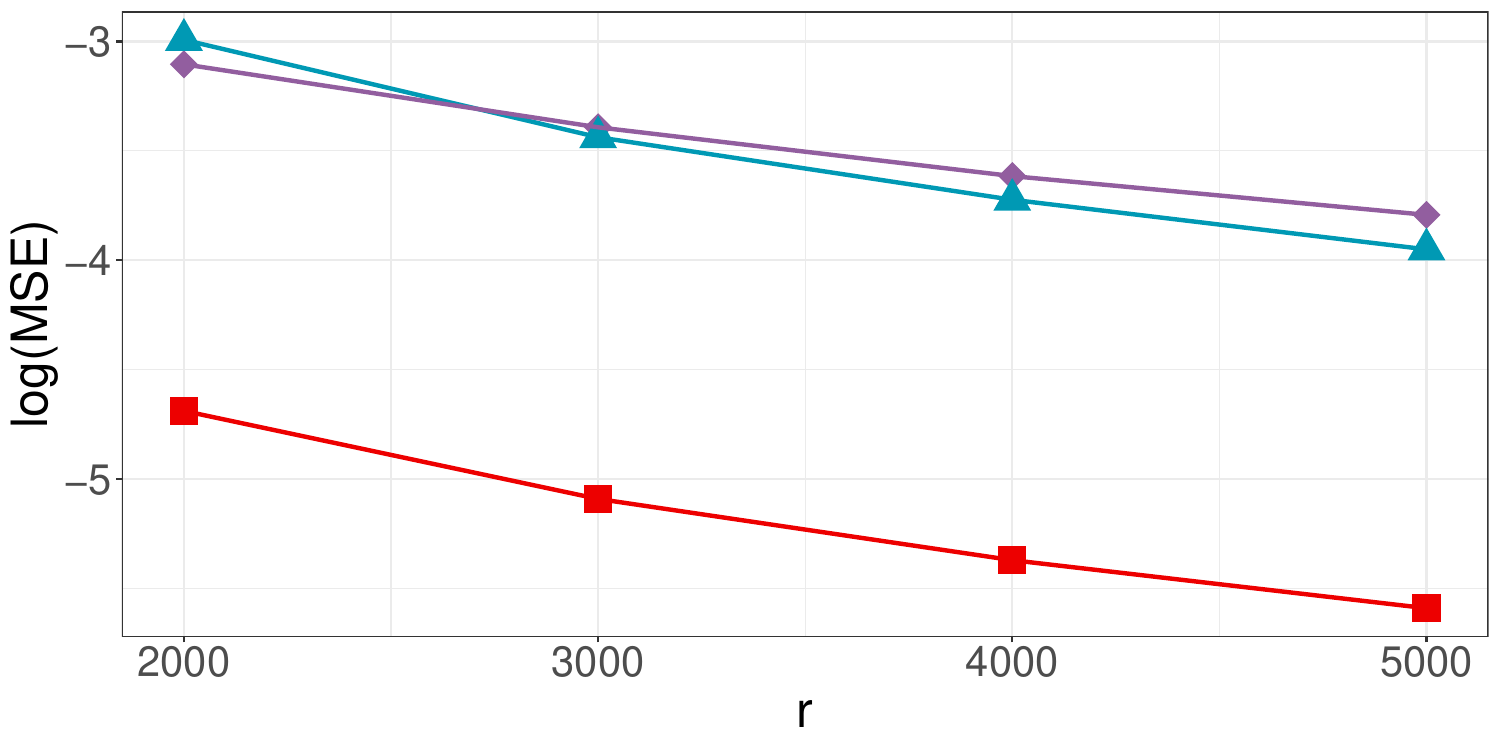}
  \caption{Case 2}
\end{subfigure}
\begin{subfigure}{.32\textwidth}
  \centering
\includegraphics[width=1\linewidth]{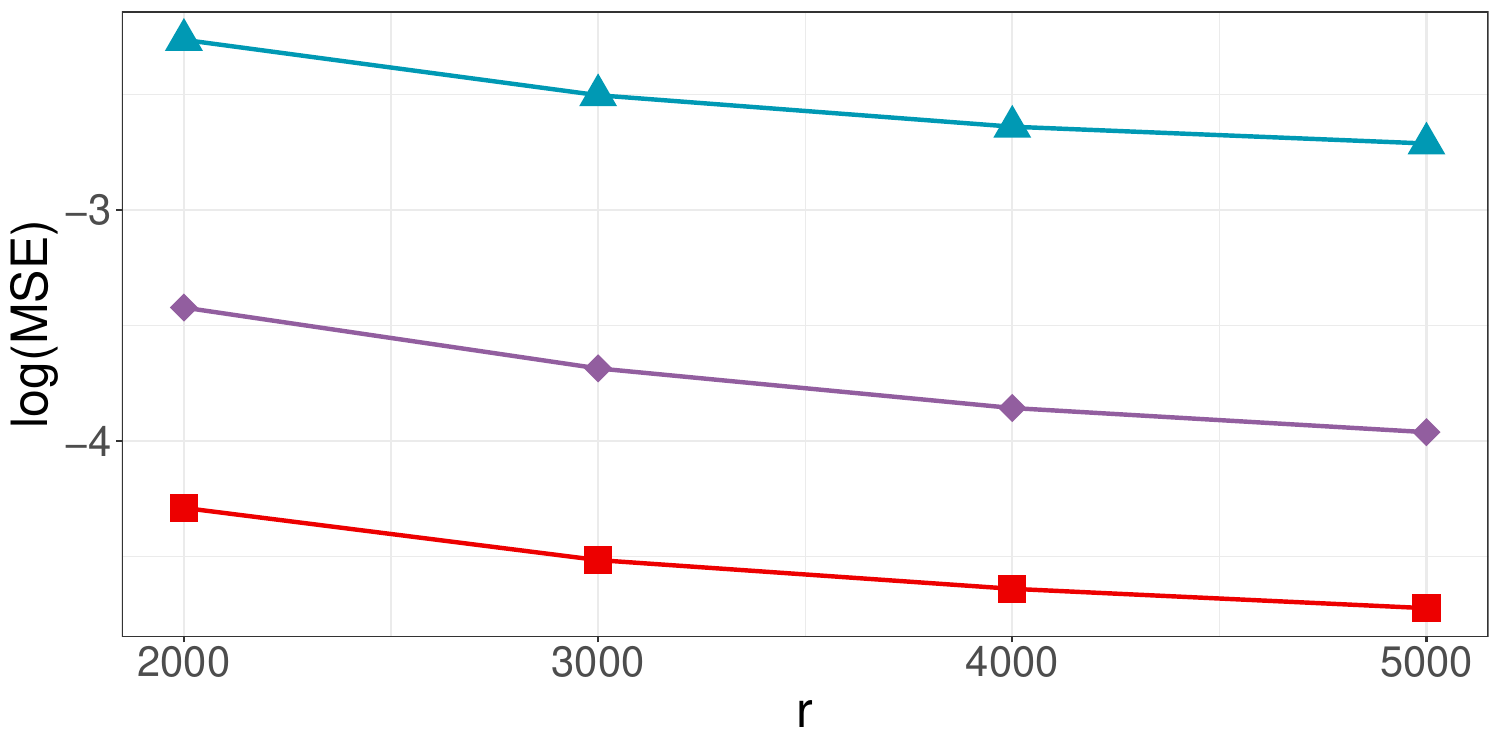}
  \caption{Case 3}
\end{subfigure}

\begin{subfigure}{.32\textwidth}
  \centering
\includegraphics[width=1\linewidth]{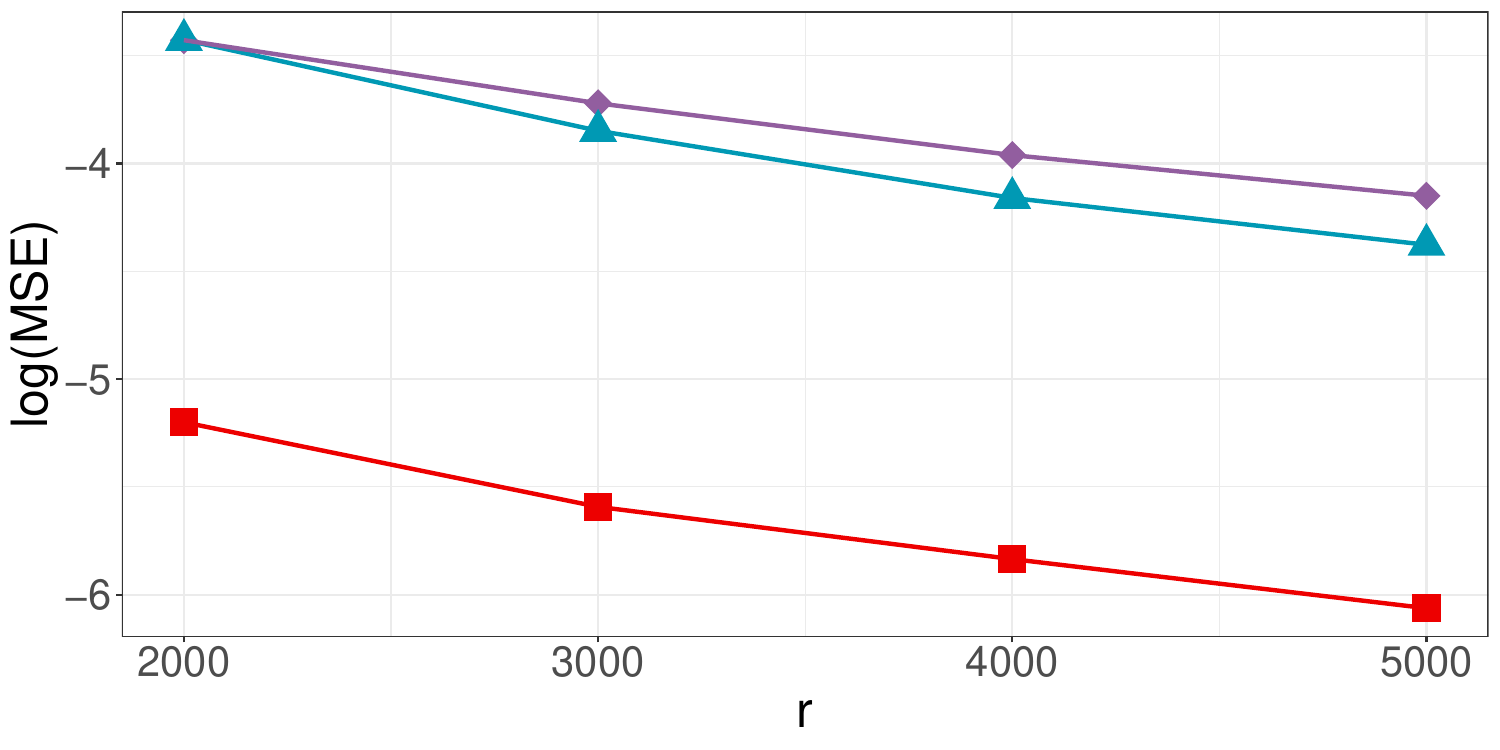}
  \caption{Case 4}
\end{subfigure}
  \begin{subfigure}{.32\textwidth}
  \centering
\includegraphics[width=1\linewidth]{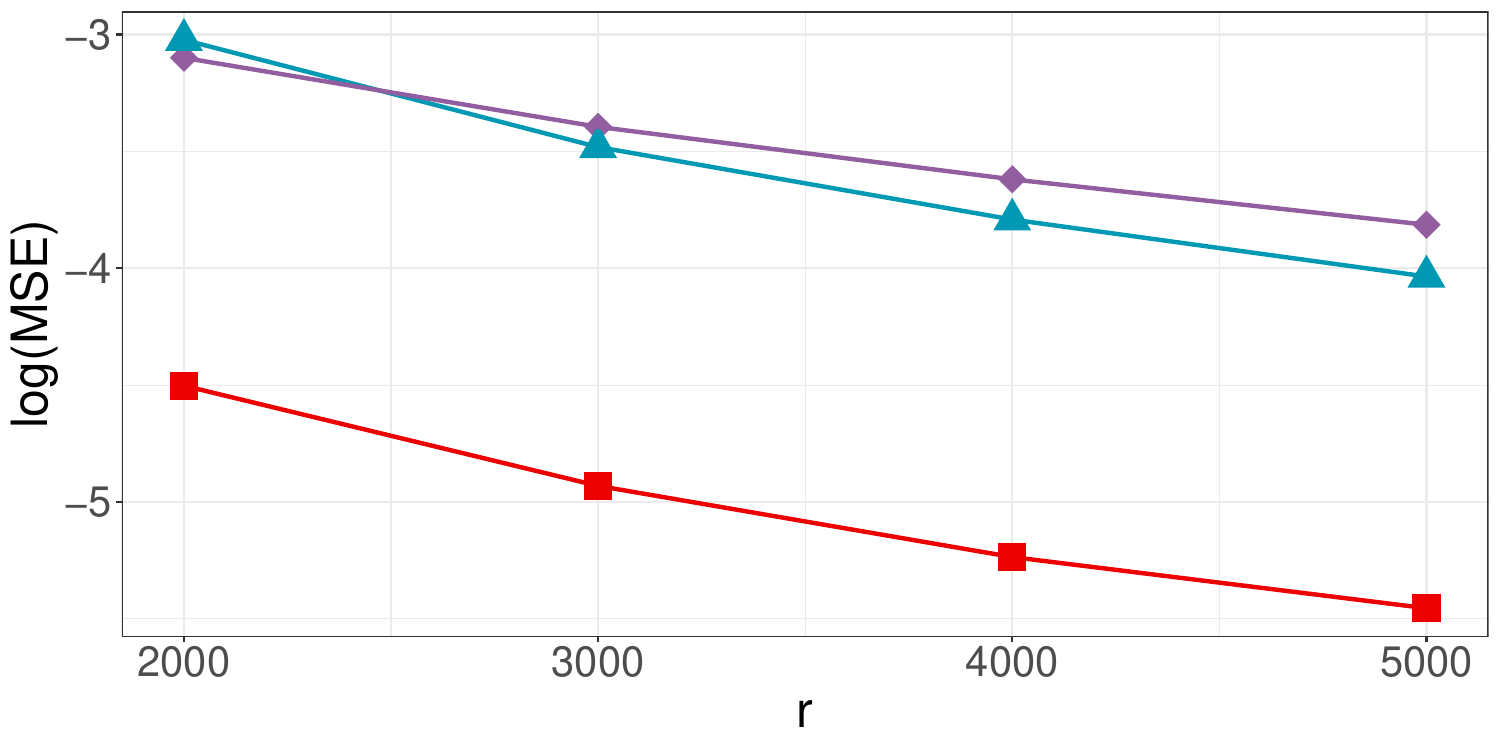}
  \caption{Case 5}
\end{subfigure}
\begin{subfigure}{.32\textwidth}
  \centering
\includegraphics[width=1\linewidth]{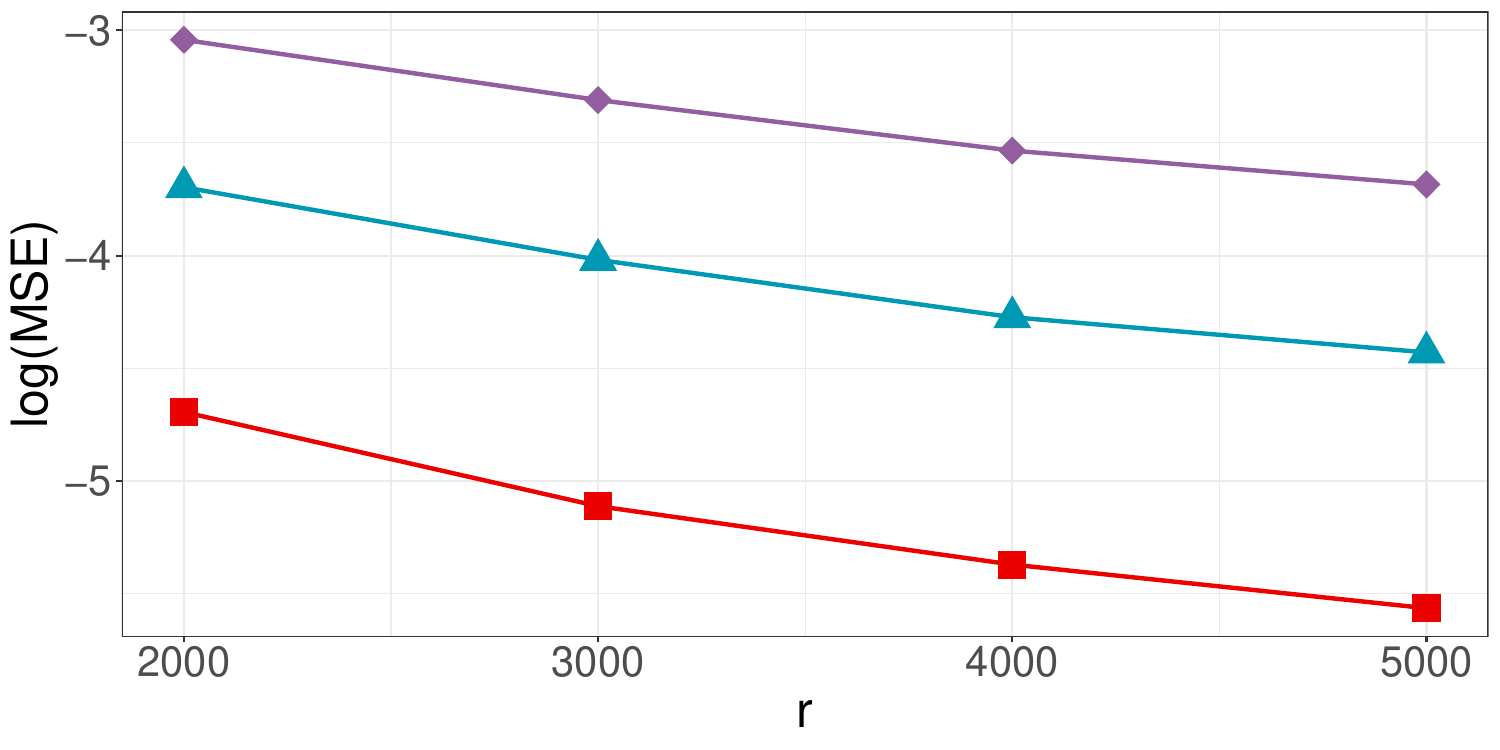}
  \caption{Case 6}
\end{subfigure}
    \caption{{The log MSE with different subsample size ${r}$
    for DWD based on UNIF ($\blacklozenge$), OSMAC ($\blacktriangle$) and MROSS  (Algorithm~\ref{alg:RBS1}, $\blacksquare$).}}
    \label{fig:simu.dwd.20}
\end{figure}

\begin{figure}[htbp]
\centering\spacingset{1}
\begin{subfigure}{.25\textwidth}
  \centering
\includegraphics[width=1\linewidth]{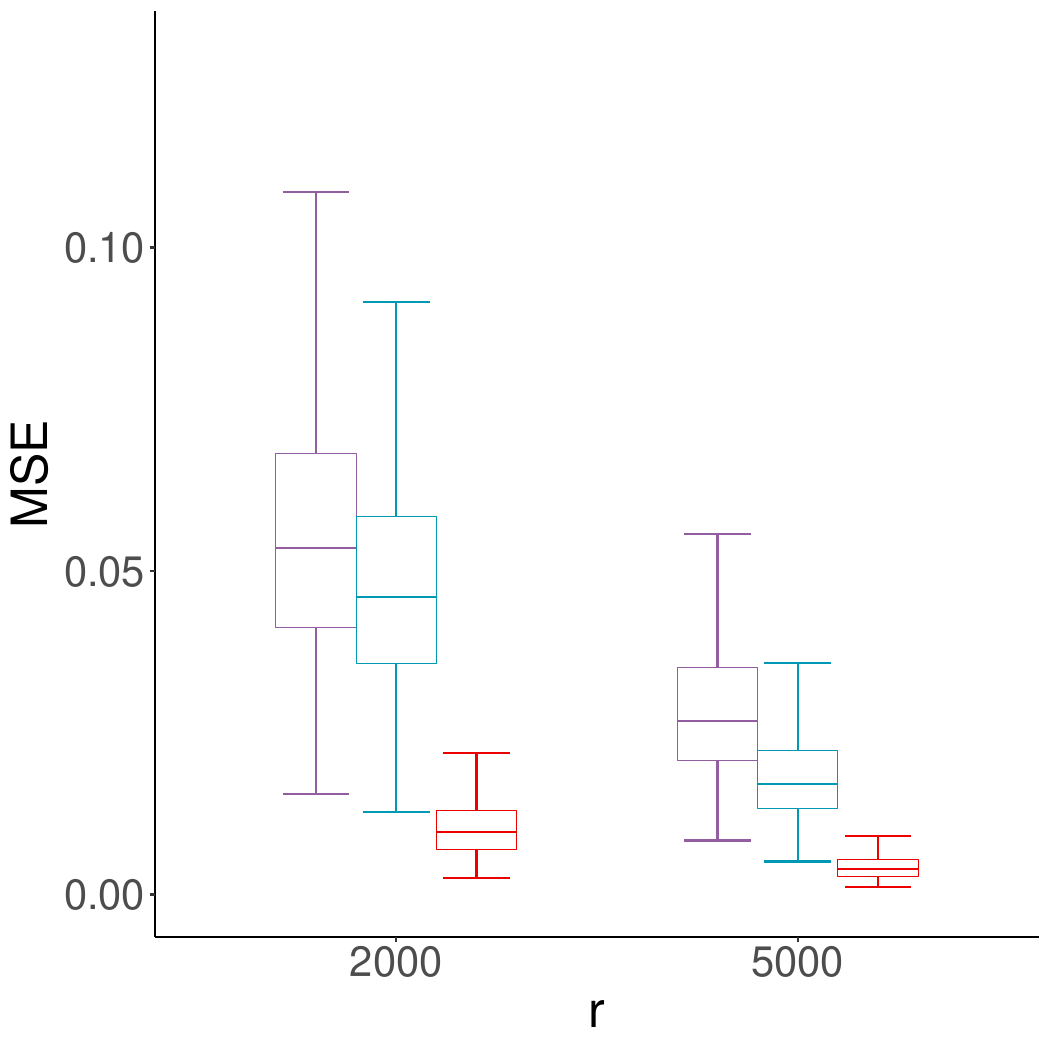}
  \caption{Case 1}
\end{subfigure}
\begin{subfigure}{.25\textwidth}
  \centering
\includegraphics[width=1\linewidth]{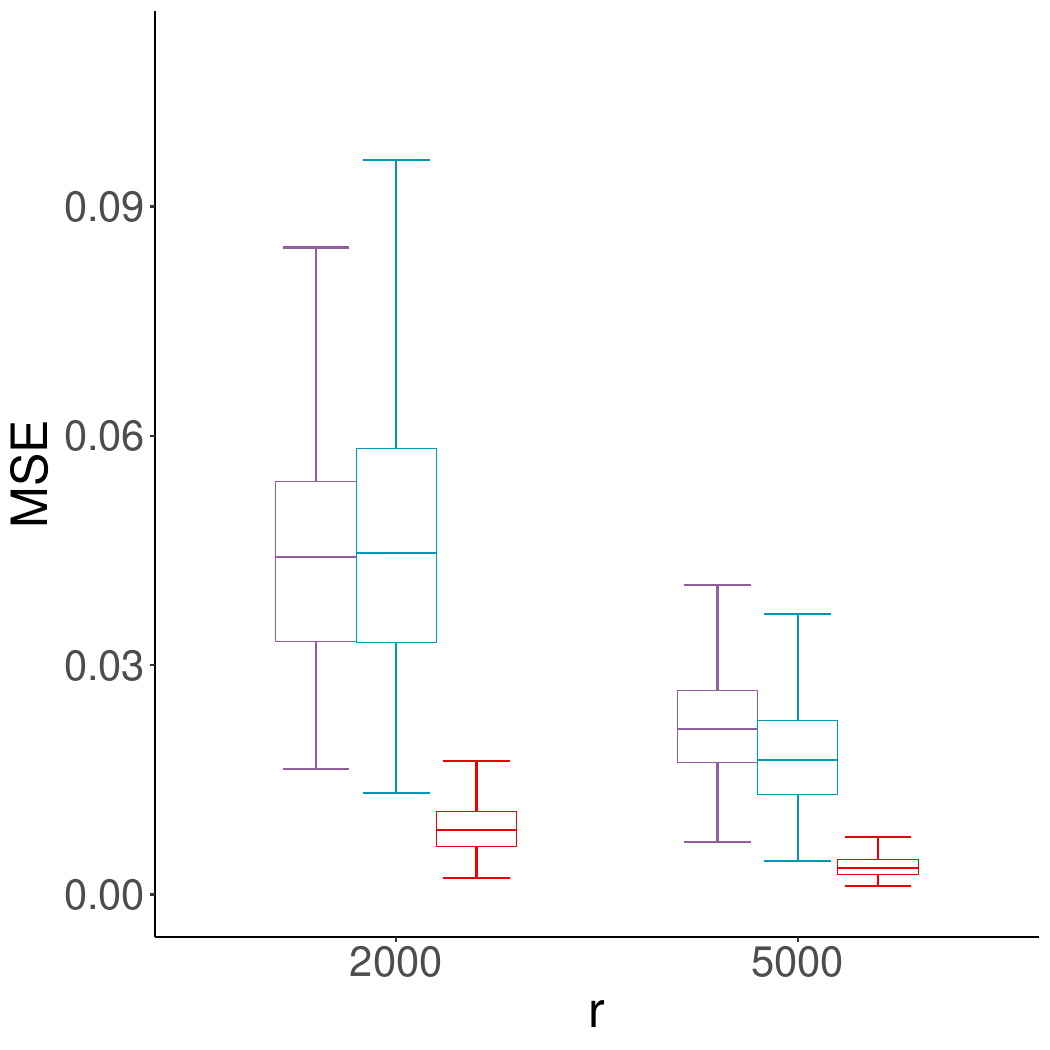}
  \caption{Case 2}
\end{subfigure}
\begin{subfigure}{.25\textwidth}
  \centering
\includegraphics[width=1\linewidth]{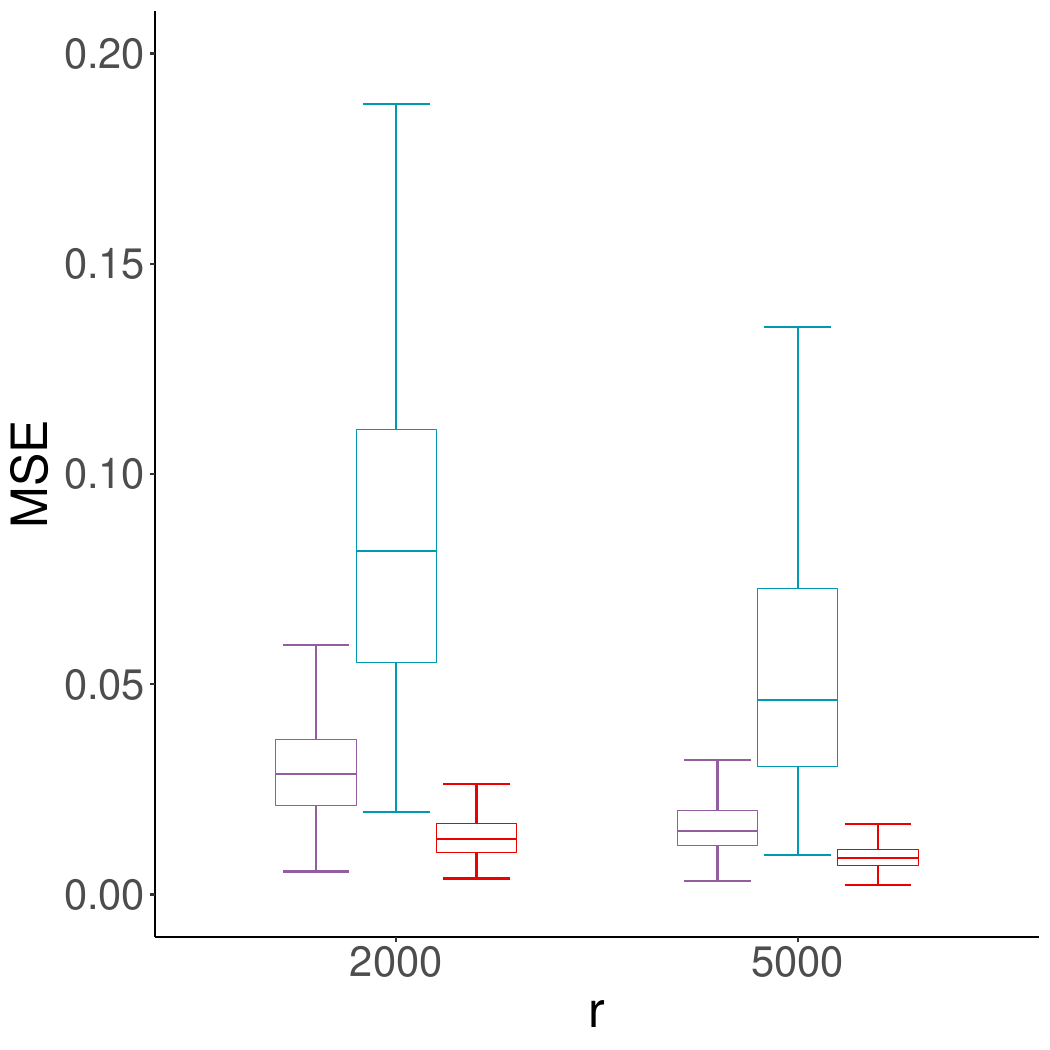}
  \caption{Case 3}
\end{subfigure}

\begin{subfigure}{.25\textwidth}
  \centering
\includegraphics[width=1\linewidth]{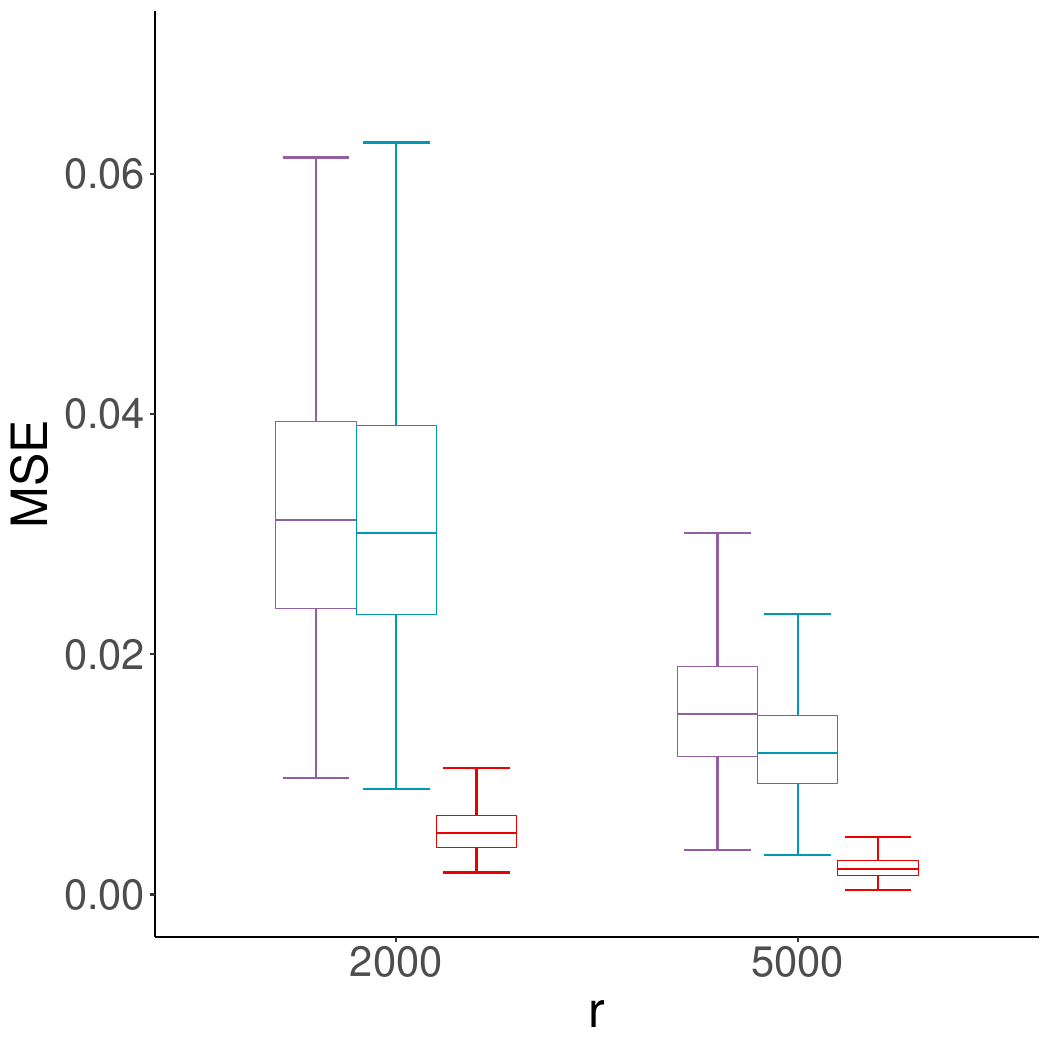}
  \caption{Case 4}
\end{subfigure}
  \begin{subfigure}{.25\textwidth}
  \centering
\includegraphics[width=1\linewidth]{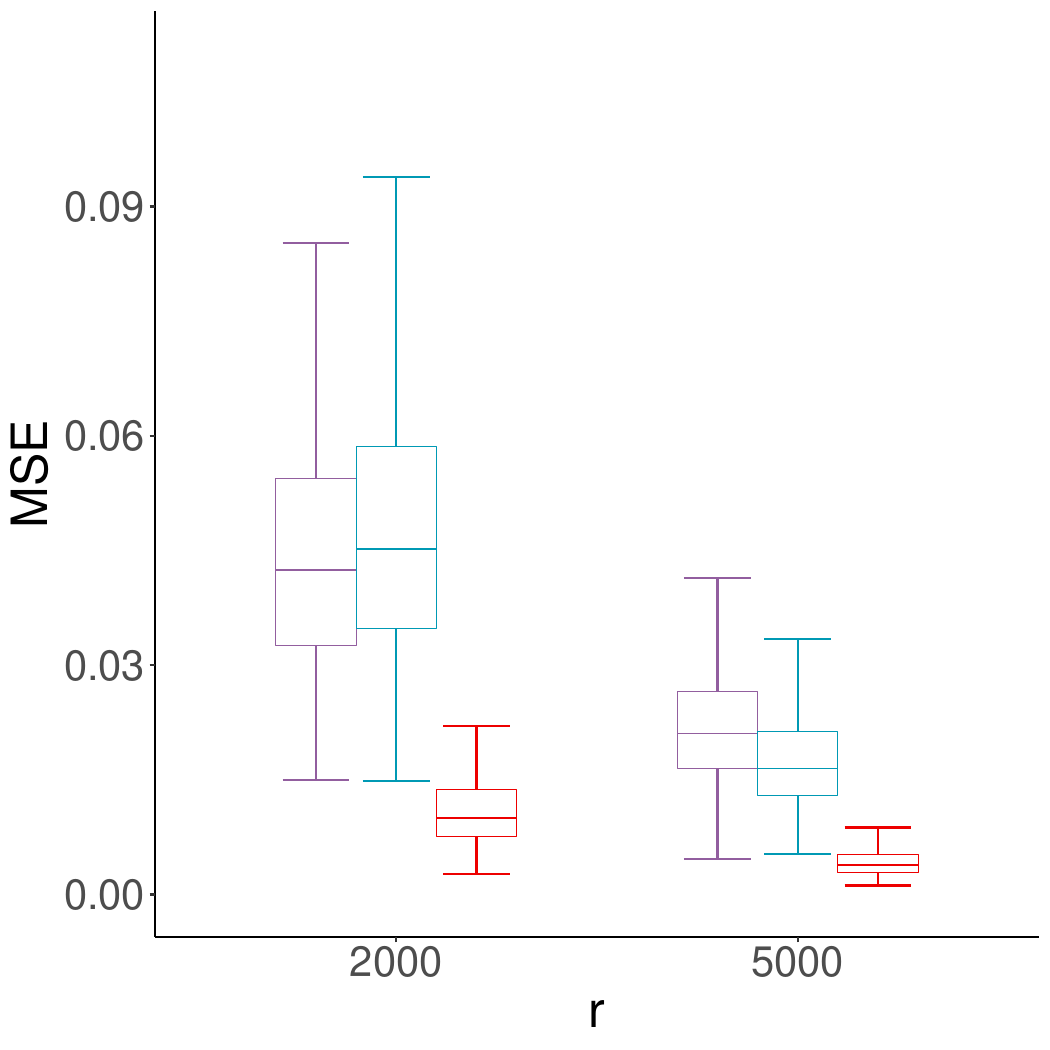}
  \caption{Case 5}
\end{subfigure}
\begin{subfigure}{.25\textwidth}
  \centering
\includegraphics[width=1\linewidth]{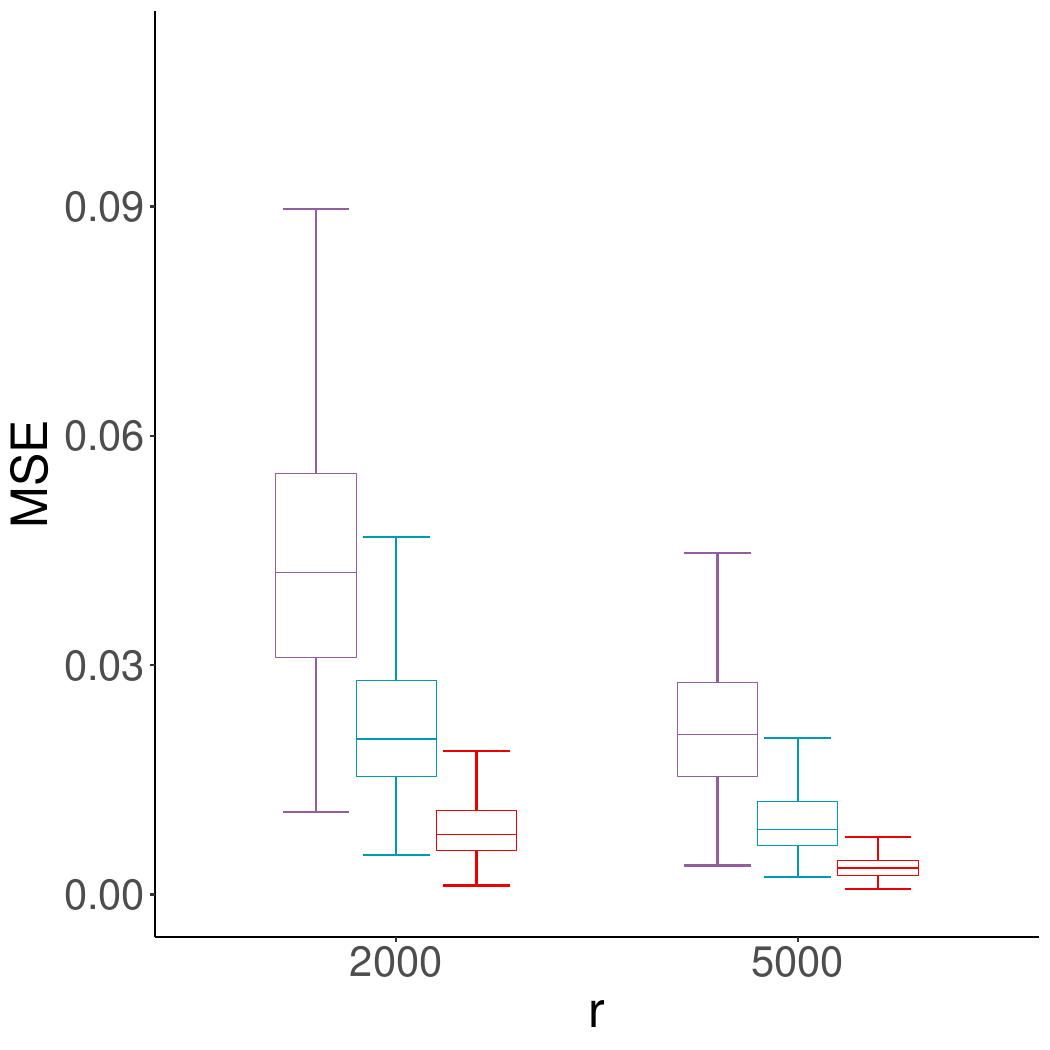}
  \caption{Case 6}
\end{subfigure}
    \caption{{A boxplot showing the MSE with ${r=2,000}$ and $5,000$
    for DWD based on UNIF, OSMAC, and  MROSS  (from left to right in each group).} }
    \label{fig:simu.dwd.boxplot}
\end{figure}

\textbf{Computing Time.} We conclude this section, by reporting the computing time for the aforementioned subsampling methods.
The 
\verb|Sys. time()|  is used to record the start and end times of the corresponding code. Each subsampling strategy has been evaluated 100 times. 
To address the advantage of Poisson sampling that researchers only need to scan the entire dataset once, we use the for-loop implement by \verb|Rcpp| \citep{Rcpp2023}. Computations were carried out on a laptop running Windows 10 with an Intel I5 processor and 8GB memory. 
Note that the computing time is also affected by the programming techniques and it has been shown that IBOSS and MSCLE have similar performance in terms of computing time \citep[see, for example][]{wang2019more,cheng2020information,wang2022maximum}.
Thus we only report the computing times for  OSMAC and Algorithm~\ref{alg:RBS1}.
Results for both logistic regression and DWD under Case~1 are reported  and the computing time for the full data set is also reported for comparison.

\begin{table}[htbp]
  \centering\spacingset{1}\footnotesize
  \caption{Average computing time (in seconds) together with the MSE  for logistic regression and DWD. The corresponding standard deviations are reported in parentheses.  
  }
    \begin{tabular}{cllcclcc}
    \toprule
          & \multicolumn{1}{c}{\multirow{2}[2]{*}{Method}}        & \multicolumn{3}{c}{$d = 8$} &        \multicolumn{3}{c}{$d = 21$} \\
          &              & \multicolumn{1}{c}{$r$} & MSE ($\times 10^{-2}$) & Time &        \multicolumn{1}{c}{$r$} & MSE ($\times 10^{-2}$) & Time \\
    \midrule
    \multirow{5}[5]{*}{logistic} 
          & OSMAC        & $~~2\times 10^3$  & 2.710(1.582) & 0.222(0.081) &        $~~2\times 10^3$  & 7.540(2.868) & 0.317(0.004) \\
          & MROSS           & $~~2\times 10^3$  & 0.065(0.050) & 0.215(0.016) &        $~~2\times 10^3$  & 0.895(0.436) & 0.320(0.006) \\
          & OSMAC        & $~~5\times 10^3$  & 1.090(0.672) & 0.229(0.014) &        $~~5\times 10^3$  & 2.988(1.131) & 0.332(0.004) \\
          & MROSS         & $~~5\times 10^3$  & 0.039(0.025) & 0.229(0.014) &        $~~5\times 10^3$  & 0.407(0.206) & 0.349(0.006) \\
\cmidrule{2-8}          
& Full         & $~~5\times10^5$ &   0.018(0.009)    &    0.616(0.075)   &        $~~5\times10^5$ &   0.077(0.028)    & 2.239(0.149) \\
    \midrule
    \multirow{5}[5]{*}{DWD} 
          & OSMAC        & $~~2\times 10^3$  & 1.573(0.910) & 0.218(0.036) &        $~~2\times 10^3$  & 4.622(1.697) & 0.309(0.010) \\
          & MROSS        & $~~2\times 10^3$  & 0.109(0.071) & 0.204(0.018) &        $~~2\times 10^3$  & 1.066(0.536) & 0.308(0.010) \\
          & OSMAC        & $~~5\times 10^3$  & 0.621(0.410) & 0.225(0.020) &        $~~5\times 10^3$  & 1.791(0.703) & 0.334(0.058) \\
          & MROSS       & $~~5\times 10^3$  & 0.045(0.027) & 0.227(0.016) &        $~~5\times 10^3$  & 0.421(0.183) & 0.350(0.070) \\
\cmidrule{2-8}          
& Full         & $~~5\times10^5$ &   0.011(0.005)    &     1.548(0.187)   &       $~~5\times10^5$ &    0.037(0.013)   & 3.932(0.200) \\
    \bottomrule
    \end{tabular}%
  \label{tab:time}%
\end{table}%

In Table~\ref{tab:time} we display the computing times of the two methods for various numbers of covariates ($d$) and different subsample sizes ($r$). Here we fixed $r_0=1,000$ and data are generated as in Case~1. For a fair comparison, we used $r+r_0$ subdata points for uniform subsampling on the entire dataset $\fu$ as in the previous simulation.
The two subsampling methods efficiently reduce the computing time compared with the direct calculation of the full data, especially for the large dimensional case.  
As discussed in Section~\ref{sec:method}, the time differences between MROSS  and OSMAC are negligible 
since most of the additional calculations are performed on the subsample or during the data scanning.

\subsection{Forest cover type dataset}\label{subsec:forest}

In this subsection, we examine the performance of MROSS  on the forest cover type dataset available on the UCI machine learning repository \citep{misc_covertype_31}.
The primary task here is to distinguish spruce/fir (type 1) from the rest six types (types 2-7).
Although there are seven forest cover types in this dataset, the one-vs-rest is one of the most commonly used methods to extend the binary classifier to a multi-class response.
Getting into the data, there are $581,012$ instances with ten quantitative features that describe geographic and lighting characteristics.

In the following, we build the subsample-based linear classifier via the logistic loss and DWD loss based on ten quantitative features. 
We randomly select 80\% of the dataset as the training set and leave the rest 20\% as the testing set to calculate prediction accuracy.

We compared our method with UNIF and OSMAC. {For the logistic loss function} the results of  MSCLE are also reported even though the true data generating model may not be appropriately described by a logistic regression model  (note that  IBOSS is a deterministic subsampling method and not suitable in this context). 
We consider the estimator from the full sample as $\bm\theta_t$ and repeat the subsampling procedures $500$ times to obtain the MSE and prediction accuracy on the test data.
Similar to the simulation studies, we fixed $r_0=1,000$ and $r$ varies from $2,000$ to $5,000$. The pilot estimate is calculated from a sample which is obtained via random splitting of the training set. The tuning parameter $\mathrm{C}_{r_0} $ is set to be $6.91$ and $5.91$ for logistic regression and DWD, respectively. The log  MSE and prediction accuracy are reported in Figure~\ref{fig:forest.logistic}. {We observe that MROSS   yields a substantial improvement compared to the other methods. In particular, the prediction accuracy is not much worse than the accuracy obtained by the classifier from the full sample.}

\begin{figure}[h!]
    \centering\spacingset{1}
\begin{subfigure}{.45\textwidth}
    \centering
\includegraphics[width=1\linewidth]{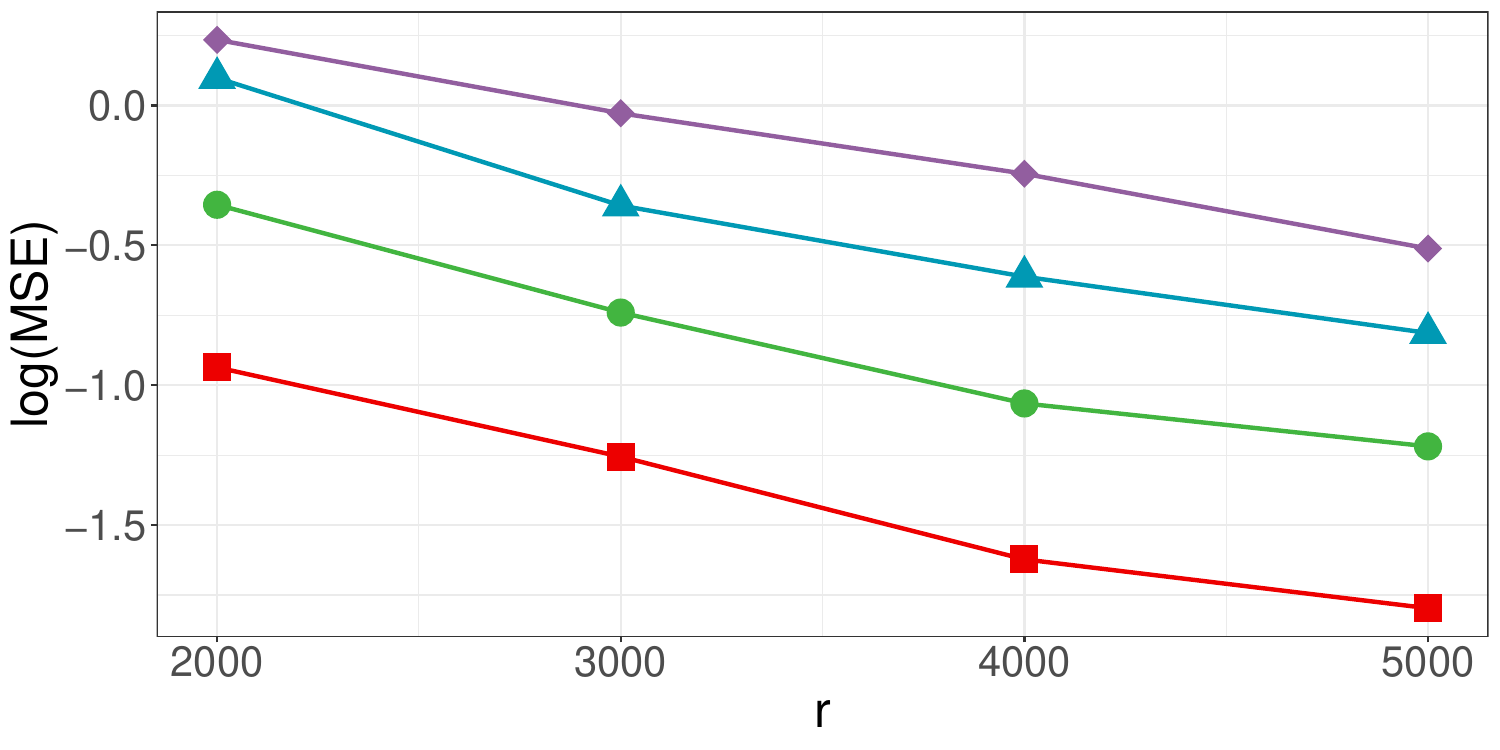}
  \caption{logistic}
\end{subfigure}
\begin{subfigure}{.45\textwidth}
     \centering
\includegraphics[width=1\linewidth]{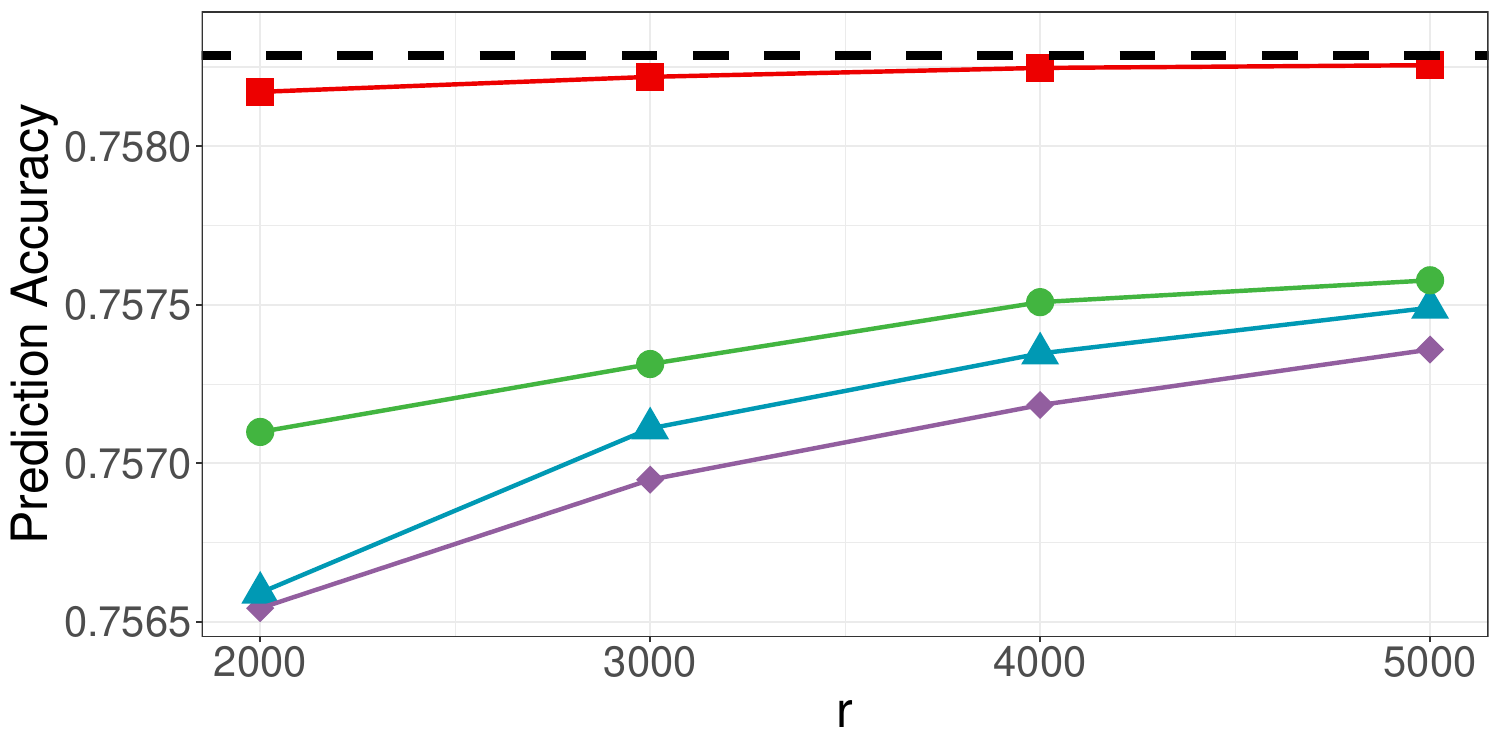}
  \caption{logistic}
\end{subfigure}

\begin{subfigure}{.45\textwidth}
    \centering\spacingset{1}
\includegraphics[width=1\linewidth]{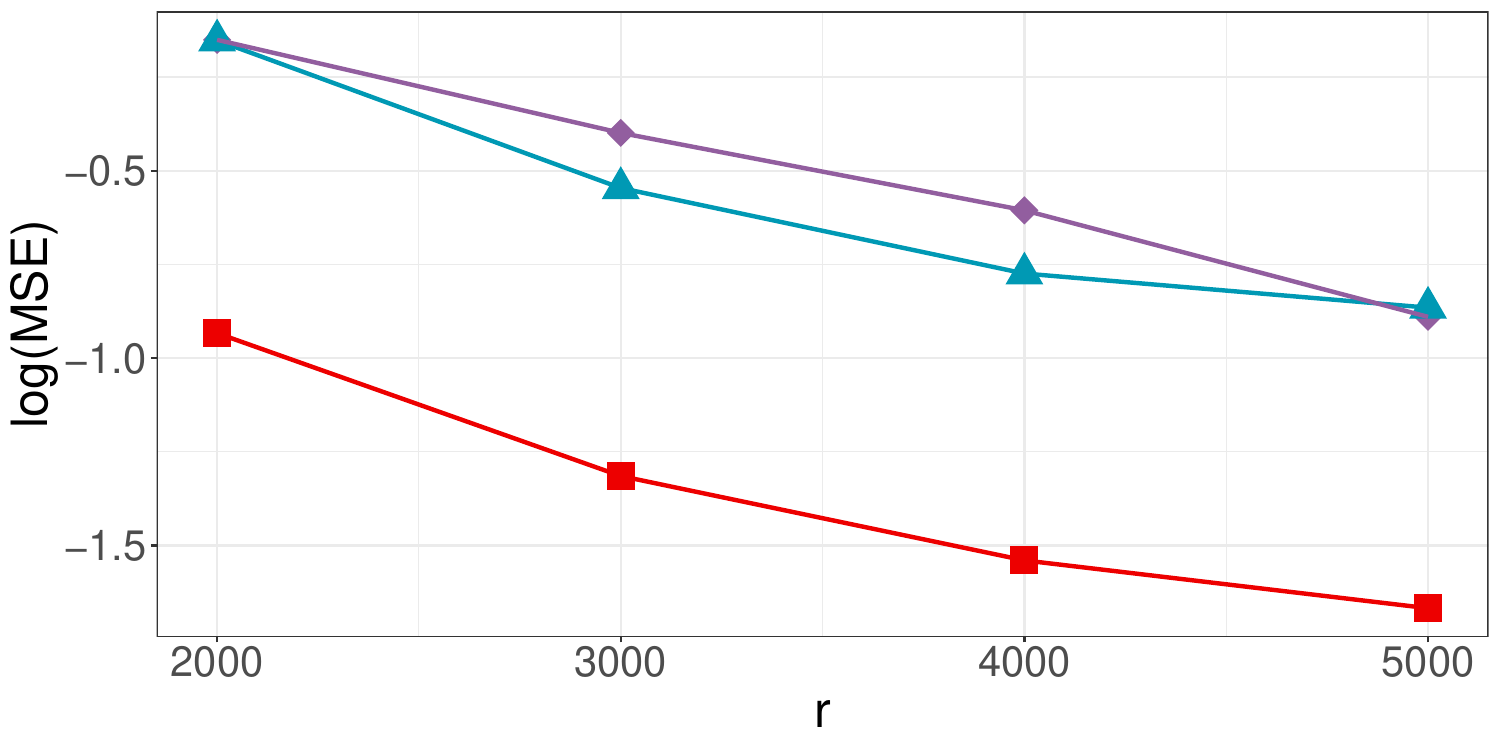}
  \caption{DWD}
\end{subfigure}
\begin{subfigure}{.45\textwidth}
     \centering
\includegraphics[width=1\linewidth]{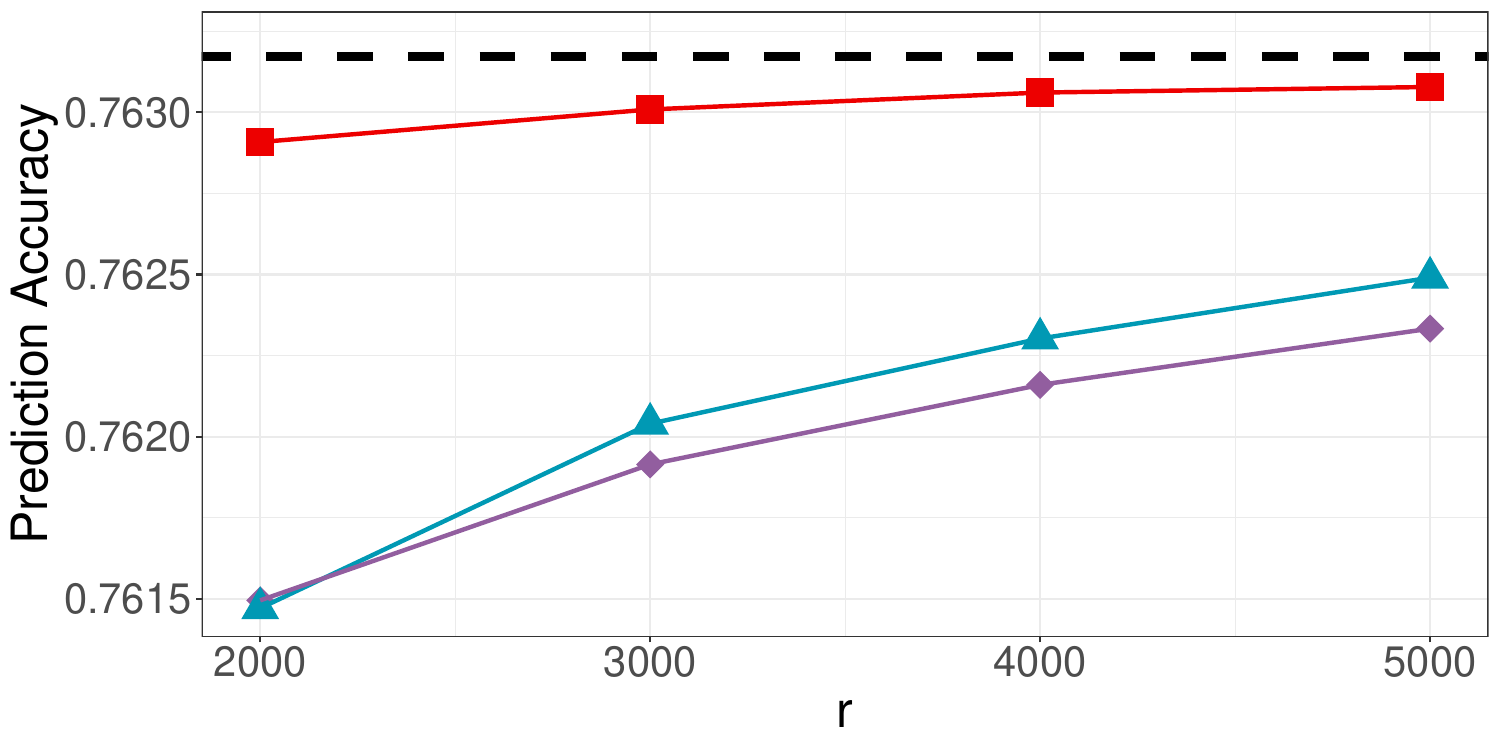}
  \caption{DWD}
\end{subfigure}
    \caption{ The log MSE and prediction accuracy of the subsample based logistic regression (top panel) and DWD (bottom panel) for cover type dataset. MROSS (Algorithm~\ref{alg:RBS1}, $\blacksquare$) is compared with UNIF ($\blacklozenge$), OSMAC ($\blacktriangle$), MSCLE ($\bullet$). The dash line stands for the prediction accuracy based on the full sample estimator.    
    }
    \label{fig:forest.logistic}
\end{figure}

\subsection{Beijing multi-site air quality dataset}
The Beijing air-quality dataset \citep{misc_beijing_multi-site_air_quality_501} includes hourly air pollutants data from 12 monitoring sites governed by the Beijing Municipal Environmental Monitoring Center. 
The primary task here is to predict the air quality in the next hour by the air pollutants collected by the monitoring sites.
Getting into the data, there are $420,768$ instances on PM 2.5 value together with ten quantitative features including some air pollutants (such as $\rm CO$, $\rm NO_2$, $\rm SO_2$) concentration, temperature, wind speed, etc.
According to the air quality standard in China, we call the air is polluted if and only if the PM2.5 concentration is greater than 75$\mu g/m^3$.

In the following, we fit the logistic regression and DWD based on the subsample obtained by different subsampling strategies. After removing all the missing values, we randomly select 80\% of the dataset as the training set and leave the rest 20\% as the testing set to calculate the prediction accuracy. 

All settings are the same as in Section~\ref{subsec:forest}.
Figure~\ref{fig:PM25.logistic} reports the MSE and prediction accuracy. Again MROSS outperforms the other methods.

\begin{figure}[h!]
	\centering\spacingset{1}
	\begin{subfigure}{.45\textwidth}
		\centering
		\includegraphics[width=1\linewidth]{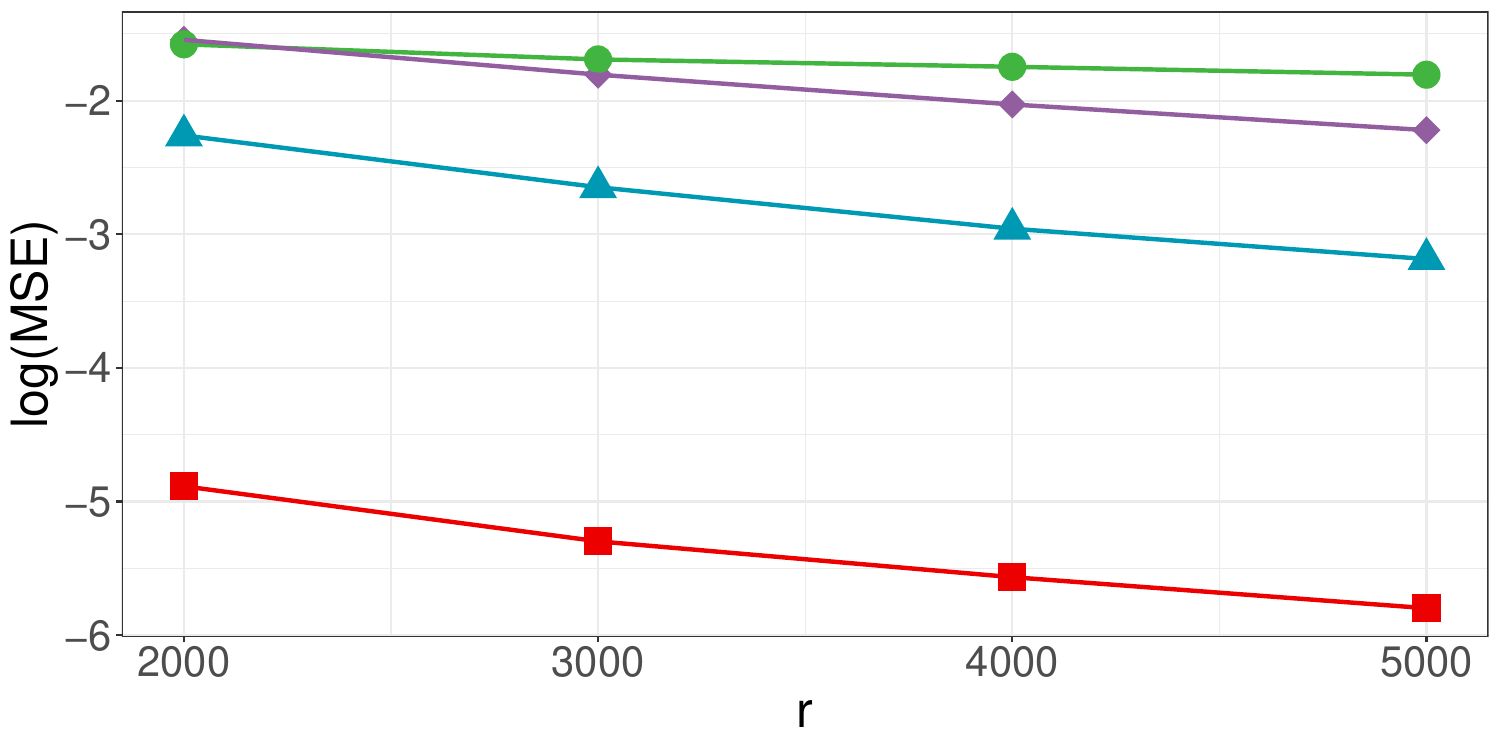}
		\caption{logistic}
	\end{subfigure}
	\begin{subfigure}{.45\textwidth}
		\centering
		\includegraphics[width=1\linewidth]{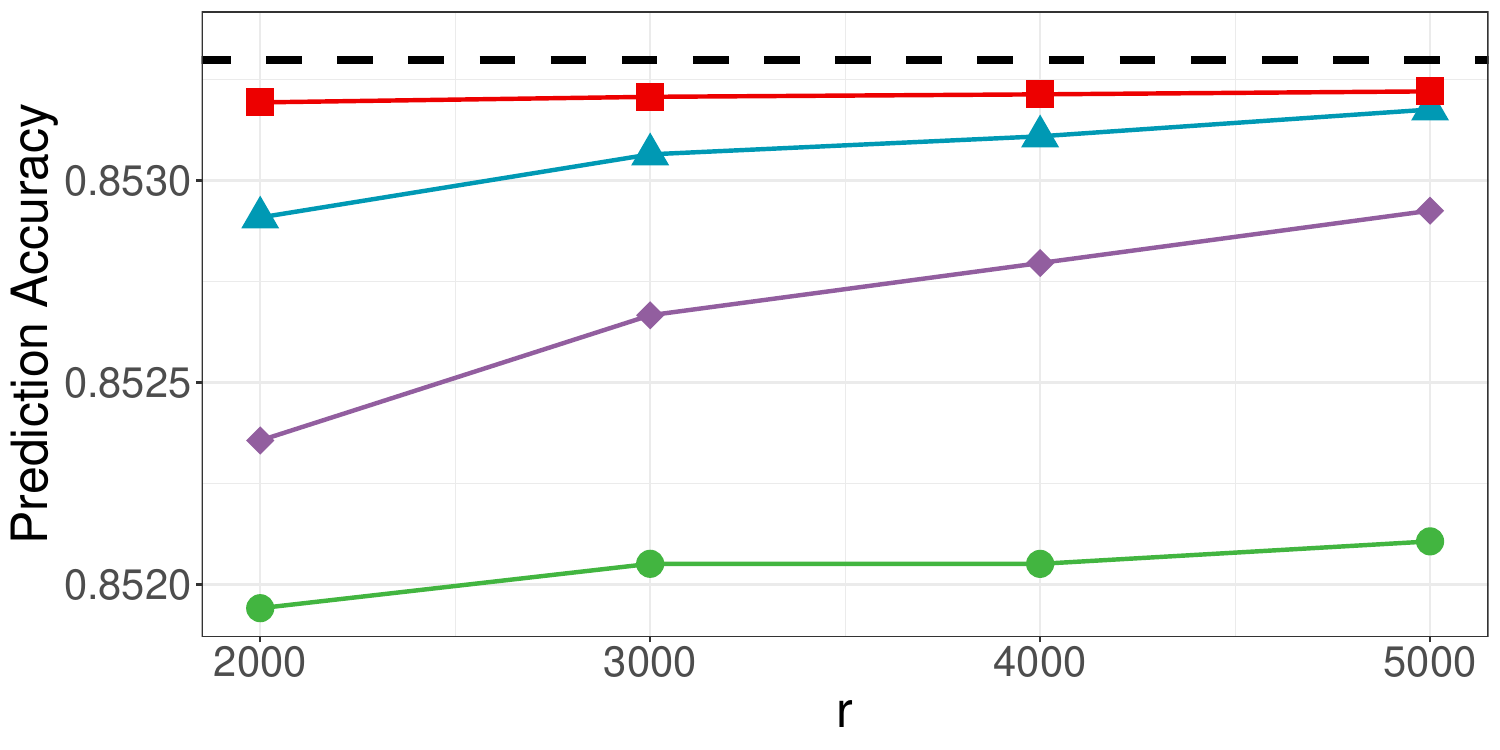}
		\caption{logistic}
	\end{subfigure}
	
	\begin{subfigure}{.45\textwidth}
		\centering\spacingset{1}
		\includegraphics[width=1\linewidth]{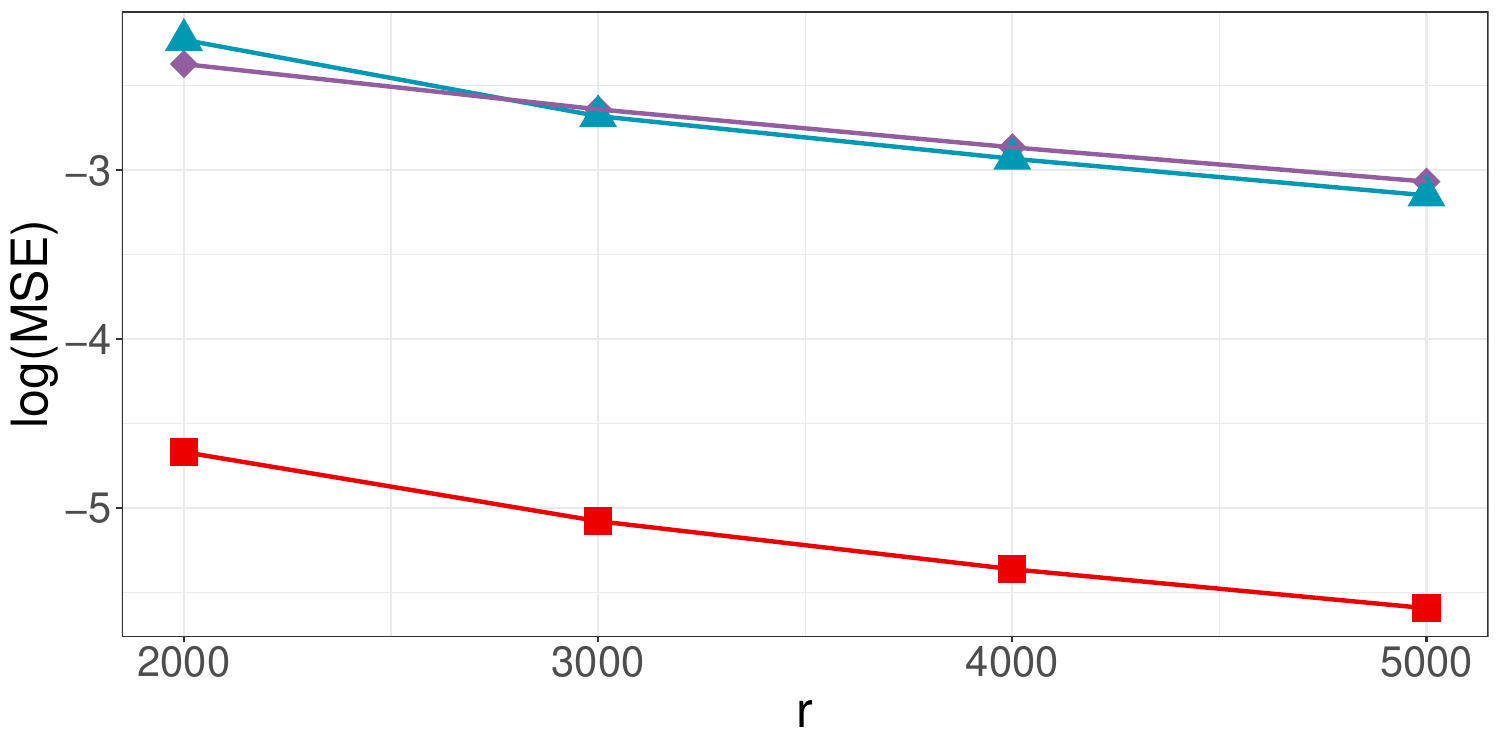}
		\caption{DWD}
	\end{subfigure}
	\begin{subfigure}{.45\textwidth}
		\centering
		\includegraphics[width=1\linewidth]{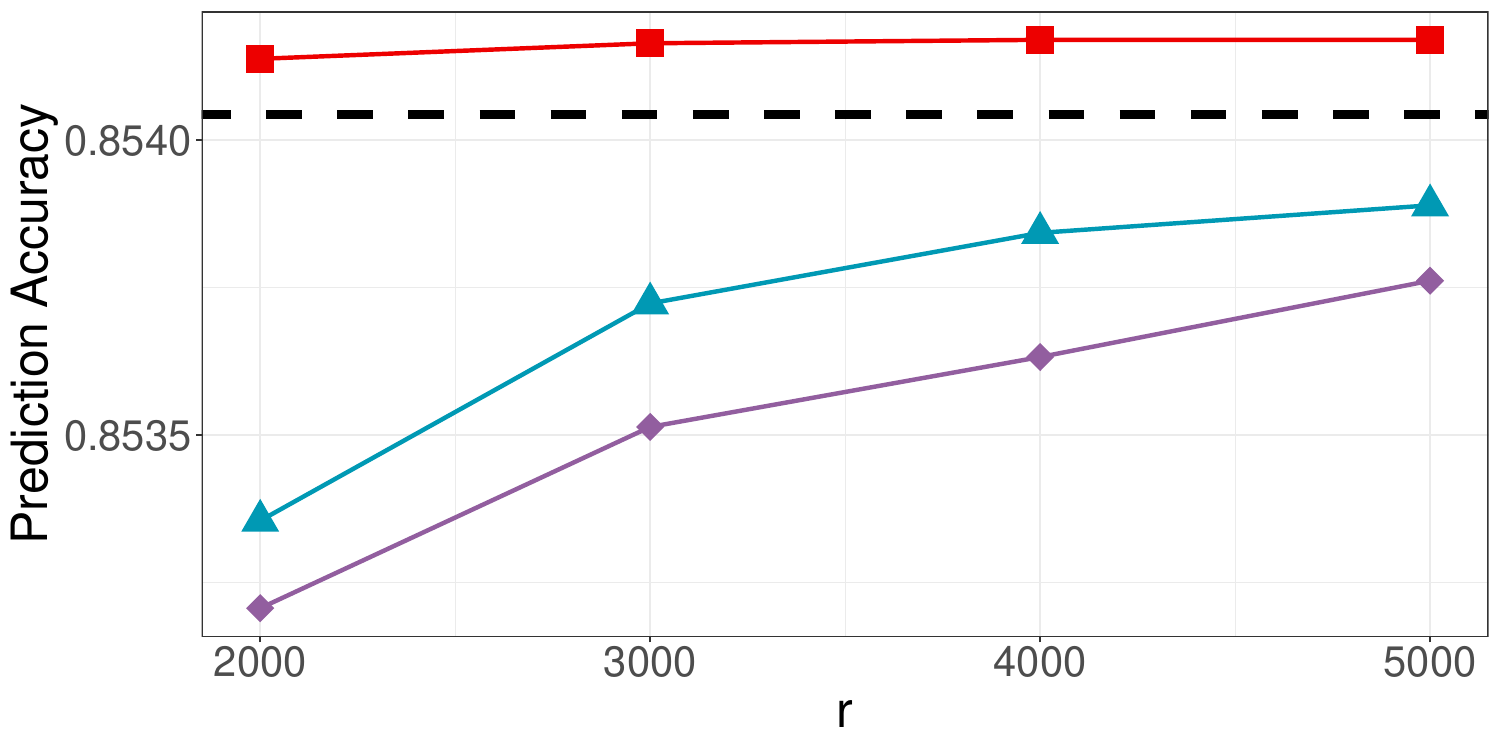}
		\caption{DWD}
	\end{subfigure}
	\caption{ The log MSE and prediction accuracy of the subsample based logistic regression (top panel) and DWD (bottom panel) for air quality dataset. MROSS (Algorithm~\ref{alg:RBS1}, $\blacksquare$) is compared with UNIF ($\blacklozenge$), OSMAC ($\blacktriangle$), MSCLE ($\bullet$). The dash line stands for the prediction accuracy based on the full sample estimator.
	}
	\label{fig:PM25.logistic}
\end{figure}

\subsection{Supersymmetric benchmark dataset}

Supersymmetry is a critical component of theoretical particle physics, such as string theory, and has the great potential to provide some solutions to the nature of dark matter \citep{Baldi2014searching}. One of the main tasks of the Supersymmetric benchmark (SUSY) dataset provided in the UCI machine learning repository \citep{misc_susy_279} is to distinguish between a signal process that produces supersymmetric particles and a background process that does not.
Within the SUSY dataset, $5,000,000$ instances are tracked and recorded including both supersymmetric and non-supersymmetric events. Each record represents a hypothetical collision between particles. There are 18 features included in the dataset, including eight kinematic properties features such as energy levels, and momenta. The other ten features are functions of the eight aforementioned features that physicists design to help discriminate between the two classes. 


The dataset has been partitioned into a training set with the first $4,500,000$ examples and a testing set with the {remaining  $500,000$ examples}.  
A logistic regression and DWD are fitted based on the subsample obtained by different subsampling strategies as in Section~\ref{subsec:forest}.
All the $18$ features have been used to obtain the classifiers.
Figure~\ref{fig:susy.logistic} reports the MSE and prediction accuracy with the same settings described in Section~\ref{subsec:forest}.

\begin{figure}[h!]
    \centering\spacingset{1}
\begin{subfigure}{.45\textwidth}
    \centering
\includegraphics[width=1\linewidth]{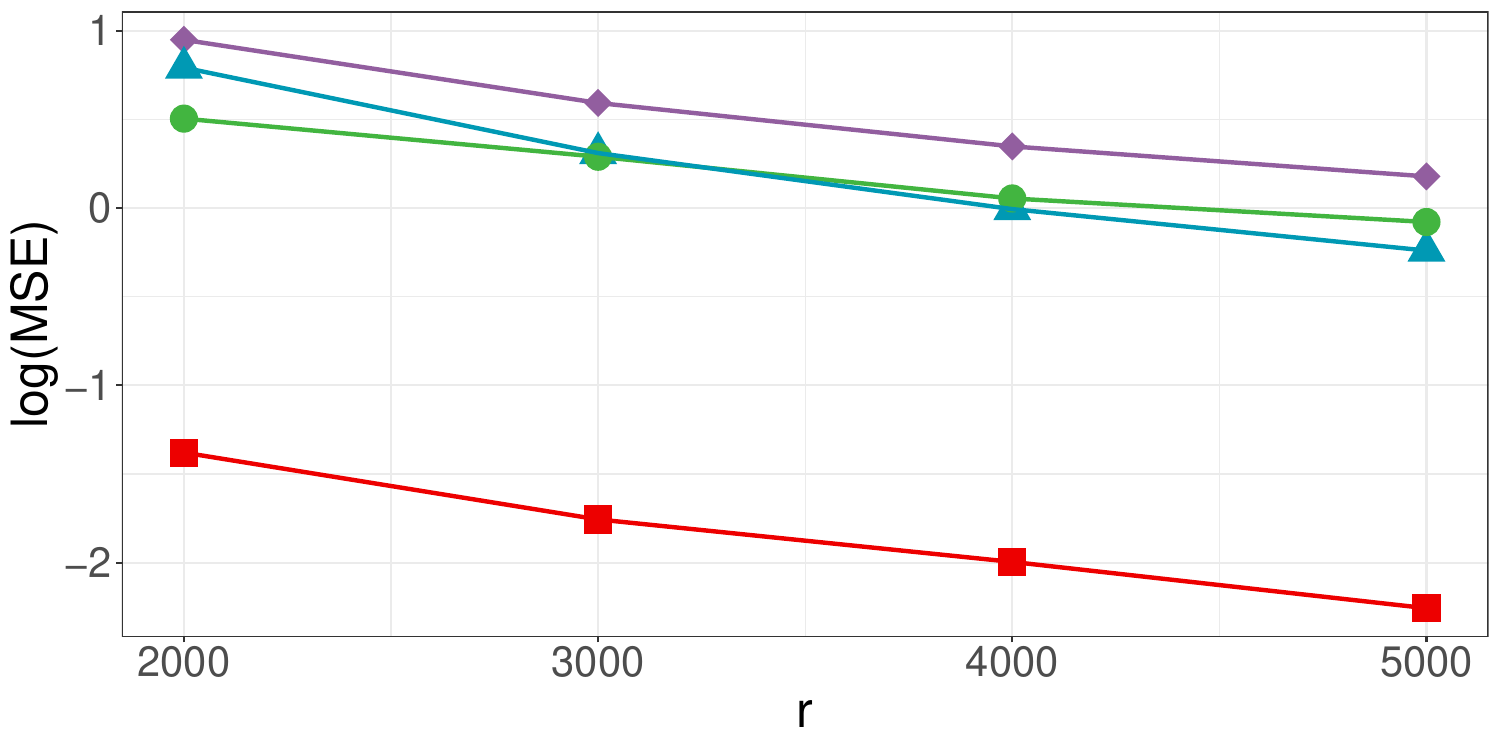}
  \caption{logistic}
\end{subfigure}
\begin{subfigure}{.45\textwidth}
     \centering
\includegraphics[width=1\linewidth]{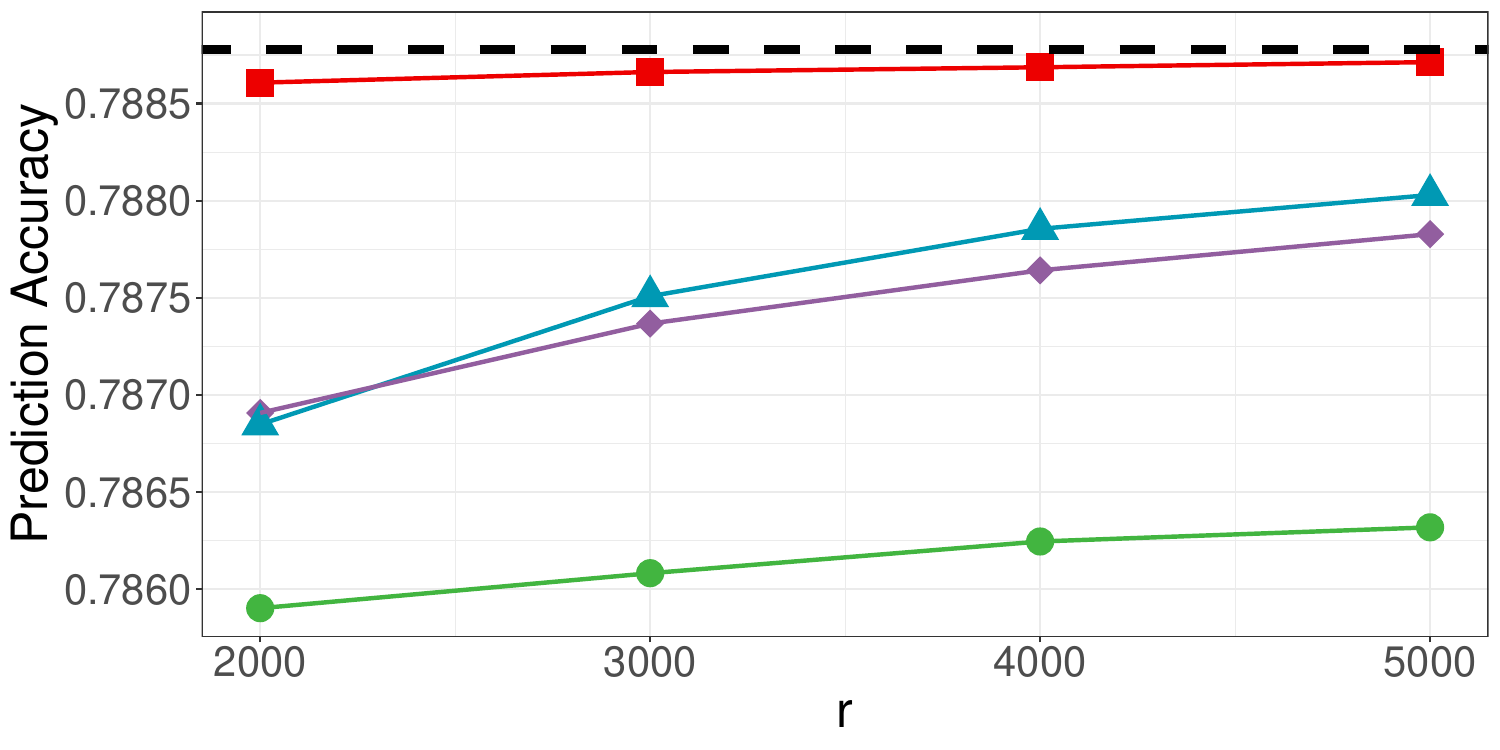}
  \caption{logistic}
\end{subfigure}

\begin{subfigure}{.45\textwidth}
    \centering
\includegraphics[width=1\linewidth]{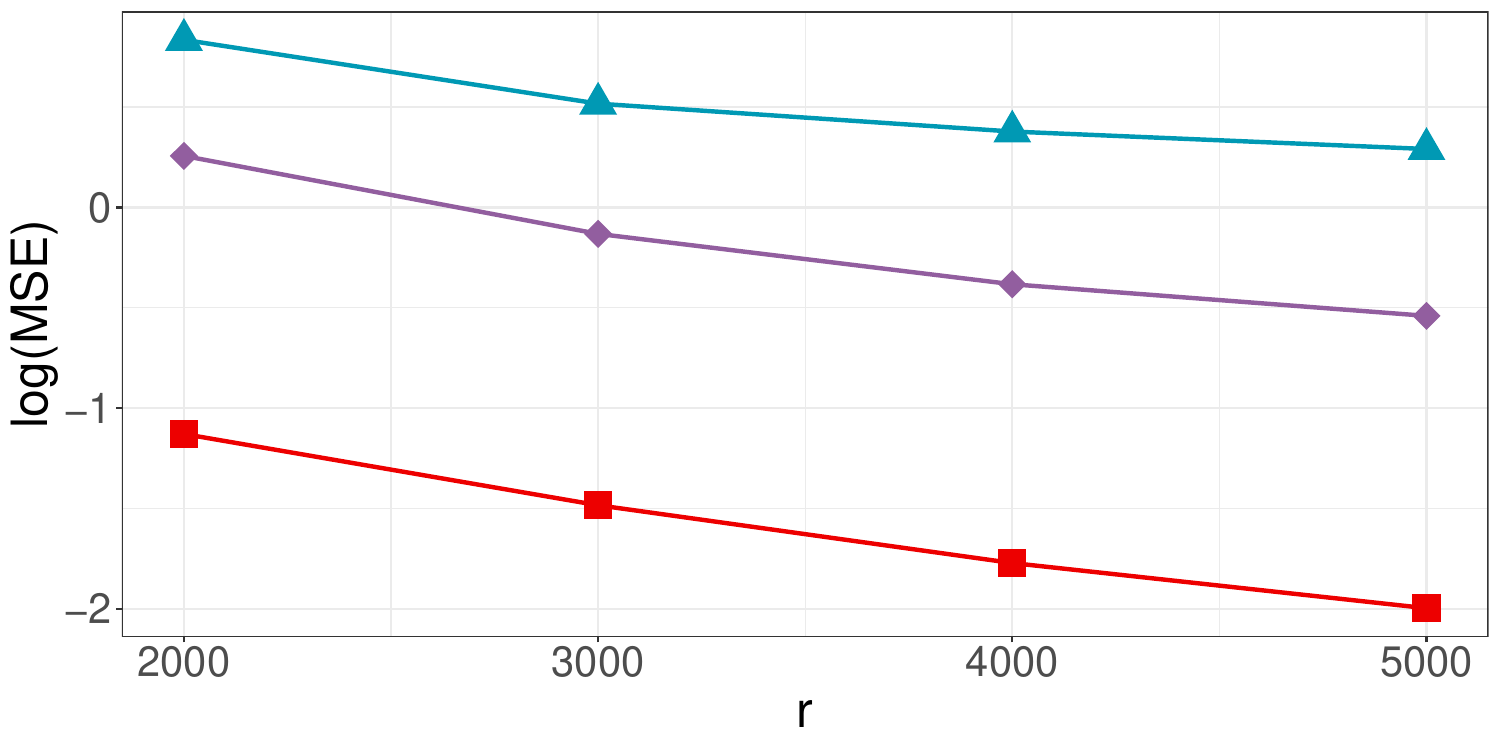}
  \caption{DWD}
\end{subfigure}
\begin{subfigure}{.45\textwidth}
     \centering
\includegraphics[width=1\linewidth]{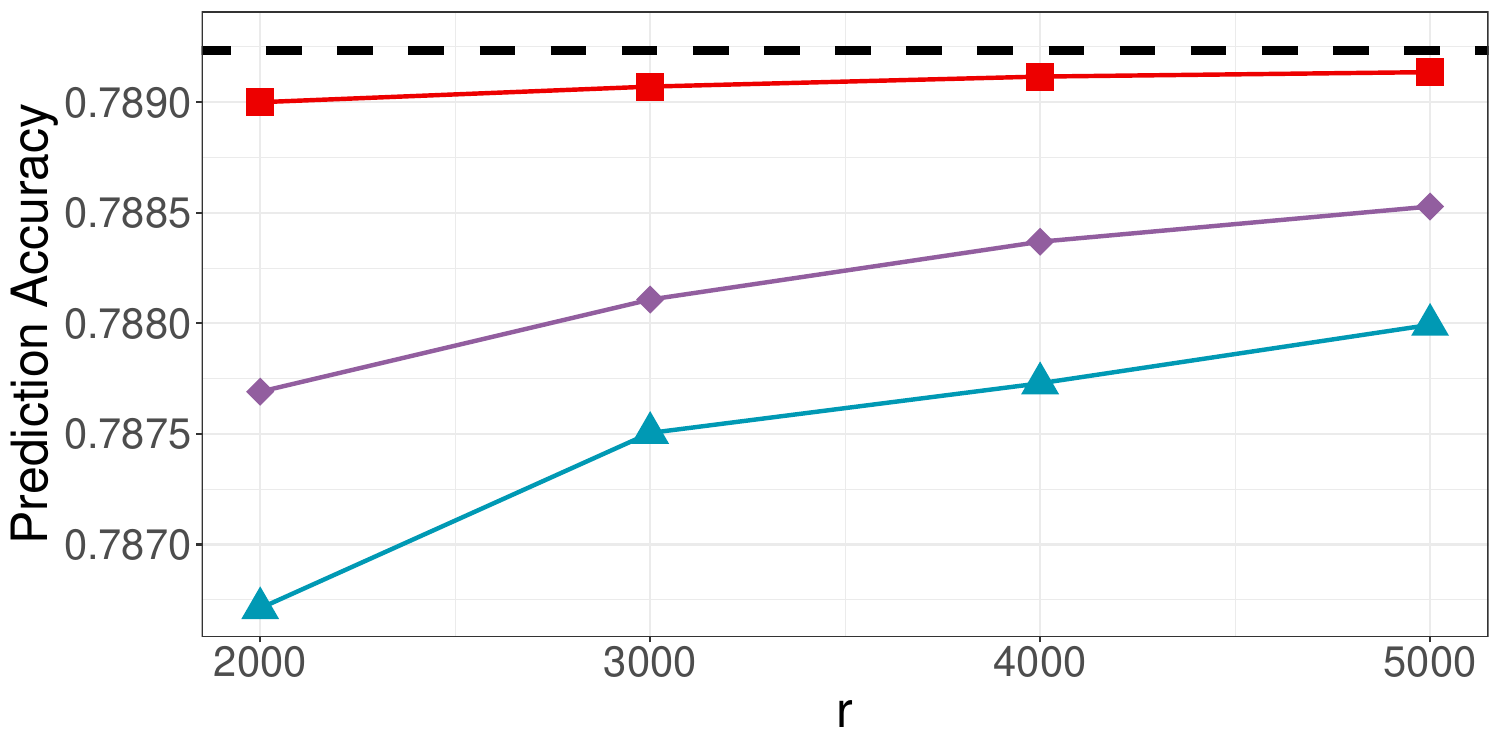}
  \caption{DWD}
\end{subfigure}
    \caption{{The log MSE and prediction accuracy of the subsample based logistic regression (top panel) and DWD (bottom panel) for the SUSY dataset. MROSS (Algorithm~\ref{alg:RBS1}, $\blacksquare$) is compared with UNIF ($\blacklozenge$), OSMAC ($\blacktriangle$), MSCLE ($\bullet$). The dash line stands for the prediction accuracy based on the full sample estimator. 
    }}
    \label{fig:susy.logistic}
\end{figure}

\section{Concluding remarks}\label{sec:conclusion}
  \def\theequation{5.\arabic{equation}}	
  \setcounter{equation}{0}

Ordinary subsampling and subdata selection aim to find the most informative data points that benefit data exploration or statistical inference (such as parameter estimation, and prediction).
Direct analysis of the selected subdata may also generate a large variance estimator due to the ignorance of the information from the unsampled data.
In this work, we go beyond the classical subsampling/subdata selection framework and involve the summary information from the unsampled data points. 
We propose MROSS, a multi-resolution optimal subsampling algorithm,  to combine the information from the selected informative data points and unsampled data which is characterized by summary statistics.
We establish convergence results and construct confidence intervals for the estimator obtained by the proposed algorithm.
Our theoretical results echo the numerical findings on both synthetic and real datasets.
It should be highlighted that our algorithm not only improves the statistical efficiency of the optimal subsample based estimator but also results in a more efficient estimator for general subsampling strategies.
The results of Section~\ref{sec:RB} can be easily generalized to  M- and Z-estimators and have the potential to solve large-scale regression problems.
Although our analysis has focused on linear classification problems, we believe that the proposed algorithm can be applied to more complex
classification problems. For example, when facing the nonlinear classification problem, one can embed the feature in the functional space spanned by some spline
bases such as B-splines. Then the idea of linear classification can be extended to the nonlinear classifier.

\bigskip

\textbf{Acknowledgements.} 
This work was partially supported by the DFG Research unit 5381 {\it Mathematical Statistics in the Information Age}, project number 460867398.
The authors would like to thank Birgit Tormöhlen, who wrote parts of this paper with considerable technical expertise. 

\bigskip

\begin{center}
{\large\bf SUPPLEMENTARY MATERIAL}
\end{center}
\vspace{-0.35cm}

All technical proofs are included in the online supplementary material.

\vspace{-0.35cm}

\bibliographystyle{apalike}
\bibliography{reference}

\end{document}